\documentclass[a4paper,fleqn,usenatbib]{mnras}

\usepackage[T1]{fontenc}
\usepackage{ae,aecompl}

\usepackage{graphicx}	% Including figure files
\usepackage{amsmath}	% Advanced maths commands
\usepackage{amssymb}	% Extra maths symbols

\usepackage{hyperref}	% Hyperlinks
\hypersetup{colorlinks=true,linkcolor=blue,citecolor=blue,filecolor=blue,urlcolor=blue}
\def\pasa{PASA}%
\def\aj{AJ}%
\def\apj{ApJ}%
\def\apjl{ApJ}%
\def\apjs{ApJS}%
\def\ao{Appl.~Opt.}%
\def\aap{A\&A}%
\def\icarus{Icarus}%
\def\mnras{MNRAS}%
\def\pasa{PASA}%
\def\pasp{PASP}%
\def\pasj{PASJ}%
\def\nat{Nature}%
\title[Protoplanets in self-gravitating discs]
{\scalebox{0.93}[1.0]{The diverse lives of massive protoplanets in self-gravitating discs}}

\author[Stamatellos, D. \& Inutsuka, S.]{
Dimitris Stamatellos$^{1}$\thanks{E-mail: dstamatellos@uclan.ac.uk}
\& Shu-ichiro Inutsuka$^{2}$
\\
$^{1}$Jeremiah Horrocks Institute for Mathematics, Physics \& Astronomy, University of Central Lancashire, Preston, PR1 2HE, UK\\
$^{2}$Department of Physics, Nagoya University, Chikusa-ku, Nagoya 464-8602, Japan\\
}

\date{Accepted 2017. Received 2017; in original form 2017}

\pubyear{2017}

\begin{document}
\label{firstpage}
\pagerange{\pageref{firstpage}--\pageref{lastpage}}
\maketitle

\begin{abstract}

Gas giant planets may form early-on during the evolution of protostellar discs, while these are relatively massive. We study how Jupiter-mass  planet-seeds (termed {\it protoplanets}) evolve in { massive, but gravitationally stable ($Q\stackrel{>}{_\sim}1.5$)},  discs using radiative hydrodynamic simulations. We find that the protoplanet initially migrates inwards rapidly, until it opens up a gap in the disc. Thereafter, it either continues to migrate inwards on a much longer timescale  or starts migrating outwards. Outward migration occurs when the protoplanet resides within a gap with gravitationally unstable edges, as a high fraction of the accreted gas is high angular momentum gas  from outside the protoplanet's orbit.  The effect of radiative heating from the protoplanet is critical in determining the direction of the migration and the eccentricity of the protoplanet. Gap opening is facilitated by efficient cooling that may not be captured by the commonly used $\beta$-cooling approximation. The protoplanet initially accretes at a high rate ($\sim 10^{-3} {\rm M_J\ yr^{-1}}$), and its  accretion luminosity could be a few tenths of the host star's luminosity, making the protoplanet easily observable (albeit only for a short time). Due to the high gas accretion rate, the protoplanet generally grows above the deuterium-burning mass-limit. Protoplanet radiative feedback reduces its mass growth so that its final mass is near the brown dwarf-planet boundary. The fate of a young planet-seed is diverse and could vary from a gas giant planet on a circular orbit at a few AU from the central star to a brown dwarf on an eccentric, wide orbit. 

\end{abstract}


%%%%%%%%%%%%%%%%%%%%%%%%%%%%%%%%%%%%%%%%%%%%%%%%%%

\section{Introduction}

Observational advances over the last 20 years have made possible the discovery of a large number of exoplanets (see the {\it ``Extrasolar Planets Encyclopaedia"}; http://exoplanet.eu). Most of these exoplanets are gas giants, as these are easier to detect with the main current  methods (radial velocity, transits) than small, rocky planets like Earth. 

 Two theoretical models have been proposed for the formation of gas giant planets, (i)  core accretion, and (ii)  gravitational fragmentation of protostellar discs. 
 
 In the core accretion model \citep{Safronov:1972a,Goldreich:1973a,Mizuno:1980a,Bodenheimer:1986a,Pollack:1996a} the cores of giant planets form in circumstellar discs by coagulation of dust particles to progressively larger objects, until a core with a few Earth masses has formed. Such a core  subsequently accretes an  envelop of gas from the disc. One of the main drawbacks of this model is that, in its standard formulation, requires a few million years to form gas giants, a timescale which may be long when compared to the observed lifetimes of circumstellar discs \citep[on the order of a few Myr;][]{Haisch:2001a,Hernandez:2008a,Muzerolle:2010a}.  Moreover, this model seems unlikely to be able to produce massive planets on wide orbits \citep{Kraus:2008a, Kraus:2014a, Marois:2008a, Faherty:2009a, Ireland:2011a,Kuzuhara:2011a,Kuzuhara:2013a, Aller:2013a,Bailey:2014a,Rameau:2013b,Naud:2014a,Galicher:2014a}, like e.g. the planets around HR8799.
 
 A second way to form gas giants is by  gravitational fragmentation of protostellar discs \citep{Kuiper:1951a,Cameron:1978a, Boss:1997a}. This model circumvents the timescale problem of the core accretion theory, as planets form on a dynamical timescale, i.e. within a few $10^3$~yr. One of the main issues relating to this model is whether protostellar discs are able to fragment or not. There are two criteria for disc fragmentation,  (i) the Toomre criterion \citep{Toomre:1964a} that postulates that the disc must be massive enough in order gravity to dominate over the local thermal and centrifugal support, and (ii) the Gammie criterion \citep{Gammie:2001a,Johnson:2003a,Rice:2005a} that asserts that the disc must be able to cool fast enough, i.e. on a dynamical timescale \cite[see][for an alternative view]{Takahashi:2016a}. There has been relative consensus that these two criteria can be  satisfied at distances $>50-100$~AU from the star hosting the disc; therefore discs can fragment at such distances \citep{Rafikov:2005a,Matzner:2005a,Whitworth:2006a,Boley:2006a,Stamatellos:2008a}. This fact, together with the discovery of exoplanets on wide orbits, may suggest two modes of gas giant planet formation: core accretion that forms planets close to the host star, and disc fragmentation that forms planets on wide orbits \citep{Boley:2009a}.
  
Many studies suggesting that planets form by disc fragmentation,  refer to the {\it initial} mass of the fragments produced. The initial fragment mass \citep[e.g.][]{Stamatellos:2009a,Stamatellos:2009d} is determined by the opacity limit for fragmentation, i.e. the minimum mass a fragment produced by fragmenting gas may have \citep[e.g.][]{Low:1976a,Rees:1976a, Silk:1977a,Boss:1988a, Masunaga:1999a,Whitworth:2006a}. However, once a fragment forms in a disc it evolves: it migrates within the disc,  accretes gas from the disc opening a gap, and inevitable increases in mass \citep{Zhu:2012a, Hall:2017a}. 
 
Previous  studies \citep{Stamatellos:2009a,Stamatellos:2009d, Kratter:2010b, Zhu:2012a} suggest that the most likely outcome of disc fragmentation is brown dwarfs.  \cite{Stamatellos:2009a} performed an  ensemble of 12 simulations of self-gravitating, fragmenting discs producing in total $\sim 100$ objects; 67\% of these objects are brown dwarfs, 30\% are low-mass hydrogen burning stars and only 3\% are planets.  Additionally, they point out that the objects that end up as planets ($<13~{\rm M_J}$) they are ejected from the disc quickly after their formation, thus  avoiding any subsequent gas accretion.  These ejected planets contribute to the suspected large population of free-floating planets. \cite{Stamatellos:2009a} also argue that objects that remain in the disc, although  they are planets when they form  (i.e. their mass is below $\sim13~{\rm M_J}$), they grow in mass to become either brown dwarfs or hydrogen-burning stars.

Another problem of the disc fragmentation theory is whether planets forming early on in protoplanetary discs are able to avoid fast inward migration towards their parent star. Previous studies suggest the giant planets that form in relatively massive protoplanetary discs migrate inwards rapidly, on a timescale reminiscent of Type I migration,  without opening up a gap in the disc \citep{Vorobyov:2005a, Vorobyov:2006a,Baruteau:2011a,Michael:2011a, Malik:2015a}. However, this has been questioned by the studies of \cite{Lin:2012b} and \cite{Cloutier:2013a} that show that planets may migrate outwards due to gravitational edge mode instabilities. \cite{Stamatellos:2015a}  showed that planets are able to open gaps as they become more massive by accreting gas from the disc; by doing so, their inward migration slows down or even changes to outward migration.

An interesting variant of the gravitational fragmentation theory is the tidal downsizing hypothesis \citep[see review by][]{Nayakshin:2017a}, in which clumps that form by disc fragmentation contract slowly while they  migrate inwards. They may eventually get tidally disrupted, losing mass and possibly leaving behind a solid core. \cite{Nayakshin:2017b} find that the final fate of a clump forming in the disc depends on how efficiently it cools and identifies two possible outcomes for the clump:  it becomes either a brown dwarf on a wide orbit or a planet within 20 AU from the central star. He concludes that even though disc fragmentation may commonly happen,  only a small fraction of stars may have giant planets on wide orbits; this is corroborated by observations \citep{Brandt:2014a,Galicher:2016a,Vigan:2017a} which show that only 1-10\% of star host gas giant planets on wide orbits.

Recent ALMA observations of the disc of the young star HL Tau \citep{ALMA-partnership:2015a} have revealed the presence of multiple gaps at mm wavelengths. These gaps may be carved either by planets \citep{Dipierro:2015b} or be due to other processes \citep{Takahashi:2014a,Gonzalez:2015b,Takahashi:2016b}. HL Tau  is an extremely young object \citep{Greaves:2008a} that shows signs of outflows in the form of jets and infall from its parent cloud. Its disc is relatively massive \citep[0.1-0.15~M$_{\sun}$;][]{Testi:2015a} and presumably is still being fed with gas by its ambient cloud. Such gaps have been observed in other discs e.g. in TW Hydra \citep{Andrews:2016a} and in  HD 163296 \citep{Isella:2016b}. Numerical simulations using resistive MHD have suggested that discs form at an early stage during star formation and that they may be relatively massive in comparison to the mass of their host stars \citep{Machida:2010c,Machida:2011c,Machida:2011a,Machida:2014a}.  ALMA  observations strongly support  that discs exist from the Class 0 phase \citep{Tobin:2016a}. They appear to be relatively massive and extended,  and they appear to have spiral arms indicative of gravitational instabilities \citep{Perez:2016a,Tobin:2016a}. The possible presence of  planets in such young discs raises  the exciting possibility that planets may form much faster than it has been previously thought and therefore their early evolution, no matter how they have formed, will occur within a relatively massive disc.

In this paper we study the evolution of a Jupiter-like planet-seed, referred to  as the {\it protoplanet}, which finds itself within a protostellar disc that is massive enough for its self-gravity to be important for its evolution. 
The assumed disc is close to being gravitationally unstable, meaning that weak spiral arms may develop, but the disc does not fragment. We use the term protoplanet even though we refer to an object that only starts off as planet in the disc (irrespective of its formation mechanism). This object may accrete mass to eventually stop being a planet and become a brown dwarf ($m>13~{\rm M_J}$). {Our aim is to investigate the whether such an object may actually survive as a planet by performing a set of numerical experiments. We expand on the work of \cite{Stamatellos:2015a}, exploring a wider parameter space. More specifically we investigate the effect of different disc viscosity, dust opacities, the orientation of the protoplanet's orbit, the protoplanet's initial orbital radius, and the effect of the radiation from the host star and the protoplanet itself. We therefore examine in detail how robust are the results of \cite{Stamatellos:2015a}. Additionally, we explore the reasons behind the different results with previous studies, comparing with simulations using a parameterized prescription for the disc cooling ($\beta$-cooling approximation).}

We describe the initial condition of the simulations performed in Section 2, and in Section 3 the hydrodynamic method used and its ingredients (radiative transfer method, opacities, radiative feedback). We present the first set of simulations performed and their results in Section 4, commenting in detail  on how fast the protoplanet grows in mass, and how its orbital properties (semi-major axis, eccentricity) change during its early evolution. In Section 5, we compare the results of our simulations with those of previous studies that employ the $\beta$-cooling approximation. In Section 6, we examine in detail the evolution of protoplanets at different radii from the central star. In Section 7, we present possibly the more realistic set of simulations, in which both radiative feedback from the planet and the host star  are taken into account.  Finally, in Section 8 we discuss the implications of this work for planet formation and evolution studies.

 \section{Computational setup}
 
 We assume a star with initial mass $M_\star=1\,{\rm M}_{\sun}$ that is attended by a protostellar disc with mass $M_{_{\rm D}}=0.1~{\rm M}_{\sun}$ and initial  radius $R_{_{\rm D}}=100$~AU. The disc is modelled by $10^6$ SPH particles. { The discs that we model are chosen so that they are not gravitationally unstable ($Q\stackrel{>}{_\sim}1.5$). Therefore, they are not expected to develop strong spiral arms (unless these are induced from the protoplanet).}
 
 The disc initial surface density is 
\begin{equation}
\Sigma_{_0}(R)=\Sigma(1 {\rm AU})\,\left(\frac{R}{\rm AU}\right)^{-1}\,,
\end{equation}
and the disc temperature
\begin{equation}\label{EQN:TBG}
T_{_0}(R)=250\,{\rm K}\,\left(\frac{R}{\rm AU}\right)^{-3/4}+10~{\rm  K}\,,
\end{equation}
where $\Sigma(1 {\rm AU})$ is determined by the disc mass and radius, and $R$ the distance from the central star measured on the disc midplane. The  above density and temperature profiles  are consistent with observations of late-phase (T Tauri) discs but the properties of discs in the early-phase are uncertain \citep{MacFarlane:2017a}.  \cite{Andrews:2009b} observed a sample of circumstellar discs in Ophiuchus and estimated a  disc surface density that drops with the distance from the host star as $p\approx 0.4-1.0 $. In Taurus-Auriga and Ophiuchus-Scorpius star formation regions,  \cite{Andrews:2007a} find a  median $p\approx 0.5$; they also find a temperature profile that drops with distance as $q\approx 0.4-0.74$. \cite{Osterloh:1995a} find that the disc temperature drops with radius, with an exponent $q\approx 0.35-0.8$. { We note though that the profile defined in Equation~\ref{EQN:TBG} is the initial disc temperature profile but as the disc evolves it acquires a temperature profile self-consistently within the radiative hydrodynamic simulation.}

The disc is allowed to relax, i.e. to evolve without a protoplanet,  for 3 outer orbital periods ($\sim 3$~kyr). A protoplanet with a mass of $M_p=1~{\rm M_J}$ is then embedded in the disc. The protoplanet's initial  orbital velocity is set the same as the orbital velocity of the local gas, i.e. Keplerian (including the contribution from the disc mass within the radius of the protoplanet). We additionally assume an initially circular orbit ($e_i=0$).

The choice of the initial disc mass ($0.1~{\rm M}_{\sun}$) is intentionally conservative. If the protoplanet has formed by disc fragmentation then numerical studies indicate that a disc needs to have mass at least a few tenths ${\rm M_{\sun}}$ \citep[e.g. the simulations of][suggest that a disc around a $0.7~ {\rm M}_{\sun}$ star needs to be more massive than $0.25~{\rm M}_{\sun}$ in order to fragment]{Stamatellos:2011d}. 

The choice of the  initial mass of the protoplanet (1~${\rm M_J}$) is also conservative. The minimum mass of an object forming by gravitational fragmentation is set by the thermodynamics of gas, as in order for a condensation to collapse, the energy delivered by compression needs to be efficiently radiated away. The energy loss of a fragment is determined by its opacity, so  this is usually referred to  as the {\it opacity limit}  for fragmentation \citep[e.g.][]{Whitworth:2006a,Whitworth:2007a}. Theoretical studies estimate a minimum mass for fragmentation to be $1-10~{\rm M_{\rm J}}$ \citep{Low:1976a,Rees:1976a,Silk:1977a,Boss:1988a, Masunaga:1999a,Boyd:2005a,Whitworth:2006a,Boley:2010b,Kratter:2010b,Forgan:2011b,Rogers:2012a}. We therefore study a system in which both the disc mass and the initial protoplanet mass are close to the lower limits that they may have, if the protoplanet has formed by disc fragmentation. However, our study is general and applies even for planets that have formed by core accretion.

\section{Computational methods}
\label{sec:methods}

\subsection{Gas hydrodynamics and radiative transfer}
\label{sec:methods.rt}

We use the SPH code {\sc seren} \citep{Hubber:2011b, Hubber:2011a} to treat the disc thermodynamics. The code  uses an octal tree to compute gravity and find neighbours, multiple particle timesteps for optimisation, and a 2$^{\rm nd}$ Runge-Kutta integration scheme. We use a time-dependent artificial viscosity \citep{Morris:1997a}  with parameters $\alpha_{\min}=0.1$, $\alpha_{\max}=1$ and $\beta=2\alpha$,  so as to reduce artificial shear viscosity. The chemical and radiative processes that determine the disc temperature are treated with the diffusion approximation of \cite{Stamatellos:2007b} and \cite{Forgan:2009b}. 
The net radiative heating rate for an SPH particle $i$ is 
\begin{equation} 
\label{eq:radcool}
\left. \frac{du_i}{dt} \right|_{_{\rm RAD}} =
\frac{\, 4\,\sigma_{_{\rm SB}}\, (T_{_{\rm A}}^4-T_i^4)}{\bar{{\Sigma}}_i^2\,\bar{\kappa}_{_{\rm R}}(\rho_i,T_i)+{\kappa^{-1}_{_{\rm P}}}(\rho_i,T_i)}\,,
\end{equation}
where $u_i$ is the specific internal energy of the  particle,  and $\rho_i$, $T_i$, its density and temperature, respectively.
The positive term on the right hand side represents heating by the various radiation sources (star, protoplanet, and pseudo-background radiation field; see  Section~\ref{sec:methods.feedback}), and ensures that the gas and dust cannot cool radiatively below the pseudo-background temperature $T_{_{\rm A}}$. $\sigma_{_{\rm SB}}$ is the Stefan-Boltzmann constant, $\bar{{\Sigma}}_i$ is the mass-weighted mean column-density, and  $\bar{\kappa}_{_{\rm R}}(\rho_i,T_i)$ and ${\kappa_{_{\rm P}}}(\rho_i,T_i)$ are suitably adjusted Rosseland- and Planck-mean opacities \citep[see][for details]{Stamatellos:2007b}. { The method takes into account compressional heating, viscous heating, heating by the background radiation field, and radiative cooling/heating.}  The method has previously applied to disc studies \citep{Stamatellos:2008a,Stamatellos:2009d,Stamatellos:2011d} and has given similar results with grid-based computational methods \citep{Boley:2006a,Cai:2008a}. The method used has been shown that may overestimate the column density through which the disc gas cools \citep{Wilkins:2012a,Young:2012a, Lombardi:2015a} by a factor of a few; however, considering that the dust opacity in discs is not known accurately enough, this limitation of the method is not critical for the results presented in this paper.  The detailed treatment of radiative transfer is important when considering how the gas fragments but it is not expected to be crucial (at least qualitatively)  for the issues relating to the mass growth and migration of protoplanets already formed in the disc that are discussed in this paper.

\subsection{Opacities}
\label{sec:opacities}
\begin{figure}
\centerline{
\includegraphics[width=0.75\columnwidth,angle=-90]{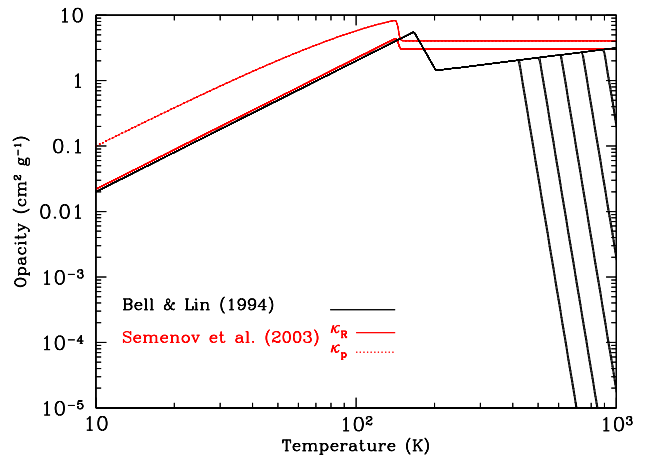}}
\caption{The two types of opacity used for the simulations presented in this paper: the Bell \& Lin (1994) opacities (black lines; different lines correspond to different densities), and the Semenov et al. (2003) opacities (red lines) that have been proposed  for protostellar discs.}
\label{fig:opacity}
\end{figure}

We use two prescriptions for the opacity the disc (see Figure~\ref{fig:opacity}):
\begin{enumerate}
\item The \cite{Bell:1994a} opacity which is parameterized as
\begin{equation} 
\label{eq:opacity}
{\kappa}(\rho,T) =\kappa_{_{\rm 0}}\ \rho^l\ T^m\,,
\end{equation}
where $\kappa_{_{\rm 0}}$, $l$, $m$ are constants that depend on
the species and the physical processes that contribute to the opacity at different
densities and temperatures (note that Planck mean and Rosseland mean opacities are assumed to be the same). For the temperatures in the discs we simulate here (up to $\sim1,000$~K) the opacity is due to dust grains that at low temperatures are coated with ice, which evaporates at higher temperatures. The dust grains  start to evaporate at around 1,000~K. 

\item The \cite{Semenov:2003a} model includes similar physical processes but the dust considered includes spherical and aggregate particles of various sizes that consist of ice, organics, troilite, silicates and iron. This composition is thought to be more appropriate for the conditions found in protoplanetary discs.
\end{enumerate}

The Semenov et al.  Rosseland mean opacity is similar to the Bell \& Lin one at temperatures up to $\sim100$~K but higher  at larger temperatures by up to a factor of 2. At such temperatures the optically thick parts of the disc cool less efficiently when using the Semenov et al. opacity. The Semenov et al.  Planck mean opacity is larger by a factor of  5 than the Bell \& Lin one, which means that the optically thin parts of the disc cool more efficiently. 

{ The above two sets of opacities are widely used in the literature. However, the composition of dust in discs is rather uncertain, especially since significant  dust growth may occur, lowering the opacity. It is unclear whether dust growth is significant at the early disc phase that we study here.}

%%%%%%%%%%%%%%%%%%%%%%%%%%%%%%%%%%%%%%%%%%%%%
%%%%%%%%%%%%%%%%%%%%%%%%%%%%%%%%%%%%%%%%%%%%%

\subsection{Sinks}

Sink particles are used to represent the central star and the protoplanet \citep{Bate:1995a}. Sink  particles interact with the rest of the computational domain  only through their gravity (and their  luminosity if their radiative feedback in considered, see below). The sink radius of the central star is  $R_{\rm sink,\star}=0.2$~AU, and the sink radius of the protoplanets is $R_{\rm sink,p}=0.1$~AU. This value is chosen to be  smaller than the Hill radius of the protoplanet, i.e. the region around it where the its gravity dominates over the gravity of the star:
  \begin{equation}
 R_{\rm sink, p}<R_{\rm H}=R\left(\frac{M_p}{3M_\star}\right)^{1/3}\,.
 \end{equation}
For example, when a Jovian planet is at 50~AU away from the central star then its Hill radius is $ R_{\rm H}\sim3.5$~AU. The Hill radius may increase as the protoplanet accretes material from the disc or it may decrease as the protoplanet moves closer to the central star. 

Gas particles accrete onto a  sink when they are within the sink radius and bound to the sink \citep{Hubber:2011a}. {The accretion rate onto the sink is computed by calculating the increase of the sink mass over a time interval that is comparable to the dynamical time of each particle at the edge of of the sink}.

\subsection{Radiative feedback from the  protoplanet and the star}
\label{sec:methods.feedback}

The radiation feedback from the protoplanet  is taken into account (in a subset of  the simulations) using the method  described in \cite{Stamatellos:2015a}. We use the method of  \cite{Stamatellos:2011a} \& \cite{Stamatellos:2012a} that invoke a pseudo-ambient radiation field  with temperature $T_{_{\rm A}}^{\rm planet}({\bf r})$  due to radiation from  the protoplanet \citep[see also][]{Stamatellos:2007b,Stamatellos:2009a}.  This pseudo-ambient temperature sets the minimum temperature that the gas can attain if it cools radiatively.
The contribution from the protoplanet is set to
 \begin{eqnarray}
 \label{eq:tplanet}
T_{_{\rm A}}^{\rm planet}({\bf r})&=&\left(\frac{L_p}{16\,\pi\,\sigma_{_{\rm SB}}\,|{\bf r}-{\bf r}_p|^2}\right)^{1/4}\,,
\end{eqnarray}
where ${\bf r}$ is the position  on the disc midplane, and  $L_p$ and  ${\bf r}_p$ are the luminosity and position of the  protoplanet, respectively. { The above approximation may overestimate the effect of the feedback in optically thick regions of the disc, that are well-shielded from the protoplanet \citep[see discussion in][]{Mercer:2017a}.}

The luminosity  of the protoplanet  is  set to
\begin{equation}
\label{eq:lplanet}
L_p=f\frac{G M_p \dot{M}_p}{R_{\rm acc}}\,,
\end{equation}
where $M_p$ is the mass of the protoplanet,  $\dot{M}_p$ is the accretion rate on to it, and $R_{\rm acc}$ the  assumed accretion radius.  $f=0.75$ is the fraction of the accretion energy that is radiated away at the photosphere of the protoplanet, rather than being expended driving jets  \citep{Machida:2006a,Offner:2009a}, or deposited within the protoplanet \citep[e.g.][]{Baraffe:2017a}. 

The exact amount of the energy radiated away from the protoplanet depends on the detailed properties of the accretion shock around the protoplanet \citep{Zhu:2015a,Marleau:2017a,Szulagyi:2017b,Mordasini:2017a}. If this protoplanet has formed by gravitational instabilities in the disc then the accretion happens onto the second hydrostatic core \citep[e.g.][]{Larson:1969a,Stamatellos:2009d}. The radius of the second core
is uncertain: $\sim~1-20~{\rm R}_{\sun}$ \citep{Masunaga:2000a, Tomida:2013a, Vaytet:2013a, Bate:2014a, Tsukamoto:2015a}.  Here we assume $R_{\rm acc}=3~{\rm R}_{\sun}$, but the results of this work are not  sensitive on this assumption. The accretion luminosity of the protoplanet is significant due to the relatively high accretion rate onto it during the initial stages of its evolution and can dominate over the stellar luminosity in the disc region around the protoplanet \citep{Owen:2014b, Stamatellos:2015a,Montesinos:2015a}, especially for the Jupiter-like planet-seeds in high-mass discs that we study here. 

If the protoplanet has formed by disc fragmentation, on top of the significant accretion luminosity, it will also radiate energy due to its contraction. This energy release is on the order of 0.1~L$_{\sun}$ \citep{Inutsuka:2010a}, which is similar to the one expected from accretion. Therefore, the protoplanet's luminosity is deemed to play an important role in its evolution  \citep{Nayakshin:2013a,Stamatellos:2015a,Benitez-Llambay:2015a}.
{ We do not take this radiation explicitly into account in the simulations presented in this paper but we have performed runs with smaller  $R_{\rm acc}$ resulting in higher luminosity for the protoplanet. These runs give qualitatively similar results to the ones we discuss in this paper.}

The radiation feedback from the star  is taken into account (in a subset of the simulations)  by invoking an additional pseudo-ambient temperature, 
 \begin{eqnarray}
 \label{eq:startemp}
T_{_{\rm A}}^{\star}({\bf r})&=&T({\rm 1AU})\,\left(\frac{R}{\rm AU}\right)^{-3/4}+10~{\rm  K}\,,
\end{eqnarray}
where $R$ is the distance from the central star on the disc midplane, and $T_0$ the temperature at $R=1$~AU from the star. The total pseudo-ambient temperature (due to the central star and the protoplanet) is then set to 
 \begin{eqnarray}
T_{_{\rm A}}^4({\bf r})&=&\left[T_{_{\rm A}}^{\star}({\bf r})\right]^4+\left[T_{_{\rm A}}^{\rm planet}({\bf r})\right]^4\,,
\end{eqnarray}
where $T_{_{\rm A}}^{\rm planet}({\bf r})$ is the contribution from the protoplanet. Note that the radiative feedback from the star  is fixed in time for simplicity, whereas the radiative feedback from the protoplanet is variable  and depends on the accretion rate of gas onto it as it moves within the disc.

 \section{Simulations of prototoplanets evolving in young discs}
 \label{sec:runs}

We follow the evolution of a protoplanet embedded in the disc for 20~kyr (approximately 20 disc outer orbital periods). The initial mass  of the protoplanet is $M_{p,i}=1~{\rm M_{\rm J}}$  and it is placed at  distance  $R_{p,i}=$ 50~AU. In this first set of simulations we ignore any radiative feedback from the central star. The effect of irradiation from the central star  will be studied in Sections 7 and in Appendix~\ref{sec:star_rt}. We perform the following  5 numerical experiments (see Table~\ref{tab:standardrun}):
 \begin{enumerate}
\item Run 1:  No radiative feedback from the central star nor the protoplanet  are included (hereafter referred to as the {\it standard run}).
\item Run 2: Radiative feedback from the protoplanet is included.
\item Run 3: Same as the standard run but with higher viscosity (viscosity $\alpha_{\min}=0.2$; $\alpha_{\min}=0.1$ in all other runs).
\item Run 4: Same as the standard run but in which  the  protoplanet has an orbit that has an initial inclination of $10^{\rm o}$  with respect to the disc midplane.
\item Run 5: Same as the standard run but using different opacities: \cite{Semenov:2003a}; for all other runs we use  the \cite{Bell:1994a} opacities (see Section~\ref{sec:opacities}).

 \end{enumerate}
 
 \begin{table}
\caption{Simulation parameters}
\label{tab:standardrun}
\centering
\begin{tabular}{@{}ccccccc} \hline
\noalign{\smallskip}
Run id &	Opacity & $\alpha_{\rm min}^1$ &  $\alpha_{p,i}^2$ & $M_{p,f}^3$ & $\alpha_{p,f}^4$ &$e_f^5$\\
	 & &      &  (AU)  & (M$_{\rm J}$)&(AU) \\
\noalign{\smallskip} 
\hline
\noalign{\smallskip}
Run 1 & BL94$^6$		&$0.1$& 40  & 24  &40 & 0.035  \\
Run 2 & BL94$^6$ 	& $0.1$ & 40  & 15& 17  &  0.0012   \\
Run 3 & BL94$^6$	& $0.2$ & 40  & 29 & 37  &0.13   \\
Run  4 & BL94$^6$ 	& $0.1$ & 40  & 25 & 36  &0.017  \\
Run 5 & SEM03$^7$ & $0.1$ & 40   & 31 &53  &0.17  \\
\noalign{\smallskip}
\hline
\end{tabular}
{$^1$}{Minimum SPH artificial viscosity.}
{$^2$}{Initial semi-major axis of the protoplanet.}
{$^3$}{Final protoplanet mass.}
{$^4$}{Final semi-major axis of the protoplanet.}
{$^5$}{Final protoplanet eccentricity.}
{$^6$}{\cite{Bell:1994a} opacity.}
{$^7$}{\cite{Semenov:2003a} opacity.}
 \end{table}

\begin{figure*}
\centerline{
\includegraphics[width=0.9\textwidth]{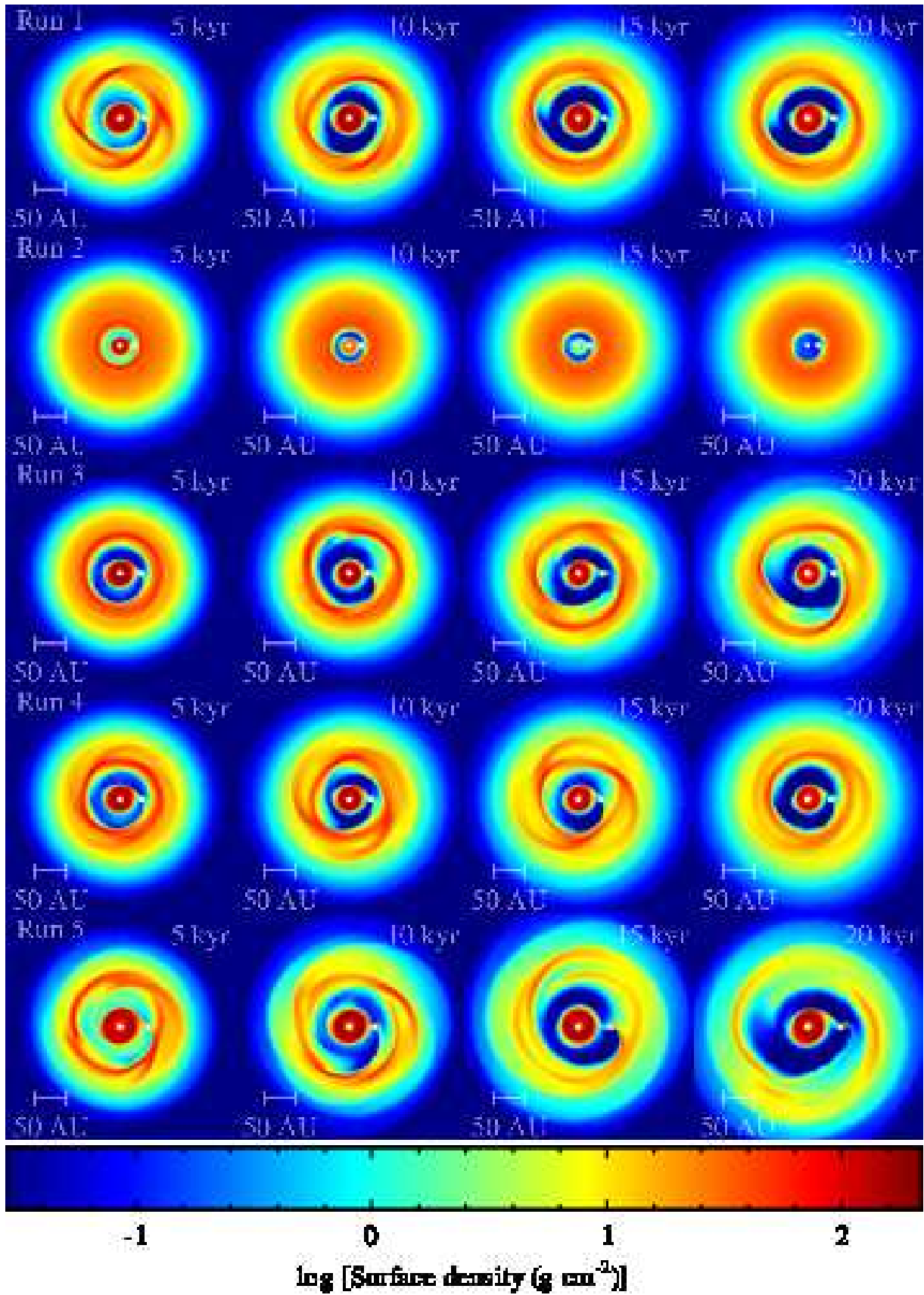}}
\caption{The evolution of the disc surface density in the 5 simulations listed in  Table~\ref{tab:standardrun}. The star and the protoplanet as depicted as wide dots. Each row corresponds to different snapshots of each run at times as stated on the graph. All runs show similar features, apart from the run in which the radiative feedback of the protoplanet is taken into account (Run 2; second row).}
\label{fig:snapshots}
\end{figure*}

 The evolution of the disc surface density for all 5 runs is shown in Figure~\ref{fig:snapshots}. We calculate the migration timescale  as 
\begin{equation}
\tau_{mig}=\alpha_p/|\dot{\alpha_p}|\,,
\end{equation}
where $\alpha_p$ is the protoplanet's semi-major axis. In all runs the protoplanet initially migrates inwards fast (see Tables~\ref{tab:standard_migration}-\ref{tab:standard_vel}), with a migration timescale of $\sim10^4$~yr, as previous studies have found \citep{Vorobyov:2006a, Baruteau:2011a,Michael:2011a, Zhu:2012a, Malik:2015a}. However, as the protoplanet migrates inwards it considerably grows in mass by accreting gas from the disc. Eventually the protoplanet is able to open up a gap and the migration slows down (for the case with radiative feedback from the protoplanet; migration timescale $\sim 10^5$~yr) or even changes to an outward direction (migration timescale $\sim 10^5$~yr) in the rest of the runs \citep[see][]{Stamatellos:2015a}. After a gap is opened up, the protoplanet grows in mass slowly, by accreting a significant amount of  material from the disc outside its orbit. An important difference of this study from previous ones \citep[e.g.][]{Baruteau:2011a,Michael:2011a} is that the protoplanet is allowed to grow in mass. This is especially significant for the case of self-gravitating discs as the protoplanet evolves in a relatively massive disc with a significant amount of gas available for accretion. In the following subsections we discuss in detail the evolution of the protoplanet within the disc in the different runs we have performed.

%%%%%%%%%%%%%%%%%%%%%%%%%%%%%%%%%%%%%%%%%%%%%
%%%%%%%%%%%%%%%%%%%%%%%%%%%%%%%%%%%%%%%%%%%%%

 %%%%%%%%%%%%%%%%%%%%%%%%%%%%%%%%%%%%%%%%%%%%%
\begin{figure}
\centerline{\includegraphics[height=0.95\columnwidth,angle=-90]{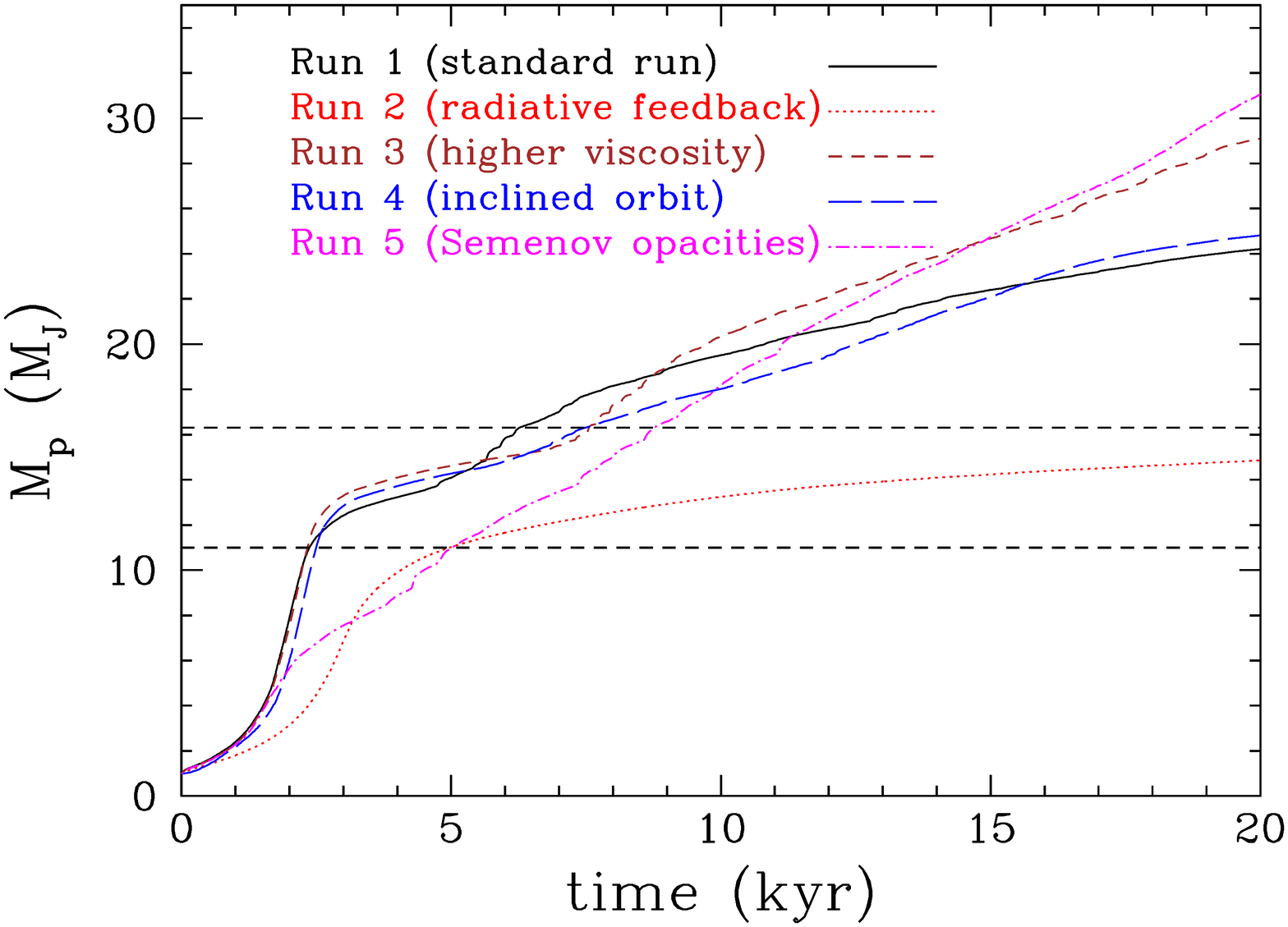}}
\caption{The evolution of  the mass $M_{\rm p}$ of the protoplanet  for the runs listed  in Table~\ref{tab:standardrun}. The increase in the protoplanet's mass is initially rapid as it opens up a gap and in most cases by the end of the simulation it has become a brown dwarf \citep[$m>11-16.3$~M$_{\rm J}$;][these limits are depicted by the horizontal dashed lines]{Spiegel:2011a}. When the radiative feedback of the protoplanet is taken into account (Run 2), its mass growth is suppressed and its mass remains within the planetary-mass regime.}
\label{fig:pmass}
\centerline{\includegraphics[height=0.95\columnwidth,angle=-90]{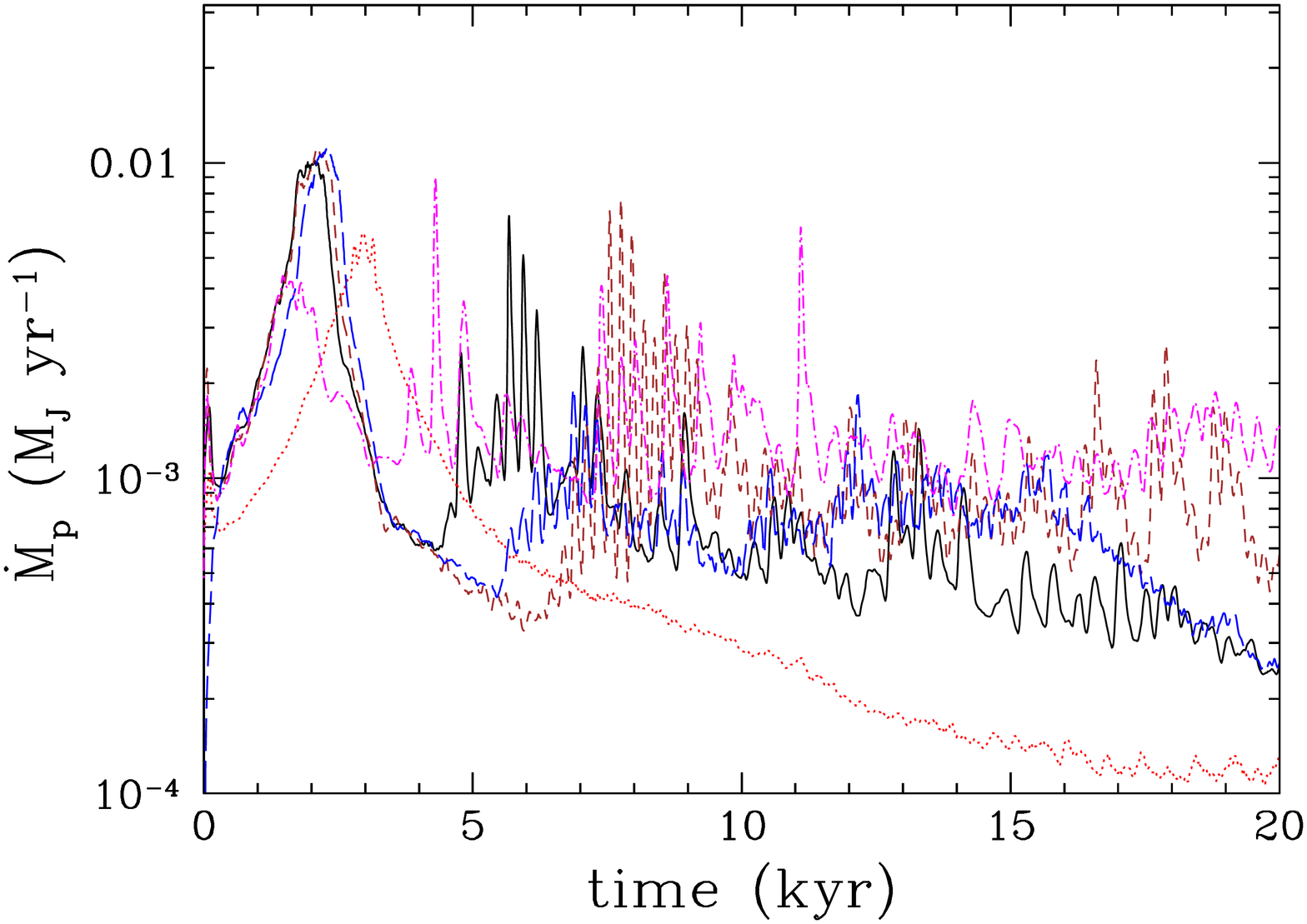}}
\centerline{\includegraphics[height=0.95\columnwidth,angle=-90]{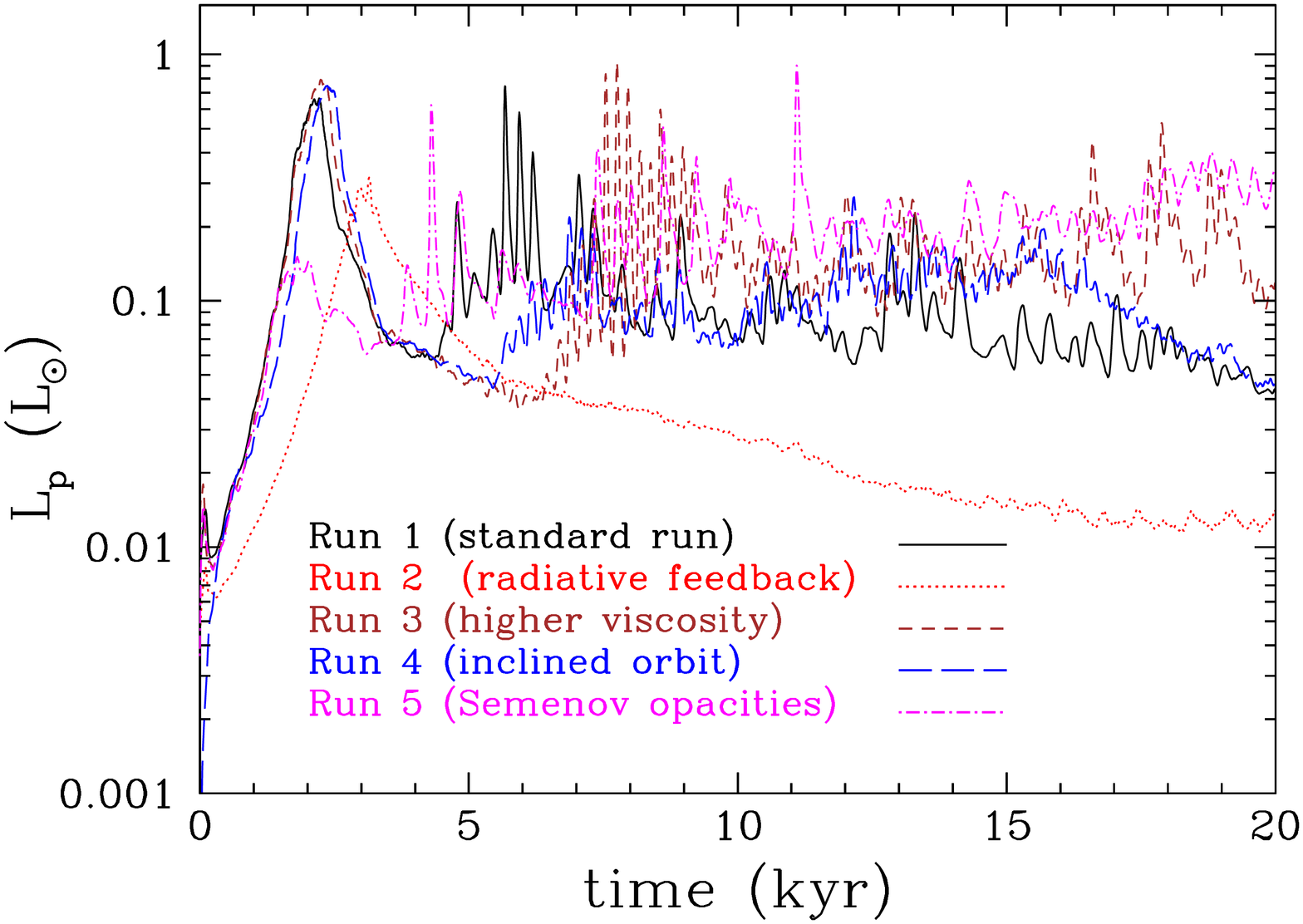}}
\caption{The mass accretion onto the protoplanet (top) and the resulting accretion luminosity (bottom). The accretion rate onto the protoplanet is high and therefore the protoplanet's luminosity may rival  that of its host star, but probably only for a short period of time. (Note that in all runs in this set, apart from Run 2, this luminosity is not fed back into the disc.)}
\label{fig:plum}
\label{fig:pacc}
\end{figure}
%%%%%%%%%%%%%%%%%%%%%%%%%%%%%%%%%%%%%%%%%%%%%
  
 %%%%%%%%%%%%%%%%%%%%%%%%%%%%%%%%%%%%%%%%%%%%%

\begin{figure}
\centerline{\includegraphics[height=0.9\columnwidth]{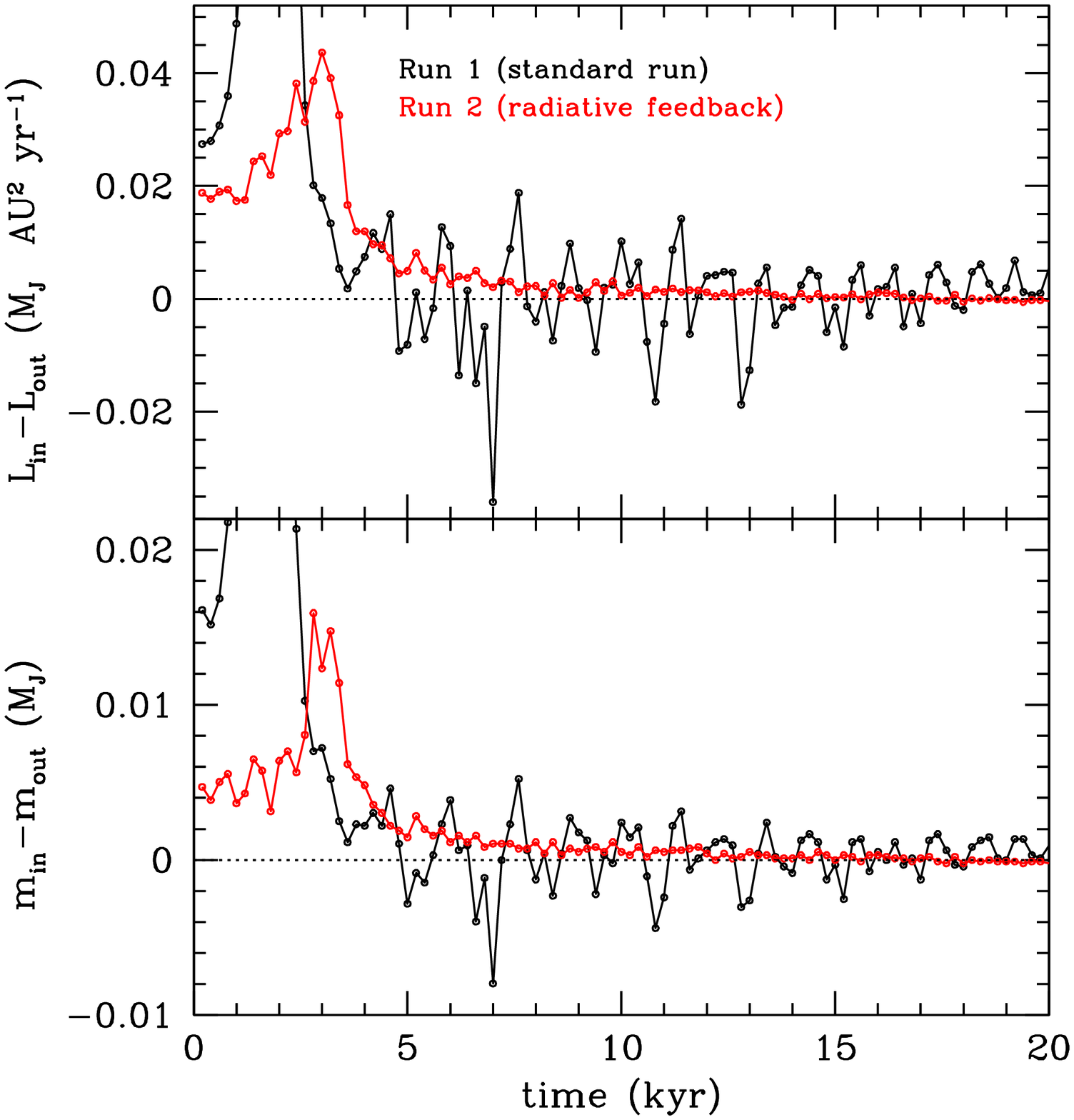}}
\caption{The difference between the mass of gas, within distance $1$ to $1.2~R_{\rm H}$  from the protoplanet, that moves towards the protoplanet from inside its orbit ($m_{\rm in}$) and from outside its orbit  ($m_{\rm out}$) (bottom) and the related  difference between their angular momenta  ($L_{\rm in}$-$L_{\rm out}$). In  the run including the protoplanets radiative feedback (Run 2, red) most of the accreted mass comes from inside the protoplanet's orbit, whereas in the case without this feedback (Run 1, black) an almost equal amount of  accreted mass comes from outside the protoplanet's orbit. {The average  angular momentum difference is positive for the radiative feedback run (for $t>5$~kyr), but negative for the standard (non-radiative feedback) run.}
After the gap opens up in the disc, in the former case the protoplanet loses angular momentum and migrate inwards, whereas in the latter case it gains angular momentum and migrates outwards, while accreting higher angular momentum gas.}
\label{fig:inout}
\centerline{\includegraphics[height=0.9\columnwidth]{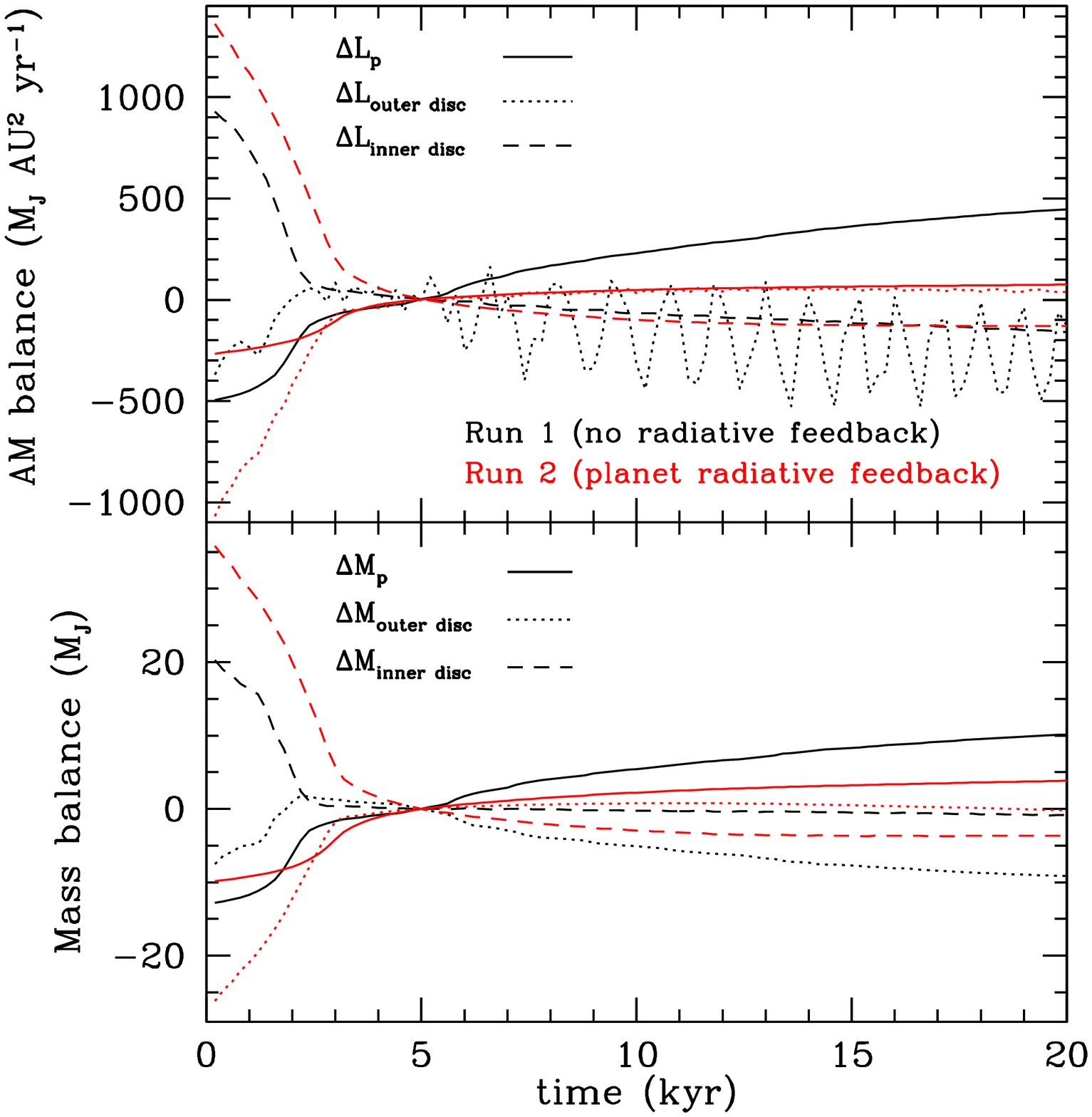}}
\caption{The change of angular momentum (top) and mass (bottom) of the planet ($\Delta L_p$, $\Delta M_p$), the inner disc (i.e. the disc inside the protoplanet's orbit; $\Delta L_{\rm inner\ disc}$, $\Delta M_{\rm inner\ disc}$), and the outer disc ($\Delta L_{\rm outer\ disc}$, $\Delta M_{\rm outer\ disc}$). The reference point is chosen at $t=5$~kyr, so as to separate the gap opening phase from the subsequent slow migration phase.}
\label{fig:ambal}
\end{figure}
%%%%%%%%%%%%%%%%%%%%%%%%%%%%%%%%%%%%%%%%%%%%%

 \subsection{Protoplanet mass growth}
 \label{sec:pmass}
 
The protoplanet grows in mass initially very fast as it opens up a gap but its mass increase slows down once the gap is opened up (see Figure~\ref{fig:pmass}). The accretion rate onto the protoplanet  (see Figure~\ref{fig:pacc}, top) is relatively high  ($\sim~10^{-4}-10^{-2}\ {\rm M_J\ yr^{-1}}$) during the gap opening phase but then it drops down; nevertheless,  accretion  continues through  streams within the gap \citep{Lubow:2006a}. Similarly high accretion rates are also seen in previous studies \citep{DAngelo:2008a,Ayliffe:2009a,Zhu:2012a,Gressel:2013b}. The resulting luminosity of the protoplanet (see Figure~\ref{fig:plum}, bottom) could rival the luminosity of the young star making its detection easier in terms of the sensitivity required.  However, this phase of high accretion lasts only for a relatively short time, making such a detection difficult. These luminosity estimates are only up to a few times higher than the luminosities estimated from hot-start models of planet formation \citep{Marley:2007a,Mordasini:2013a,Mordasini:2017a}. The value of the luminosity depends on the accretion radius, which in our model is set assuming that gas accretion happens onto the second core formed after the temperature at centre of the protoplanet-precursor clump reaches $\sim 2000$~K and molecular hydrogen dissociates \citep[e.g. see][]{Stamatellos:2009d}. { The accretion rate onto the protoplanet decreases after the gap is opened up but thereafter it shows many spikes, some with periodicity similar to the orbital period of the planet at each specific time, indicating periodic interactions with spiral structures in the disc that drive gas accretion onto the protoplanet. }Such spikes are absent when the radiative feedback of the planet is taken into account  (Run 2) as strong spiral features are suppressed. { There is a delay of $\sim2-4$~kyr between the gap opening and the gap edges becoming gravitationally unstable, driving accretion onto the protoplanet. The delay is longer for protoplanets that are closer to the central star, i.e. in a warmer region of the disc, as more gas needs to accumulate for the gap edges to become unstable.}
The accretion rate onto the protoplanet is generally lower when the protoplanet's radiative feedback is included in the simulation as seen in previous studies \citep{Nayakshin:2013a, Stamatellos:2015a,Garate:2017a}.
 
 The way that  gas is accreted onto the protoplanet depends on the dynamical state of the protostellar disc. In Figure~\ref{fig:inout} we plot the difference between the angular momentum of the gas entering the protoplanet's Hill sphere from inside the protoplanet's orbit ($L_{\rm in}$), and the angular momentum of the gas entering the Hill sphere from outside the protoplanet's orbit ($L_{\rm out}$). To calculate these we use the mass of gas within distance $1<r<1.2~R_{\rm H}$\footnote{ The same result is obtained for a different outer limit, e.g. if we consider the  gas within  distance $1<r<1.5~R_{\rm H}$ from the protoplanet.} from the protoplanet that moves towards the protoplanet from inside its orbit ($m_{\rm in}$), and the gas that moves towards the protoplanet from outside its orbit  ($m_{\rm out}$) (Figure~\ref{fig:inout}, bottom). During the initial gap opening phase, in both cases, gas is accreted onto the protoplanet mostly from the inner disc.  After the initial gap opening phase ($\sim5$ kyr), we see that for the run including the protoplanet's radiative feedback (Run 2, red), in which the disc is gravitationally stable because of the effect of the protoplanet's feedback,  most of of the accreted gas comes from inside the protoplanet's orbit. The total accreted gas in this case has lower  angular momentum than the protoplanet, so that the protoplanet's angular momentum decreases. On the other hand,  in the case without the feedback (Run 1, black),  the disc is gravitationally unstable and  a high fraction  of the accreted gas (up to almost 50\%)  comes from outside the protoplanet's orbit. This gas has higher angular momentum than the protoplanet.
 
This can also be seen in Figure~\ref{fig:ambal} where we plot the change of  angular momentum (top) and mass (bottom)  of the protoplanet ($\Delta L_p$, $\Delta M_p$), the inner disc (i.e. the disc inside the protoplanet's orbit; $\Delta L_{\rm inner\ disc}$, $\Delta M_{\rm inner\ disc}$), and the outer disc ($\Delta L_{\rm outer\ disc}$, $\Delta M_{\rm outer\ disc}$). In Run 1 the protoplanet increases in mass while the outer disc mass decreases, whereas in Run 2, in which the protoplanet's feedback is included, the outer disc mass remains almost steady (after 5 kyr) whereas the inner disc mass decreases as the mass of the protoplanet increases.
 
Apart from the run with the protoplanet's radiative feedback, there are only relatively subtle differences in the accretion onto the protoplanets in the other runs. As expected, for a higher disc viscosity the gap opens up at a later time \citep{Crida:2006a}, but thereafter the accretion rate into the protoplanet is higher, resulting in a higher final mass. 

{ The gap opens up faster in the run with the \cite{Semenov:2003a} opacities, which are generally larger than the  \cite{Bell:1994a} opacities, and the accretion rate thereafter is larger by a factor of $\sim 2$ than in the other runs.  This is because the \cite{Semenov:2003a} opacities correspond to more efficient cooling \citep[see][]{Nayakshin:2017b}, and therefore: (i) the disc is cooler, enabling the quick opening of the gap, and (ii) the thermal pressure of the circumplanetary disc is lower and gas accretion is not opposed as much as in the other runs. The dependence of the accretion rate and the final mass of the object on the assumed opacity seems to be different from what is seen in \cite{Nayakshin:2017b}. However, what matters is not the opacity itself but the cooling of the gas around the protoplanet. \cite{Nayakshin:2017b} employs an opacity that regulates gas cooling both in optically thin and optically thick regimes (i.e. the Planck-mean and Rosseland-mean opacities are the same), whereas the \cite{Semenov:2003a} opacity account of the differences between these regime.  The higher Planck-mean opacity of \cite{Semenov:2003a} results in more efficient cooling in the optically thin regions (see Equation~\ref{eq:radcool};  in the optically thin limit the denominator  on the right hand side is dominated by the $\kappa_P^{-1}$ term). Therefore as in  \cite{Nayakshin:2017b} our results mean that the gas accretion onto the protoplanet with more efficient cooling is higher. Similar results are also found by \cite{Ormel:2015a,Ormel:2015b} when studying accretion onto massive solid planet cores.}

We see that in all cases, apart from the one in which the protoplanet radiative feedback has been included, the protoplanet's final mass (i.e. the mass at the end of the hydrodynamic simulation; $t=20$~kyr) is in the brown dwarf regime ($>16.3~{\rm M_J}$). The protoplanet's mass in the case in which its radiative feedback is taken into account is at the borderline between the planetary and brown dwarf regime. However, it is expected that the protoplanet will continue to accrete gas and eventually become a brown dwarf (unless the disc dissipates by other processes, like e.g. photo-evaporation, on a faster timescale). Nevertheless, when the protoplanet's radiative feedback is taken into account  the mass growth of the protoplanet has been suppressed and, at the end of the simulation, its mass is lower by a factor of $\sim 2$ than in the other runs.

\subsection{Protoplanet migration}

%%%%%%%%%%%%%%%%%%%%%%%%%%%%%%%%%%%%%%%%%%%%%
\begin{figure}
\centerline{\includegraphics[height=0.92\columnwidth,angle=-90]{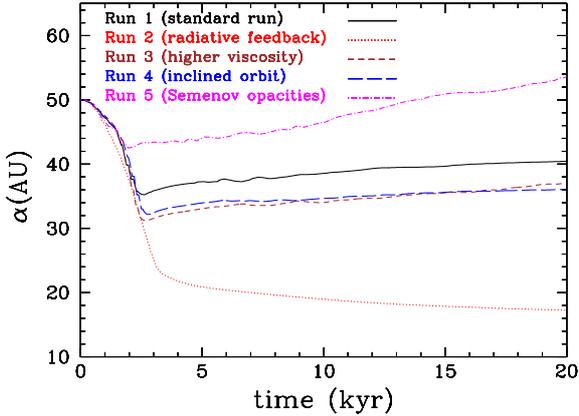}}
\caption{The semi-major axis evolution of an initially 1-M$_{\rm J}$ protoplanet in a 0.1-M$_{\odot}$ disc. The protoplanet initially migrates inwards on a Type I migration timescale ($\sim10^4$~yr) but once a gap opens up and the inward migration stops and turns into and outward migration, except for the case when the radiative feedback of protoplanet is taken into account. In the latter case the migration continues inwards but at a Type II timescale ($\sim10^5$~yr). (The details of the runs are listed in Table~\ref{tab:standardrun}).}
\label{fig:psemi}
\end{figure}

\begin{figure}
\centerline{
\includegraphics[height=0.92\columnwidth]{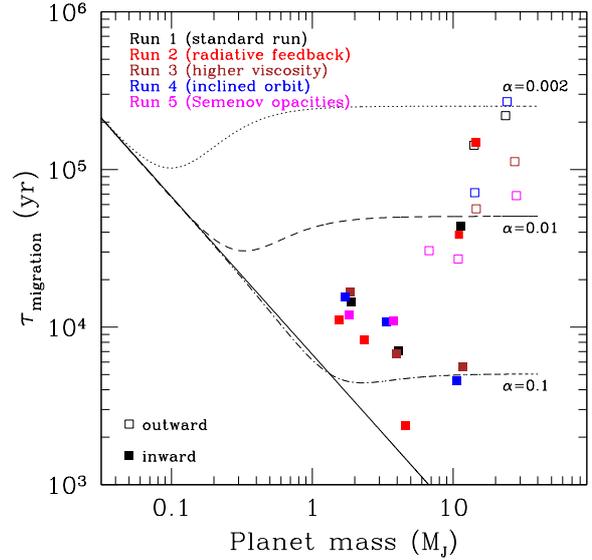}
}
\caption{The migration timescales at different times during the runs listed  in Table~\ref{tab:standardrun}. Filled boxes correspond to inward migration and empty boxes to outward migration. The solid line corresponds to the Type I migration timescale, whereas the dashed  lines correspond to a combination of Type I and Type II migration assuming different disc viscosity parameter $\alpha$ (see text for details). Once the gap is opened up,  interactions of the protoplanet with the gravitationally unstable gap edges lead to outward migration. When the radiative feedback from the planet is taken into account the gap edges are hotter, therefore they are stable and the inward migration continues but at a longer timescale.}
\label{fig:tmig_a}
\end{figure}

The protoplanet initially interacts strongly with the disc resulting in  fast inward migration, similar to the Type I migration for low-mass planets in low-mass discs. The migration timescale is only $10^4$~yr and it is similar for all 5 runs (see Figures~\ref{fig:psemi}, \ref{fig:tmig_a} and Tables~\ref{tab:standard_migration}, \ref{tab:standard_vel}). However, once a gap opens up the migration stalls; this happens at different times for each run. When the planet orbit plane is not the same with the disc midplane (Run 3) it is more difficult for the protoplanet to open up a gap. In the same way, the more viscous is the disc (Run 4) the more difficult is to open up a gap \citep[e.g][]{Crida:2006a, Malik:2015a}.  In both cases there is delay in opening up a gap and the protoplanet migrates further inwards (but only by $\sim 4$ AU compared with the other runs). In the four runs without radiative feedback from the protoplanet,  the protoplanet switches to a slow outward migration. At the end of the simulations (20~kyr) the protoplanet's semi-major axis has moved to  40, 38, and 36 au, for Runs 1, 2, and 3, respectively, from an initial semi-major axis of 50 au. In Run 5 with the higher opacity (faster cooling) \citep{Semenov:2003a} the gap opens up quicker and the protoplanet starts migrating outwards  reaching a semi-major axis of 53~AU at the end of the simulation.

The case that it is distinctly different is the one in which the radiative feedback from the planet is taken into account (Run 2). In this case the opening up of the gap is more difficult as the protoplanet heats its circumplanetary disc but also  the host protostellar disc \citep{Marley:2007a,Nayakshin:2013a,Benitez-Llambay:2015a, Montesinos:2015a,Stamatellos:2015a,Szulagyi:2017b}.  Therefore the gap opens at a later time ($\sim1$ kyr later) and the protoplanet migrates closer to central star. Its inward migration continues after the gap is opened up but at a much longer timescale ($\sim10^5$~yr; typical of Type II migration).  At the end of the simulations (20~kyr) the protoplanet is at 18~AU from the central star. 

\cite{Lin:2012b} and \cite{Cloutier:2013a} have shown that when a planet resides in a gap 
that has gravitationally unstable edges, gas is brought into the planet's co-orbital region by the spiral arms, resulting in a positive torque that pushes the planet outwards. This is indeed what we see here in the runs without the radiative feedback from the protoplanet. In these runs, a significant amount of the gas accreted onto the  protoplanet is high angular momentum material  coming from  outside the protoplanet's  orbit (see Figs.~\ref{fig:inout}-\ref{fig:ambal}). On the other hand, when the radiative feedback from the protoplanet is taken into account the temperature at the edges of the gap is raised, they are not gravitationally unstable anymore, and the protoplanet accretes a higher fraction of lower angular momentum gas from the disc inside its orbit.

Contrary to  previous studies of \citep{Baruteau:2011a,Michael:2011a, Malik:2015a}, we allow the protoplanet to accrete gas and grow in mass. By growing in mass, it is more capable to open up a gap. In previous studies mass is accumulated in the circumplanetary disc and therefore contributed to strongly couple the protoplanet with the gaseous disc in which it is formed. This behaviour is also seen in the simulations of \cite{Nayakshin:2013a}. However, the crucial  difference with previous studies is that  we use a more detailed prescription for the radiative transfer in the disc. Our prescription allows the circumstellar and circumplanetary material to moderate the efficiency of its cooling as it becomes more or less dense, whereas the studies of  \cite{Baruteau:2011a} and \cite{Malik:2015a} use the $\beta$-cooling approximation, in which the cooling time is proportional to the local orbital period of the gas, i.e. ${t_{\rm cool}}=\beta\ \Omega_K^{-1}(R)$. In Section~\ref{sec:betacooling} we discuss this issue in detail.

The  estimated migration timescales ($\tau_{\rm mig}={\alpha_p}/|\dot{\alpha_p}|$) are shown in Table~\ref{tab:standard_migration}, and the migration rates ($v_{\rm mig}=\dot{\alpha_p}$) inTable~\ref{tab:standard_vel}. These values are calculated at 5 different times during the simulations.  The migration timescales for all 5 runs are plotted against the protoplanet mass in Figure~\ref{fig:tmig_a}. On the same graph we plot the estimated migration timescales for Type I and Type II migration as calculated by \cite{Ward:1997a} and \cite{Tanaka:2002a} \citep[see also][] {Bate:2003a}. 

The  Type I migration rate  estimated  by \cite{Tanaka:2002a} is
\begin{equation}
v_{I}= -(2.7+1.1\alpha_\Sigma)\, \frac{M_p}{M_\star}\, \frac{R_p^2\Sigma_p}{M_\star}\left(\frac{H}{R_p}\right)^{-2} R_p\Omega_p\,,
\end{equation}
where $\alpha_\Sigma$ is the exponent of the disc surface density profile ($\Sigma\propto R^{-\alpha_\Sigma}$), $R_p$ the orbital radius of the planet, $\Sigma_p$ the surface density of the disc at the position of the planet,  $H_p$ the disc thickness at the position of the planet, and $\Omega_p$ the planet's angular frequency.
 The rate of migration of the Type II case is set by the viscous evolution of the disc \citep[e.g.][]{Bate:2003a}:
 \begin{equation}
 v_{II}=-\frac{3}{2}\ \alpha \left(\frac{H_p}{R_p}\right)^2 R_p \Omega_p\,,
 \end{equation}
 where $\alpha$ is the disc viscosity parameter. { We note though that the actual Type II migration timescale may vary from the above approximation \citep{Crida:2007b,Duffell:2014a,Durmann:2015a,Durmann:2017a}. For example, \cite{Durmann:2015a,Durmann:2017a} find that due to gas crossing through the gap Type II migration may be faster or slower.}
 The total rate of migration, combining the Type I and Type II regimes, is therefore
\begin{equation}
v=\frac{v_I}{1+(M_p/M_t)^3}+\frac{v_{II}}{1+(M_t/M_p)^3}\,.
\end{equation}
$M_t$ is the transition mass between Type I and Type II migration \citep{Ward:1997a} and is set to 
\begin{equation}
M_t=0.4\alpha^{2/3}(H/R_p)^{-1/3}\,.
\end{equation}
The migration timescale is therefore
\begin{equation}
\tau_{I,II}=\frac{R_p}{|v_{I,II}|}\,.
\end{equation}

We calculate the migration timescales from the above equations assuming a  planet at $R_p=5.2$~AU in a disc with  $\alpha_\Sigma=0.5$,  $\Sigma_p=75\ {\rm g\ cm^{-2}}$, and  $H_p/R_p=0.05$. In Figure~\ref{fig:tmig_a}, we plot 3 cases that correspond to different disc viscosity ($\alpha=0.1,0.01,0.002$). { We note that these calculations are for planets in  low-mass discs at specific regimes: \cite{Tanaka:2002a} study 3D discs including both Lindbland and corotational resonances assuming an isothermal disc, whereas  \cite{Ward:1997a}  study a  2D disc ignoring corotational resonances. { These calculations are not applicable for the cases we examine in this paper. Therefore,  these analytic solutions  are plotted  in the graph in Figure~\ref{fig:tmig_a} merely for reference, so as to put our results in context with the classic picture of Type I and Type II migration.}}

%%%%%%%%%%%%%%%%%%%%%%%%%%%%%%%%%%%%%%%%%%%%%

The migration timescales in Figure~\ref{fig:tmig_a} are plotted for 5 different times in the simulation: 0.7, 1.5, 2.5, 5 and 18~kyr. As the mass of the protoplanet increases with time, the same graph may be used to track the evolution of the migration timescale with time. In all runs the protoplanet initially migrates inwards on timescales similar to the Type I migration timescale of a $1-{\rm M_J}$ planet, i.e. $\sim (1-2)\times10^4$~yr. As the protoplanet moves towards the host star and grows in mass, the migration timescale becomes shorter (down to $\sim 3\times10^3$~yr), but once the gap opens up the migration slows down and eventually the protoplanet starts moving outwards in all runs apart from the case when its radiative feedback is taken into account. In this case,  inward migration  continues albeit at a much longer timescale, similar to the Type II migration case ($\sim 2\times10^5$~yr). 

%%%%%%%%%%%%%%%%%%%%%%%%%%%%%%%%%%%%%%%%%%%%%
\begin{figure}
\centerline{
\includegraphics[height=0.95\columnwidth,angle=-90]{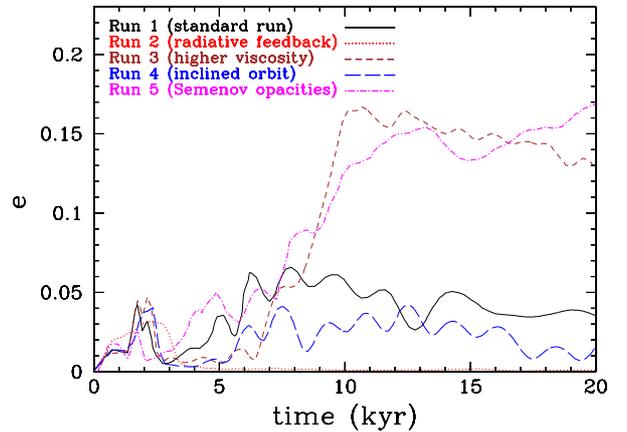}}
\caption{The protoplanet eccentricity increases with time in all runs apart from the case with the protoplanet feedback. The eccentricity growth is due to interactions with the gravitationally unstable gap edges. When the protoplanet radiative feedback is taken into account the gap edges are stable and the orbit of the protoplanet ends up almost circular.}
\label{fig:pecc}
\end{figure}
%%%%%%%%%%%%%%%%%%%%%%%%%%%%%%%%%%%%%%%%%%%%%
 
\subsection{Eccentricity}

The eccentricity of the protoplanet grows as the protoplanet interacts with the gravitationally unstable disc (see Figure~\ref{fig:pecc}). The growth pattern is rather stochastic with the eccentricity growing up to $\sim 0.15$. Even if the eccentricity after this initial growth period dampens with time due to secular interactions with the disc, these initial interactions can provide the seed for  subsequent eccentricity growth \citep{Goldreich:2003a,Duffell:2015a}. Therefore, disc-planet interactions, while the disc is still relatively massive, may help explain the observed high eccentricities   of giant planets in systems that are thought to contain only one planet \citep{Wright:2011a,Dunhill:2018a}. In the case when the radiative feedback of the protoplanet is taken into account (Run 2), the  eccentricity  initially  grows during the gap opening phase but then it is  dampened quickly, { as there is no strong stochastic driving}, so that the protoplanet ends up in a nearly circular orbit. Therefore, interactions within a gravitationally unstable disc is necessary for eccentricity growth to occur.

\subsection{Protoplanet on an inclined orbit}

\begin{figure}
\centerline{\includegraphics[height=0.95\columnwidth,angle=-90]{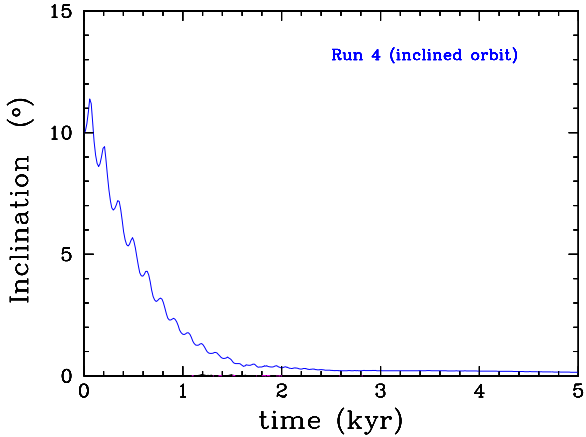}}
\caption{The evolution of the protoplanet's orbit inclination with respect to the disc midplane, for Run 4. The protoplanet's orbit quickly aligns with the disc midplane.}
\label{fig:theta}
\end{figure}

{ The protoplanet in the run in which its orbit is inclined  (Run 4, blue lines) shows a mass growth that  resembles closely the standard run (see Figure~\ref{fig:pmass}). This is because the protoplanet's orbit inclination gets smaller quickly as the protoplanet crosses through the disc midplane. Within $\sim 2$~kyr ($\sim 5$ orbits) the protoplanet's orbit has been aligned with the disc midplane (inclination has become zero,  see Figure~\ref{fig:theta}). Therefore, the long-term evolution of a protoplanet is not significantly different if the protoplanet initially forms on an inclined orbit in a relatively massive protostellar disc. }

%%%%%%%%%%%%%%%%%%%%%%%%%%%%%%%%%%%%%%%%%%%%%
%%%%%%%%%%%%%%%%%%%%%%%%%%%%%%%%%%%%%%%%%%%%%

%%%%%%%%%%%%%%%%%%%%%%%%%%%%%%%%%%%%%%%%%%%%%
\begin{figure}
\centerline{\includegraphics[width=0.95\columnwidth]{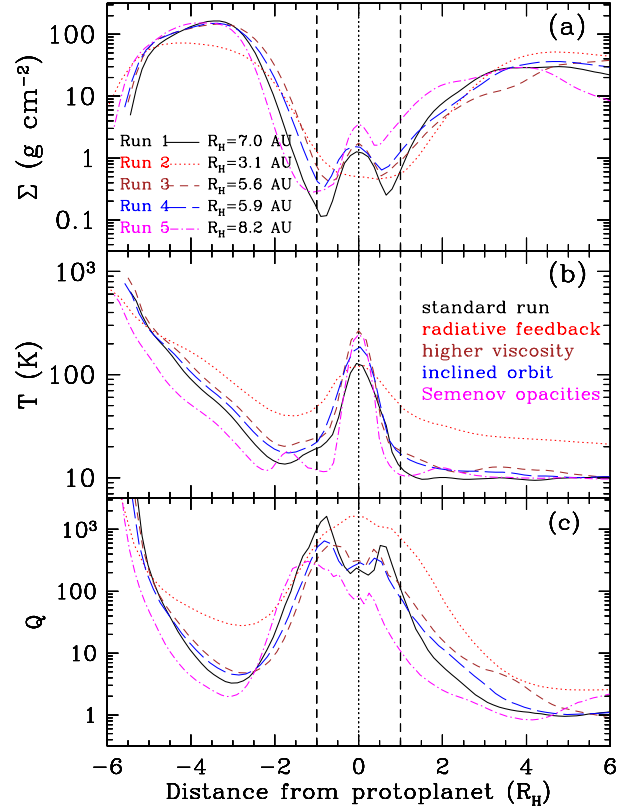}}
\caption{The surface density (a), the temperature at the disc midplane (b), and the Toomre Q parameter (c), in the protoplanets corotational frame,  as a function of the distance from the protoplanet, in units of its Hill radius, for the simulations listed in Table~\ref{tab:standardrun}.  The protoplanet is at $r=0$, whereas negative values correspond to the direction towards the central star. All simulation snapshots are at time $t=8$~kyr, i.e. after a gap has been established in the discs. The values of the Hill radius of the protoplanet for each run are also shown on the top graph.}
\label{fig:gap_a}
\end{figure}
%%%%%%%%%%%%%%%%%%%%%%%%%%%%%%%%%%%%%%%%%%%%%

\subsection{Gap opening}

The surface density, the temperature profile and the Toomre parameter Q for the region around the protoplanet for all five runs are shown in Figure~\ref{fig:gap_a} (on the protoplanet's corotational frame). The snapshots correspond to $t=8$~kyr, i.e. after the gap has been opened up in the disc. The gap size  is a few Hill radii  and rather similar for all runs (note though that the size of the Hill radius is different in each run, as the protoplanet is on different orbits in different runs). The increase of surface density towards the protoplanet within the Hill radius corresponds to the circumplanetary disc (see section below). 

{ The surface density at the boundaries of the Hill radius is asymmetric (Figure~\ref{fig:gap_a}a). In Run 2 (which includes the protoplanet's radiative feedback) the surface density at the inner boundary of the protoplanet's Hill radius (e.g. at $-1.5~{\rm R_H}$) is higher than the surface density at outer boundary of the Hill radius (e.g. at $+1.5~{\rm R_H}$)  whereas in all other runs the outer boundary surface density is higher}. This is consistent with the point made in the previous subsection, i.e. that  a large fraction of gas is accreted onto the protoplanet from outside its orbit when the gap edges are gravitationally unstable. The asymmetry is more pronounced for Run 5 (higher opacity run) which also exhibits shorter outward migration timescales (see Table~\ref{tab:standard_migration} and Figure~\ref{fig:tmig_a}). The disc temperature   (Figure~\ref{fig:gap_a}b) at the run with the protoplanet feedback is higher than in the other runs, which results in stabilizing the disc edges as the Q parameter is larger than~$\sim 3$ (Figure~\ref{fig:gap_a}c). The temperature of the disc around the protoplanet is lower for the higher opacity run (Run 5) which helps opening up the gap fast and results in higher gas accretion onto the protoplanet.

\subsection{Circumplanetary discs}

\begin{figure}
\centerline{\includegraphics[width=0.95\columnwidth]{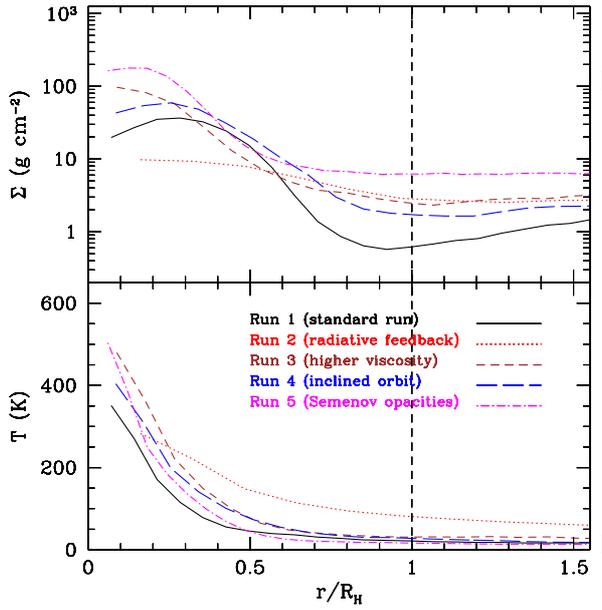}}
\caption{The properties  of the circumplanetary disc in the simulations listed in Table~\ref{tab:standardrun} at $t=8$~kyr. Azimuthally averaged surface density $\Sigma$  (top) and temperature $T$ (bottom) as a function of the distance from the protoplanet (in units of its Hill radius).}
\label{fig:cpd_a}
\end{figure}

\begin{figure}
\centerline{\includegraphics[height=0.95\columnwidth,angle=-90]{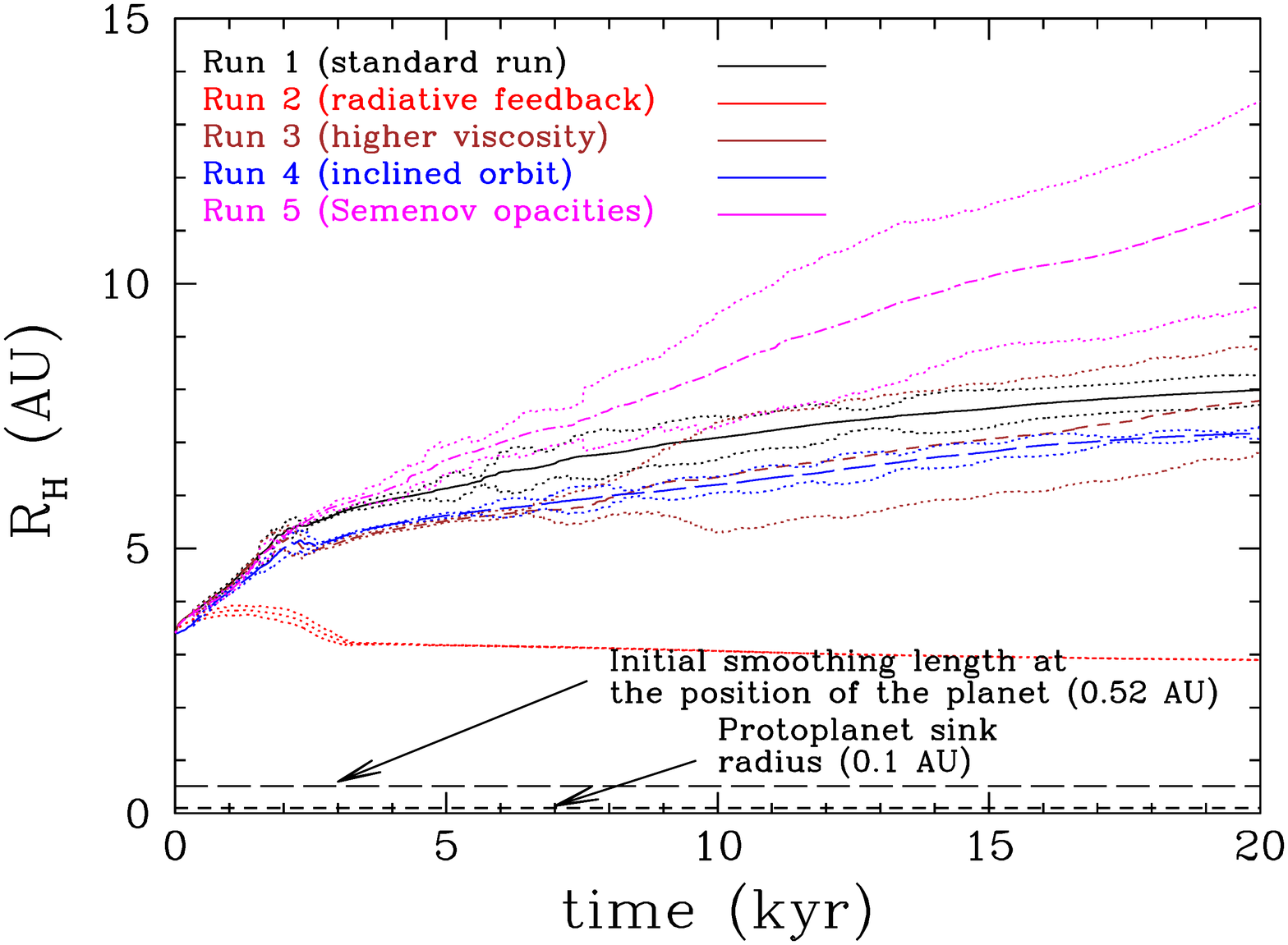}}
\caption{The evolution of the Hill radius of the protoplanet. The solid lines correspond the values calculated using the semi-major axes of the protoplanet, whereas the dotted lines correspond to the periastron and apoastron of the protoplanet's orbit. The actual Hill radius of each protoplanet varies between these two extremes.}
\label{fig:res}
\centerline{\includegraphics[height=0.95\columnwidth, angle=-90]{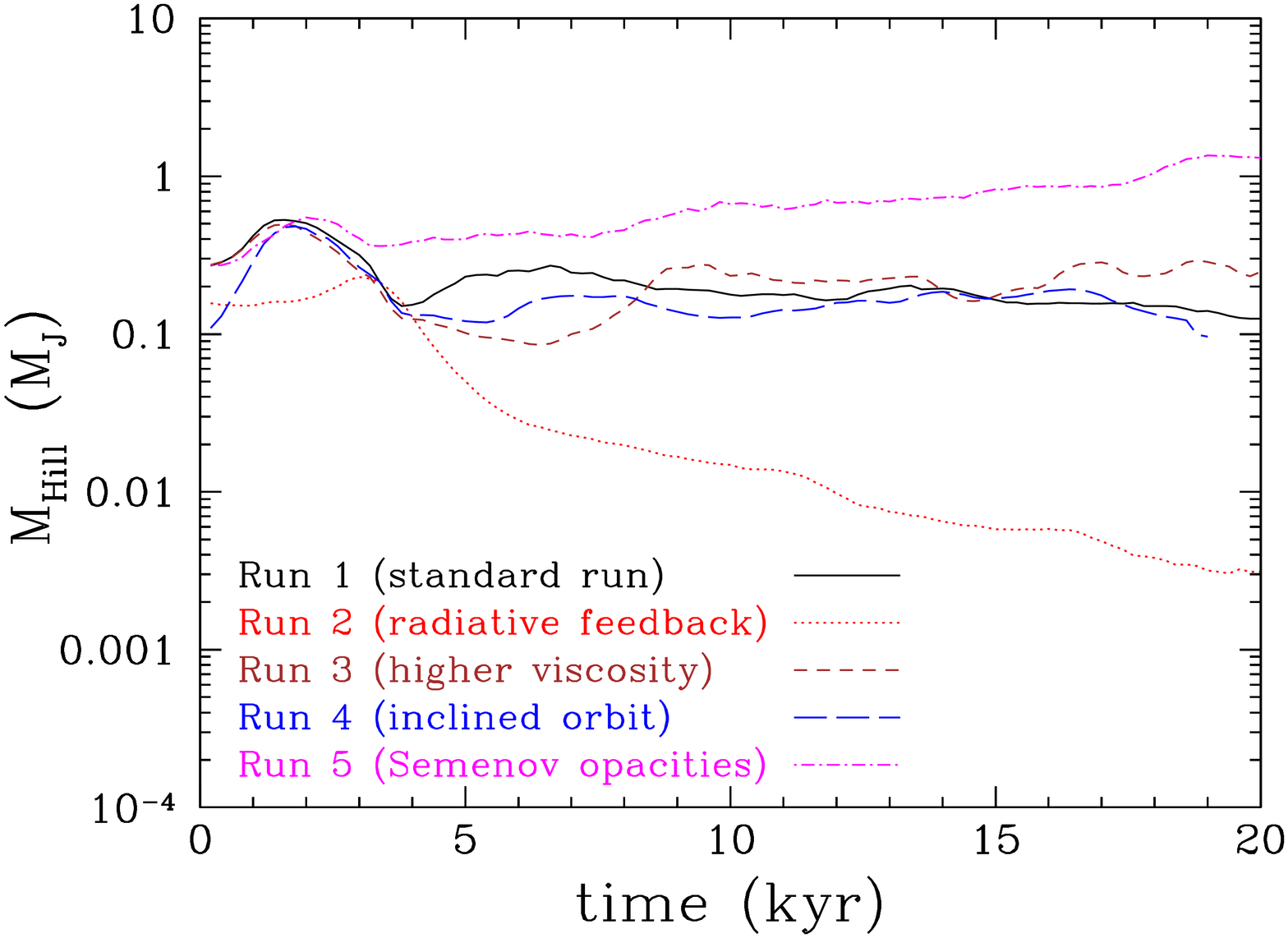}}
\caption{The mass within the protoplanet's Hill radius as a function of time for the simulations listed in Table~\ref{tab:standardrun}.}
\label{fig:hill_a}
\end{figure}

The properties of the circumplanetary discs in all 5 runs are shown in  Figure~\ref{fig:cpd_a} and the time evolution of the Hill radius and the mass within it for the protoplanet in each run are shown in Figures~\ref{fig:res} and \ref{fig:hill_a}. The typical circumplanetary disc mass is about $0.1~{\rm M_J}$, which means that the disc is resolved by $\sim1,000$ SPH particles, i.e. approximately 7 smoothing lengths (assuming a nearly 2D disc). The sink radius of the protoplanet is set to 0.1~AU so it is always smaller than the Hill radii of the protoplanets at all times, in all runs (see Figure~\ref{fig:res}). { The circumplanetary discs are therefore just resolved and the presented properties near the planet are probably resolution dependent. We expect that the surface density and the temperature near the planet are underestimated.}  In Run 2 (with protoplanet feedback) the circumplanetary disc is smaller ({with mass  of $0.01~{\rm M_J}$}; see Figure~\ref{fig:hill_a}) as the protoplanet is closer to the parent star; therefore the circumplanetary disc is just resolved in this case (only  by $\sim2$ smoothing lengths).

 %%%%%%%%%%%%%%%%%%%%%%%%%%%%%%%%%%%%%%%%%%%%%
%%%%%%%%%%%%%%%%%%%%%%%%%%%%%%%%%%%%%%%%%%%%%

{ Despite the above drawback, by comparing the derived properties we see that in Run 2, with the protoplanet radiative feedback,  the circumplanetary disc is hotter  than in the other runs, as expected.  Radiative feedback from the protoplanet together  with the gas thermodynamics \citep{Gressel:2013b} are important in determining the properties of circumplanetary discs. Further studies with higher resolution are needed for more secure results. Additionally, the presence of magnetic field may also play a significant role \citep{Gressel:2013b,Fujii:2014a} but we do not examine this case here. 

We note that we do not find temperatures in the circumplanetary discs  higher than a few hundred Kelvin in contrast to the simulations of \cite{Szulagyi:2017d},  in which they find that temperatures close to the protoplanet rise up to  a few thousand  Kelvin.  However, their simulations are able to resolve the region down to $10^{-3}\times{\rm R_{Hill}}$, which can indeed become very hot. {We note though that the  \cite{Szulagyi:2017d} simulations do not include the effect of molecular hydrogen dissociation at $\sim 2,000$~K  (nor the ionisation of hydrogen and the first and second ionisation of helium), so they may overestimate the temperatures in the inner  region. When we compare the temperatures farther out in the circumplanetary disc (e.g. at $0.5\times{\rm R_{Hill}}$) we see that the temperatures that we find here are lower than  the ones in  \cite{Szulagyi:2017d}.}

  %%%%%%%%%%%%%%%%%%%%%%%%%%%%%%%%%%%%%%%%%%%%%
\begin{figure}
\centerline{
\includegraphics[height=0.95\columnwidth,angle=-90]{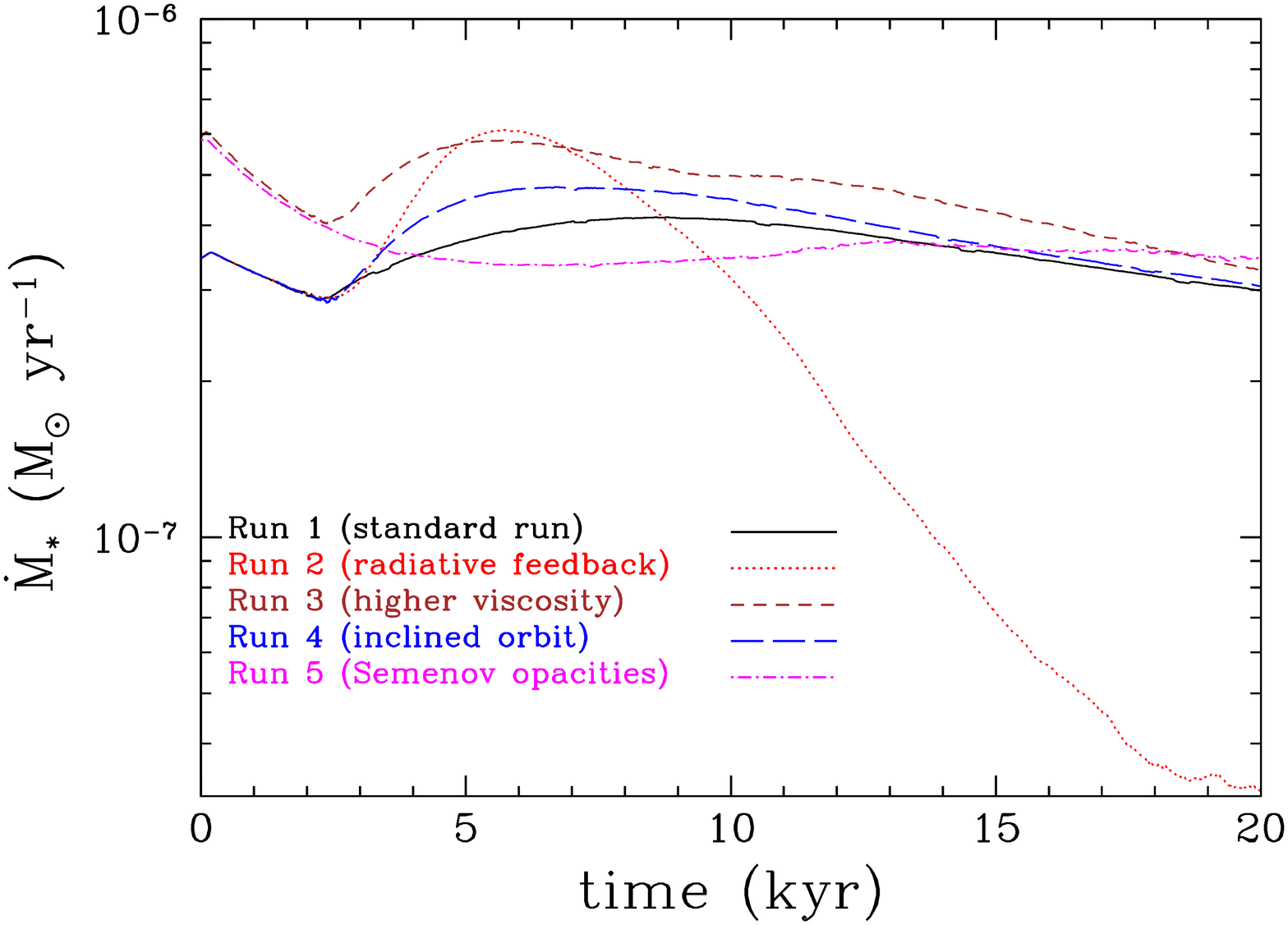}}
\centerline{\includegraphics[height=0.95\columnwidth,angle=-90]{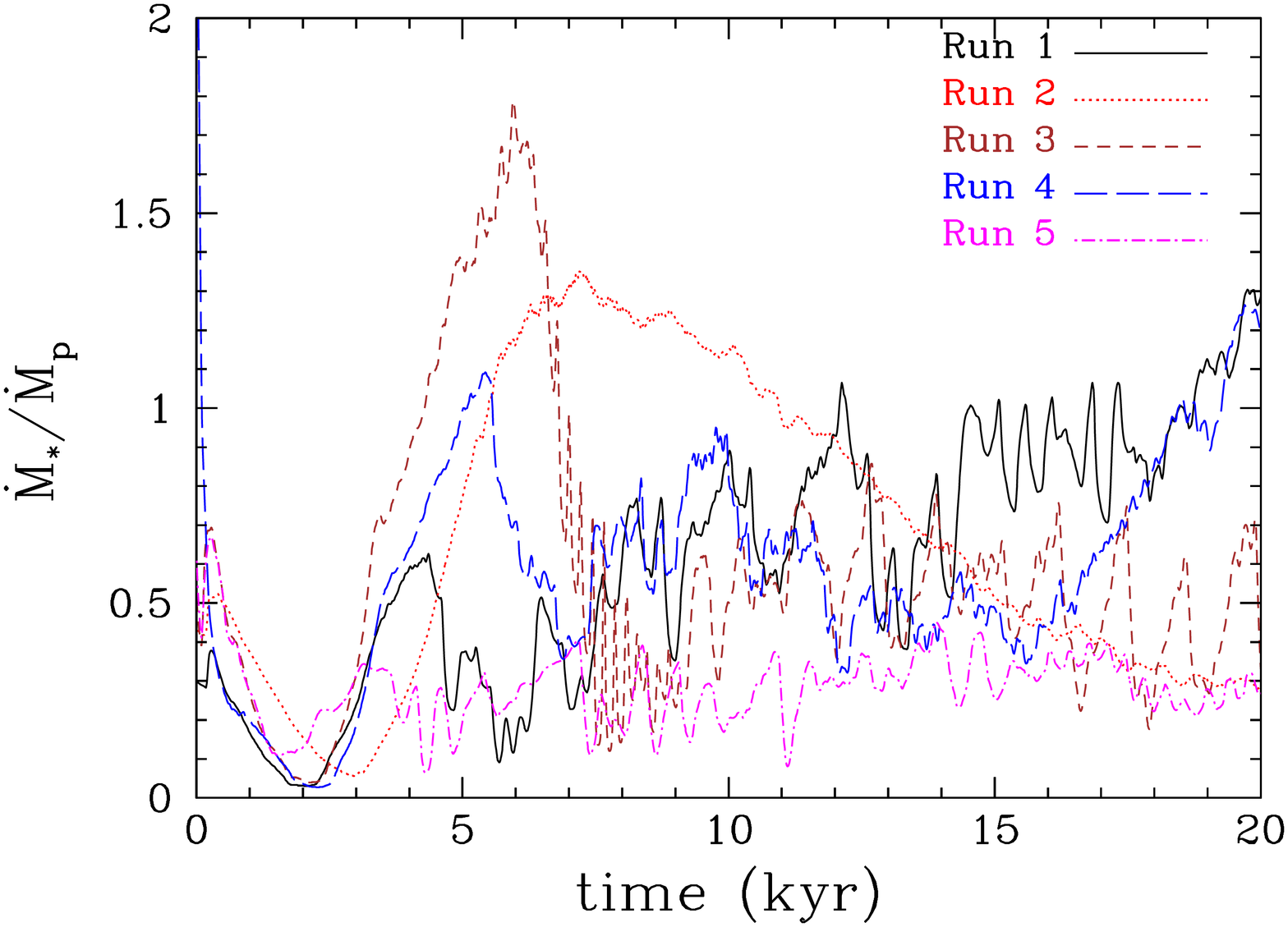}}
\caption{The accretion rate onto the central star (top) and the ratio of the accretion rate onto the star to the accretion rate onto the protoplanet (bottom). The accretion rate is higher for higher viscosity (Run 3) or for higher opacity  (Run 5).}
\label{fig:paccstar}
\label{fig:paccratio}
\end{figure}
  %%%%%%%%%%%%%%%%%%%%%%%%%%%%%%%%%%%%%%%%%%%%%
  
\subsection{Accretion onto the central star}

The presence of a protoplanet in the disc regulates the accretion rate onto the star (see Figure~\ref{fig:paccstar}). Interestingly, the accretion rate onto the protoplanet during the initial stages of its evolution that we model here, is comparable to the accretion rate onto the star itself. During the gap opening phase it can be even ten times higher. This is consistent with previous studies of planets embedded in lower-mass discs \citep{Nelson:2000a,Lubow:2006a,Owen:2014b}. During this phase the accretion rate onto the star decreases but as the protoplanet moves closer to the star and once the gap opens up, the accretion rate onto the star  increases by a factor of a few, as the protoplanet acts to drive accretion onto the star; the closer the protoplanet orbit to the star  and the higher its mass, the higher the accretion rate onto the star becomes. At this stage, and for a few thousand years, the accretion rate onto the star may become higher than the accretion rate onto the protoplanet. In Run 5 (with the Semenov opacities) the protoplanet stays sufficiently away from the star so that the accretion rate onto the star is not affected.

 In Run 2 (with protoplanet feedback) the protoplanet migrates closer to the star than in the other runs, and  the accretion rate onto the star increases significantly until a cavity forms around the star; thereafter, the accretion rate drops considerably as the presence of the protoplanet starves the star from gas.  
At the same time the protoplanet experiences similar gas starvation as it resides within the same cavity. This stage can be thought as similar to the transition disc phase \citep[see review by][]{Espaillat:2014a}. However, the accretion rate onto the protoplanet is still higher than the accretion onto the star  by a factor of 2. In this case the star-protoplanet system behaves as a  low mass-ratio binary system with a circumbinary disc, in which the secondary component (i.e. the protoplanet) accretes more than the primary component (i.e. the star) \citep{Artymowicz:1994a,Artymowicz:1996a}. In such systems, secondaries increase in mass faster than primaries, resulting in an almost equal-mass binaries \citep[e.g.][]{Satsuka:2017a}. However, in the case of a planetary-mass companion, its mass cannot become comparable to that of the star; even if an unlikely high accretion rate of $\sim10^{-4} {\rm M_J\ yr^{-1}}$ is maintained for $\sim 10^5$~ yr, then the protoplanet's mass will increase only by $10~{\rm M_J}$. 
 
\subsection{The role of the accretion rate onto the protoplanet}
{ The actual accretion rate onto the protoplanet is important as it regulates the strength of its radiative feedback, which it turn determines whether the edges of the gap opened by the protoplanet are  gravitationally unstable (so that the protoplanet migrates outwards), or gravitationally stable (so that protoplanet migrates inwards).  The luminosity of the protoplanet, $L_p$,  is proportional to the accretion rate, ${\dot{M}_p}$, onto the protoplanet (see Equation~\ref{eq:lplanet}), therefore the temperature due to the presence the protoplanet  (see Equation~\ref{eq:tplanet}) scales as $T_{_{\rm A}}^{\rm planet}(r) \propto {\dot{M}_p}^{1/4}$. The Toomre parameter, $Q$ that determines whether the gap edges are  gravitationally unstable is $Q(r) \equiv {\kappa c_{s}(r)}/{\pi G \Sigma(r)}$, where $\kappa$ is the epicyclic frequency, and $c_s$ the sound speed (we assume that the  distance, $r$, is measured from the protoplanet). Hence, $Q\propto \left[T_{_{\rm A}}^{\rm planet}(r)\right]^{1/2}$ or equivalently 
\begin{equation}
Q\propto {\dot{M}_p}^{1/8}\,.
\end{equation}
Therefore, there is only a weak dependence of the Toomre parameter on the accretion rate onto the protoplanet. If we assume that the actual accretion rate is 10 times lower than the one we estimate in the simulations we present here, then the Q value is lower only by a factor of $\sim1.3$ (assuming that the other parameters remain the same). Then the Q value in the outer edge of the gap in the run with the protoplanet radiative feedback (see Figure~\ref{fig:gap_a}, red line) will drop from 3  to 2.3, i.e. the gap edge will still be gravitationally stable ($Q>1.5$). We conclude the value of the accretion rate onto the protoplanet is not critical, at least qualitatively, regarding the effect of radiative feedback on the migration of the protoplanet.}

\section{Comparison with $\beta$-cooling studies}
\label{sec:betacooling}

%%%%%%%%%%%%%%%%%%%%%%%%%%%%%%%%%%%%%%%%%%%%%
\begin{figure}
\centerline{\includegraphics[height=0.95\columnwidth,angle=-90]{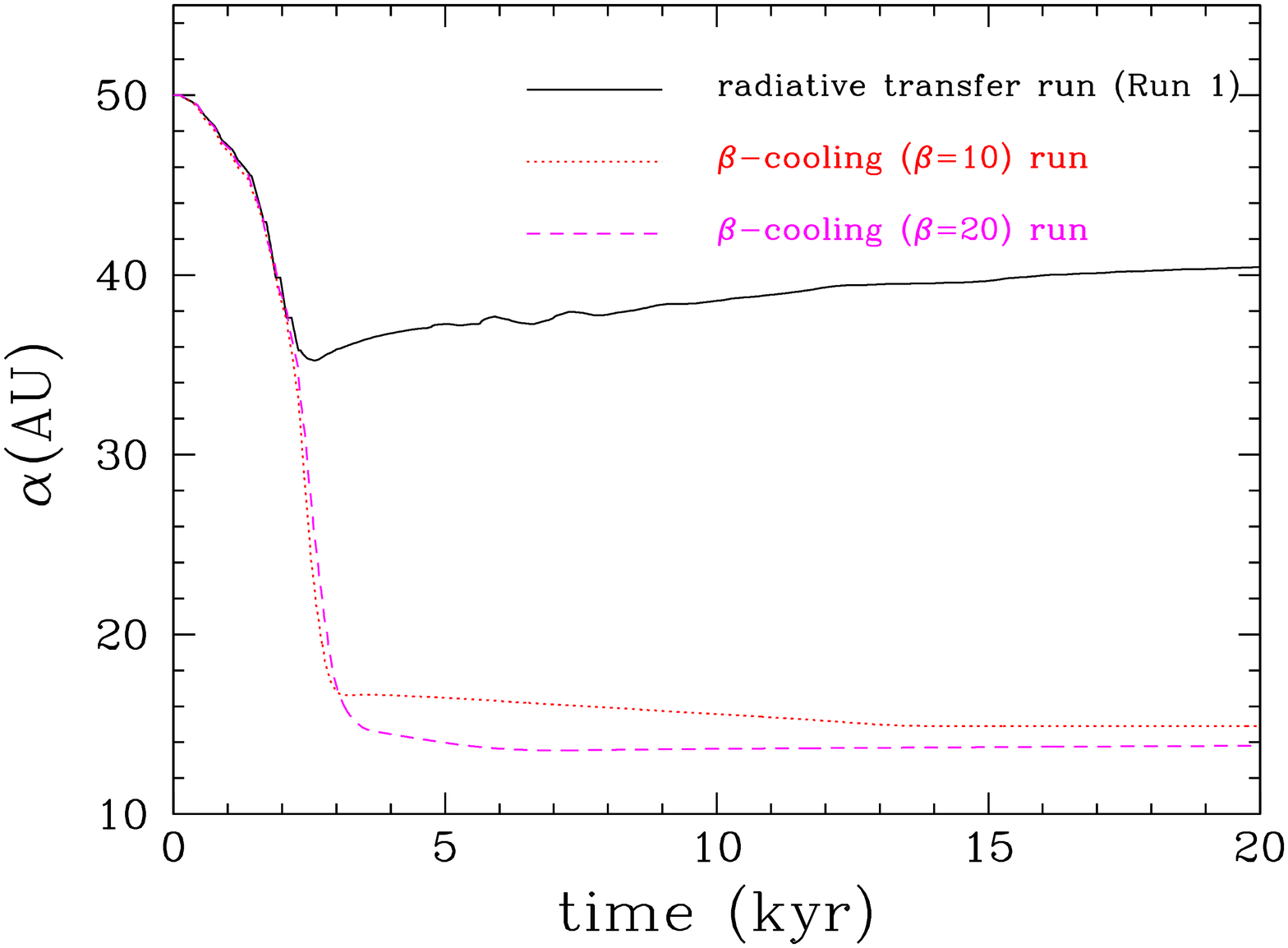}}
\centerline{\includegraphics[height=0.95\columnwidth,angle=-90]{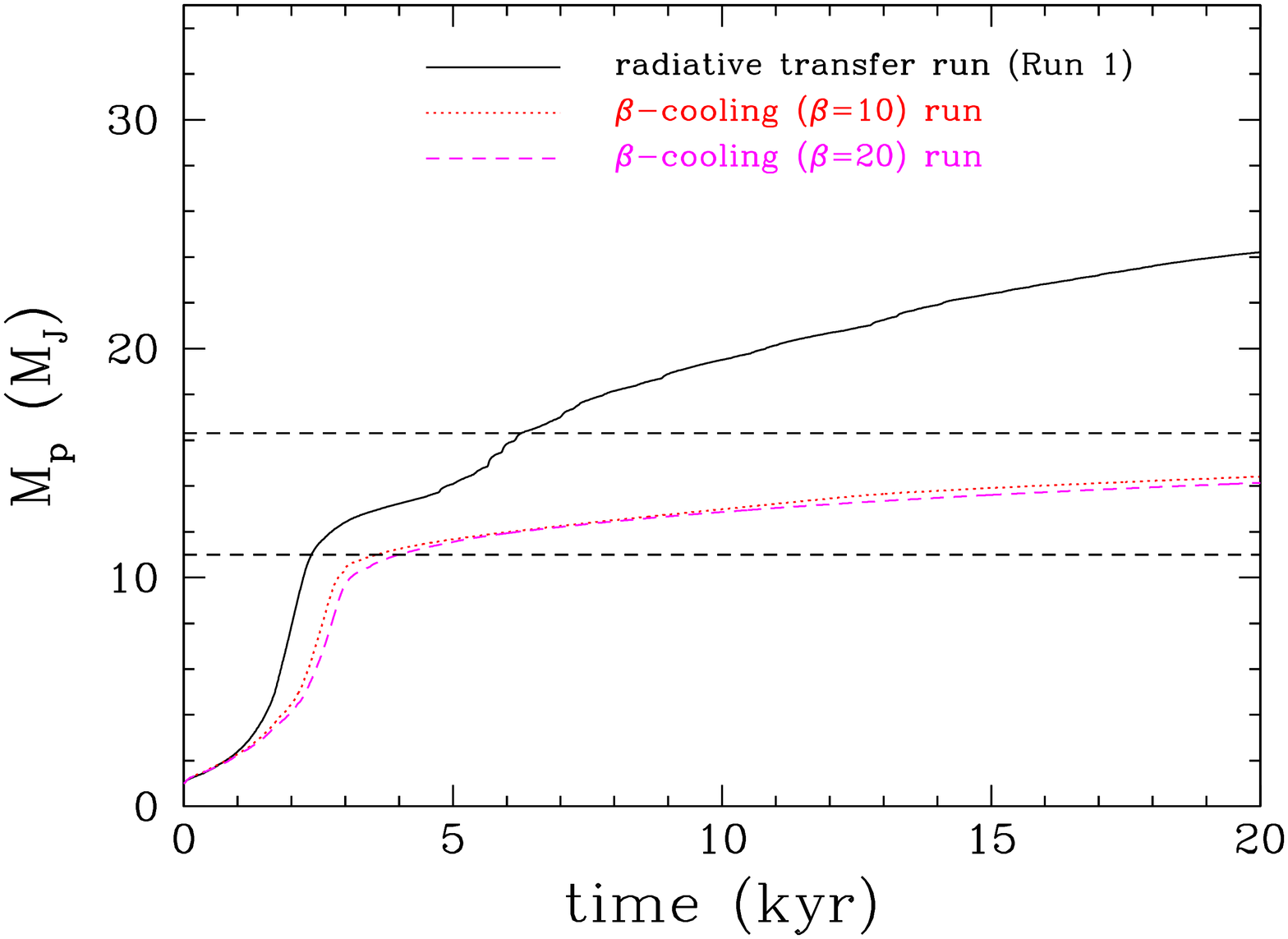}}
\caption{The evolution of a protoplanet in a disc in a run with radiative transfer (Run 1) and in two runs using the  $\beta$-cooling approximation. Semi-major axis (top) and protoplanet mass (bottom) are plotted against time. In the $\beta$-cooling runs, the  cooling times are long, the protoplanet does not open up a gap and migrates fast in the inner disc region close to the star.}
\label{fig:betaruns}
\centerline{\includegraphics[height=0.95\columnwidth,angle=-90]{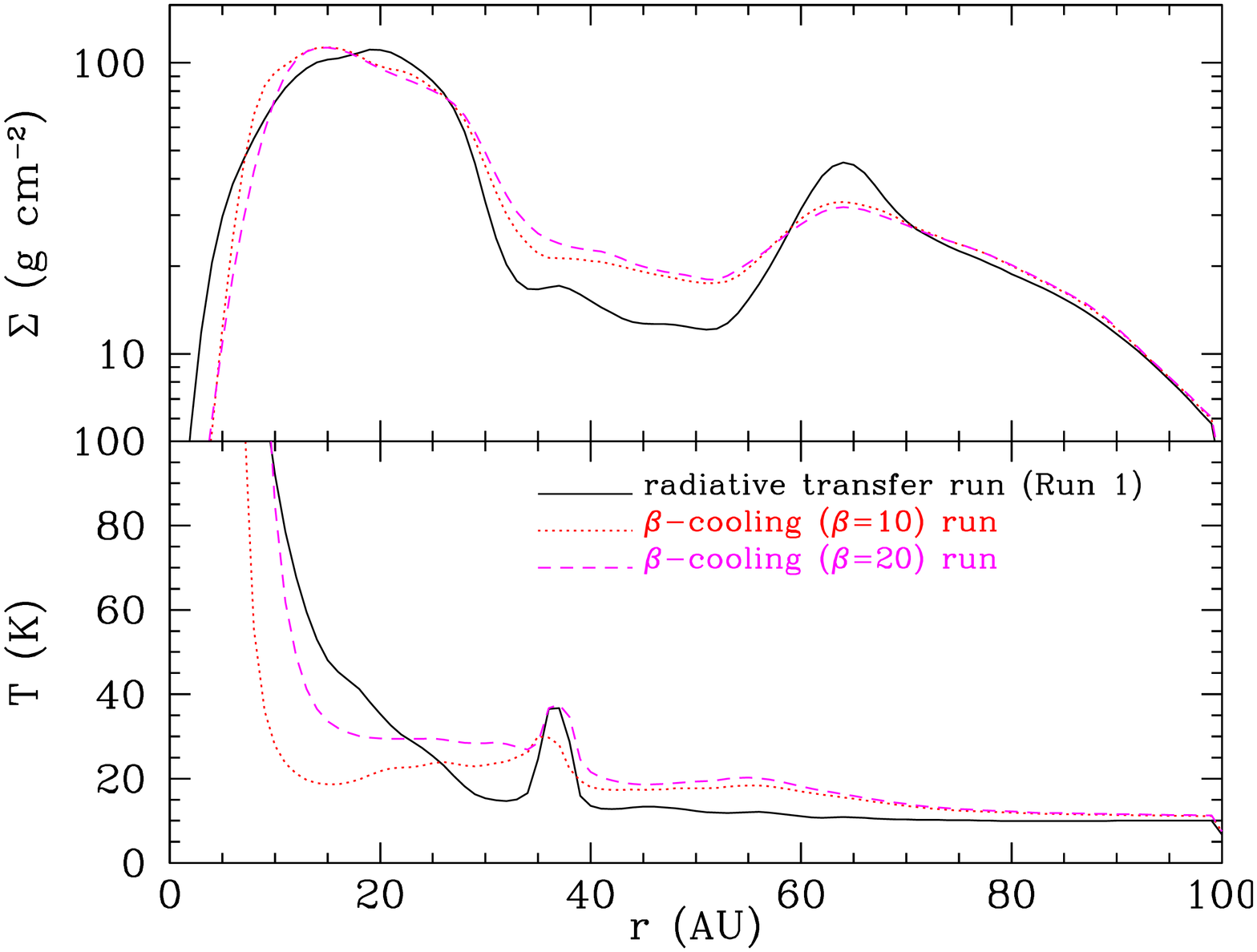}}
\caption{Azimuthally averaged disc surface density (top) and temperature (bottom), for the 3 runs in Figure 15 at $t=2$~kyr (i.e. during the initial gap opening phase). The protostellar disc is hotter in the runs that use the $\beta$-cooling approximation, so that gap opening is more difficult than in the run with a more detailed radiative transfer (Run 1). The outer regions of the protoplanetary disc (peak at $\sim37$ AU in the temperature plot) are also hotter due to inefficient cooling, limiting gas accretion onto the protoplanet.}
\label{fig:betaruns.temp}
\end{figure}
%%%%%%%%%%%%%%%%%%%%%%%%%%%%%%%%%%%%%%%%%%%%%

We compare the results of the simulations presented here with previous studies that employ the $\beta$-cooling approximation, in which the cooling time in the disc is proportional to the local orbital period. In this case the specific internal energy of each SPH particle is  set to
\begin{equation}
u = \frac{k_B T(R)}{\mu m_H\ (\gamma-1)}\,
\end{equation}
where $\mu=2.45$ is the mean molecular weight, and  $\gamma=7/5$ the adiabatic exponent.
The cooling rate of each particle is set to 
\begin{equation}
\left.\frac{du}{dt}\right|_{\rm cool}=-\frac{u}{t_{\rm cool}}\,,
\end{equation}
where
\begin{equation}
{t_{\rm cool}}=\beta\ \Omega_K^{-1}(R)\,,
\end{equation}
$R$ is the distance on the disc midplane, and 
\begin{equation}
\Omega_K(R)=\left(\frac{G M_\star}{R^3}\right)^{1/2}\,.
\end{equation}

We perform simulations with $\beta=10$ and $\beta=20$, i.e. relatively long cooling times.  These are similar to the ones used before in these type of studies: \cite{Baruteau:2011a}  use $\beta=15, 20, 30$,  whereas \cite{Malik:2015a} use  
$\beta=30$. Otherwise, the parameters of the simulations were the same as in Run 1. { We note that considering these long cooling times the discs are gravitationally stable and they are not expected to show any spiral structure in the absence of the protoplanet}. We compare the simulation we performed with the results in Run 1 in which radiative transfer is treated self-consistently with the method of \cite{Stamatellos:2007b} (see Section~\ref{sec:methods.rt}). 

{ We see  (Figure~\ref{fig:betaruns}) that in the runs that use the $\beta$-cooling approximation the protoplanet is not able to open up a gap and the migration is fast, as previous studies have found, and stops only once the protoplanet has reached the inner cavity around the star. During migration the accretion onto the protoplanet is much lower than in the runs in which the radiative transfer is treated in more detail.  
The reason for this is that in the $\beta$-cooling runs (that as in previous studies use rather large $\beta$ values; 15, 20, 30) the actual cooling is too slow compared with the cooling provided by the more detailed method, so that the protostellar disc is hotter (see Figure~\ref{fig:betaruns.temp}, bottom) and therefore the opening of a deep gap is not possible \citep{Crida:2006a}. Additionally, the cooling in the outer circumplanetary disc is inefficient, resulting in slower accretion of gas onto the protoplanet, limiting its mass growth.  { We discuss in detail the effect of cooling on the mass growth of the protoplanet and on gap opening in Section~\ref{sec:pmass}}.

Therefore, the detailed treatment of radiation transfer in the disc and the vicinity of the protoplanet, i.e. how the circumstellar and circumplanetary discs heat and cool, is  important for determining the migration and mass growth of a protoplanet evolving in a young, massive disc.

\section{The effect of the protoplanet's orbital radius (without radiative feedback from protoplanet/star}
\label{sec:sruns}

%%%%%%%%%%%%%%%%%%%%%%%%%%%%%%%%%%%%%%%%%%%%%
\begin{figure*}
\centerline{
\includegraphics[width=0.9\textwidth]{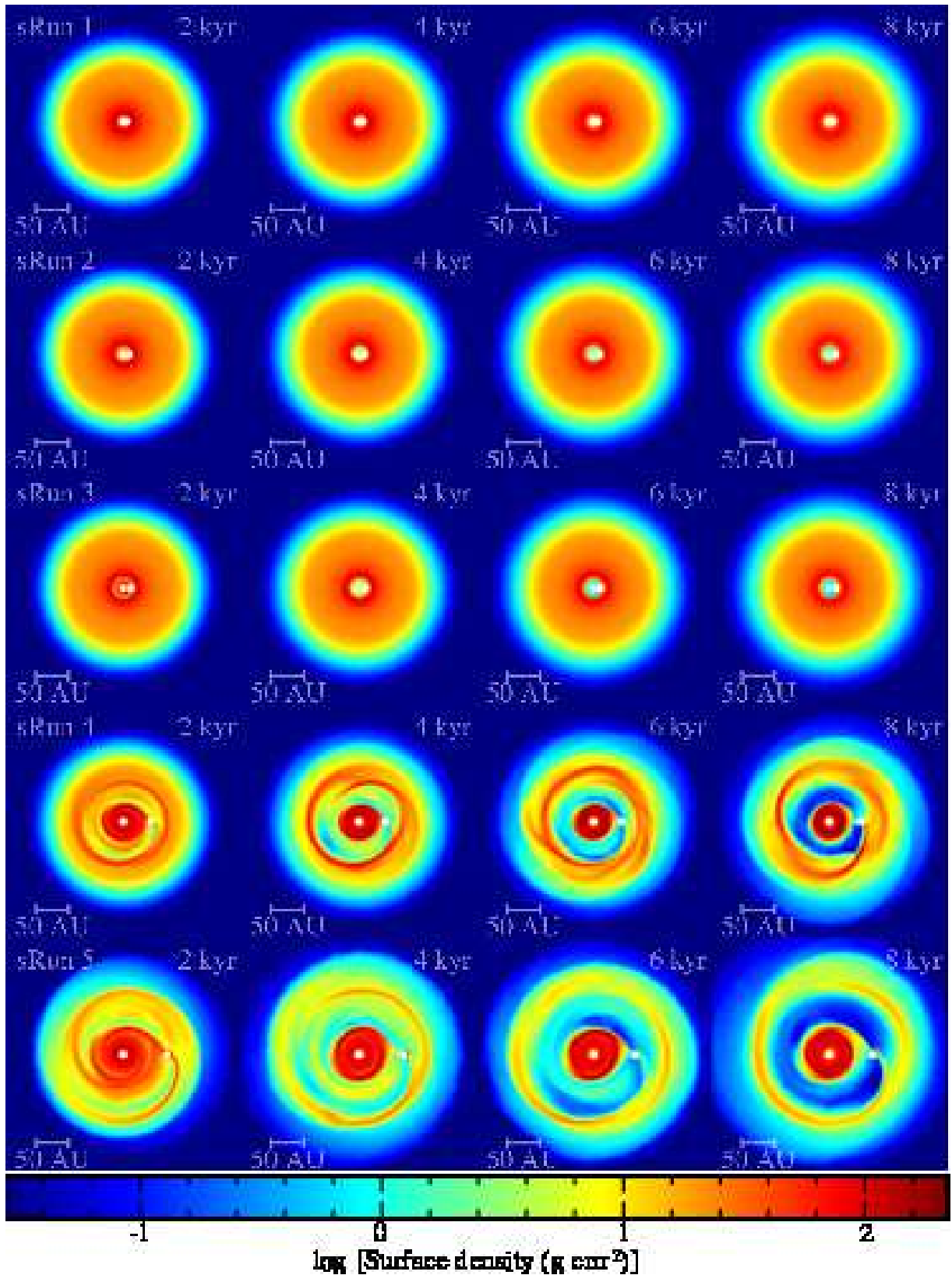}}
\caption{The evolution of a protoplanet in a 0.1-M$_{\sun}$ protostellar disc. The protoplanet is placed at different initial orbital radii within the disc: 5, 10, 20, 40 and 80 AU (top to bottom row; sRun1, ..., sRun5, respectively). There is a difference in the migration pattern depending on the position of the protoplanet in the disc. }
\label{fig:snapshots2}
\end{figure*}
%%%%%%%%%%%%%%%%%%%%%%%%%%%%%%%%%%%%%%%%%%%%%

{  We now examine  the evolution of Jupiter-mass protoplanets that are embedded at different orbital radii within protostellar discs (see Figure~\ref{fig:snapshots2}).} The details of the 5 runs (hereafter referred to as {\it sRuns}) are shown in Table~\ref{tab:srun}. A protoplanet is initially placed at 5, 10, 20,  50 and 80 AU from the central star at a circular orbit. The opacities used for these runs are the ones by  \cite{Semenov:2003a}.  For these runs we do not include radiative feedback from the star nor the protoplanet. The migration timescales for each run and the associated migration velocities are shown in Tables~\ref{tab:srun_migration} and ~\ref{tab:srun_vel}.

%%%%%%%%%%%%%%%%%%%%%%%%%%%%%%%%%%%%%%%%%%%%%
\begin{figure}
\centerline{\includegraphics[height=1\columnwidth,angle=-90]{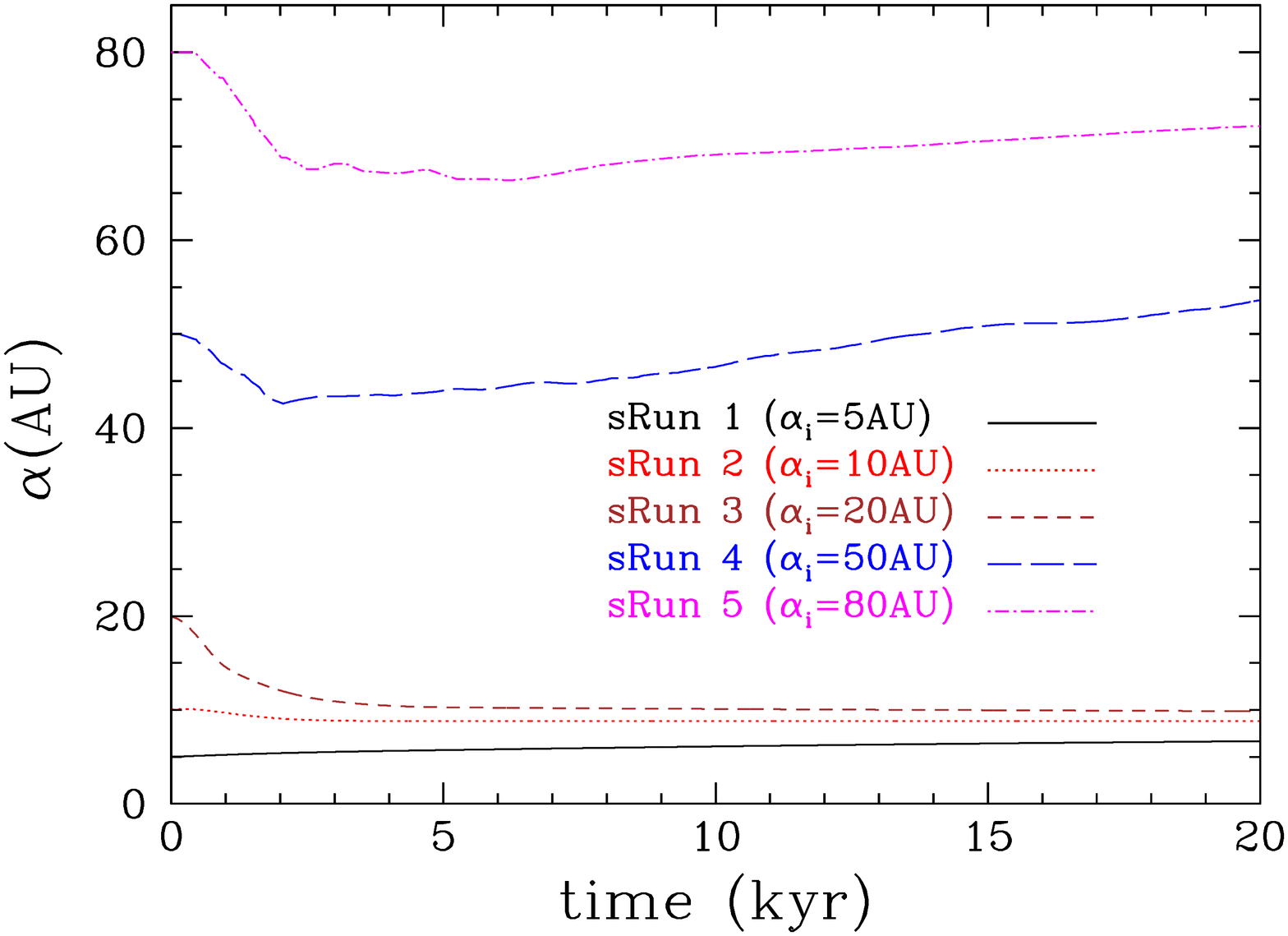}}
\caption{The semi-major axis evolution of a 1-M$_{\rm J}$ protoplanet in a 0.1-M$_{\odot}$ disc. When the protoplanet is  placed in the outer disc region, it migrates inwards on a Type I migration timescale. However, when the gap opens up, inward migration stops and it reverses into an outward migration. On the other hand, when the protoplanet is placed in the inner disc region ($\stackrel{<}{_\sim} 20$~AU), it migrates inwards initially  on a Type I migration timescale and once the gap is opened up the migration slows down and occurs on a Type II  timescale.}
\label{fig:srun.psemi}
\centerline{\includegraphics[height=1\columnwidth]{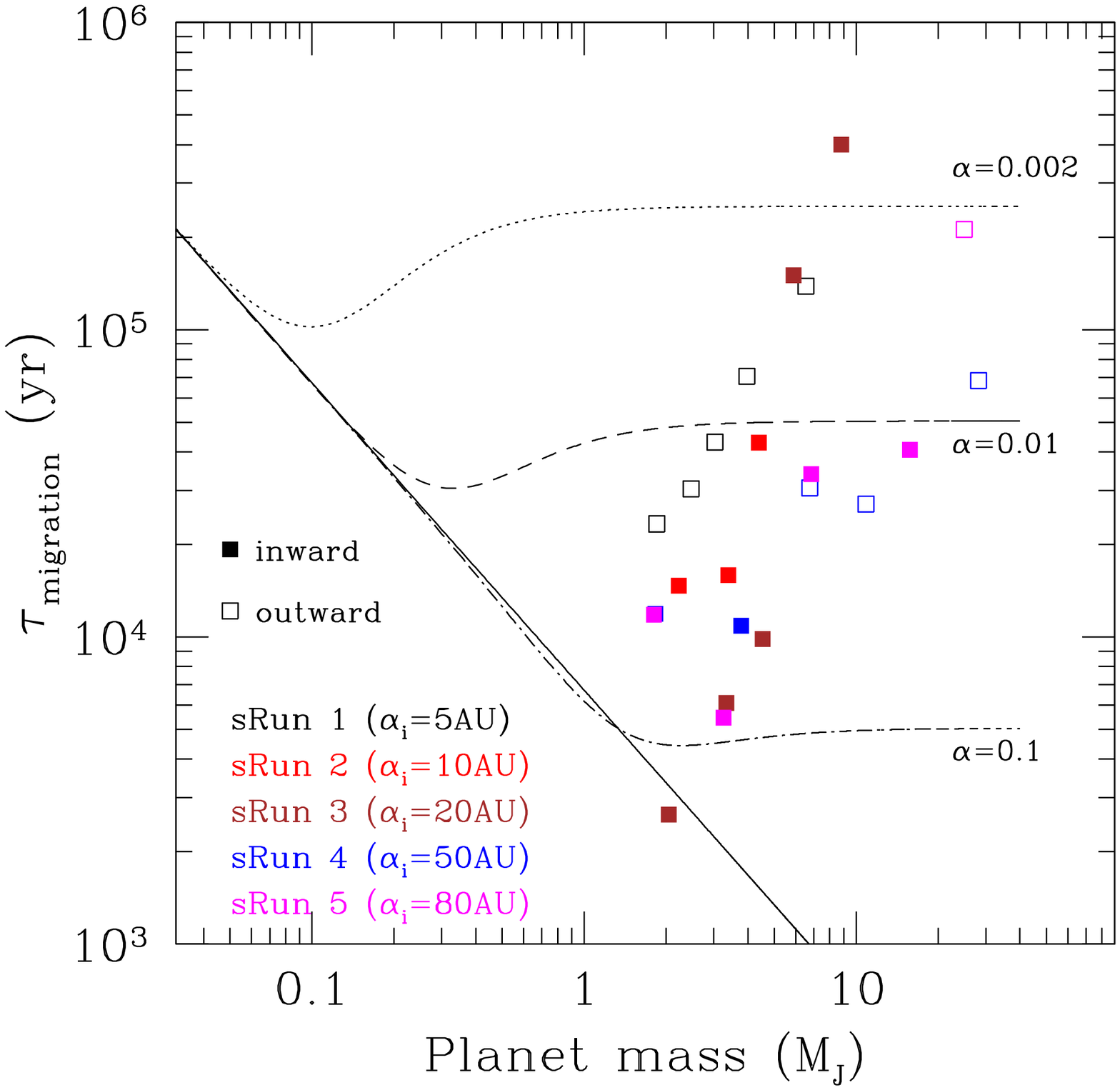}}
\caption{The migration timescales of protoplanets placed at different orbital radii within a protostellar disc. Protoplanets initially migrate inwards. If the protoplanet is in the unstable outer disc region the migration stops and changes  outwards  once a gap is opened up. Protoplanets that are in the inner disc region ($\stackrel{<}{_\sim} 20$~AU) continue to migrate inwards. Lines correspond to analytical calculations as in Figure~\ref{fig:tmig_a}. Filled boxes correspond to inward migration and empty boxes to outward migration.}
\label{fig:tmig_b}
\end{figure}
%%%%%%%%%%%%%%%%%%%%%%%%%%%%%%%%%%%%%%%%%%%%%

 %%%%%%%%%%%%%%%%%%%%%%%%%%%%%%%%%%%%%%%%%%%%%
\begin{figure}
\centerline{
\includegraphics[height=0.95\columnwidth,angle=-90]{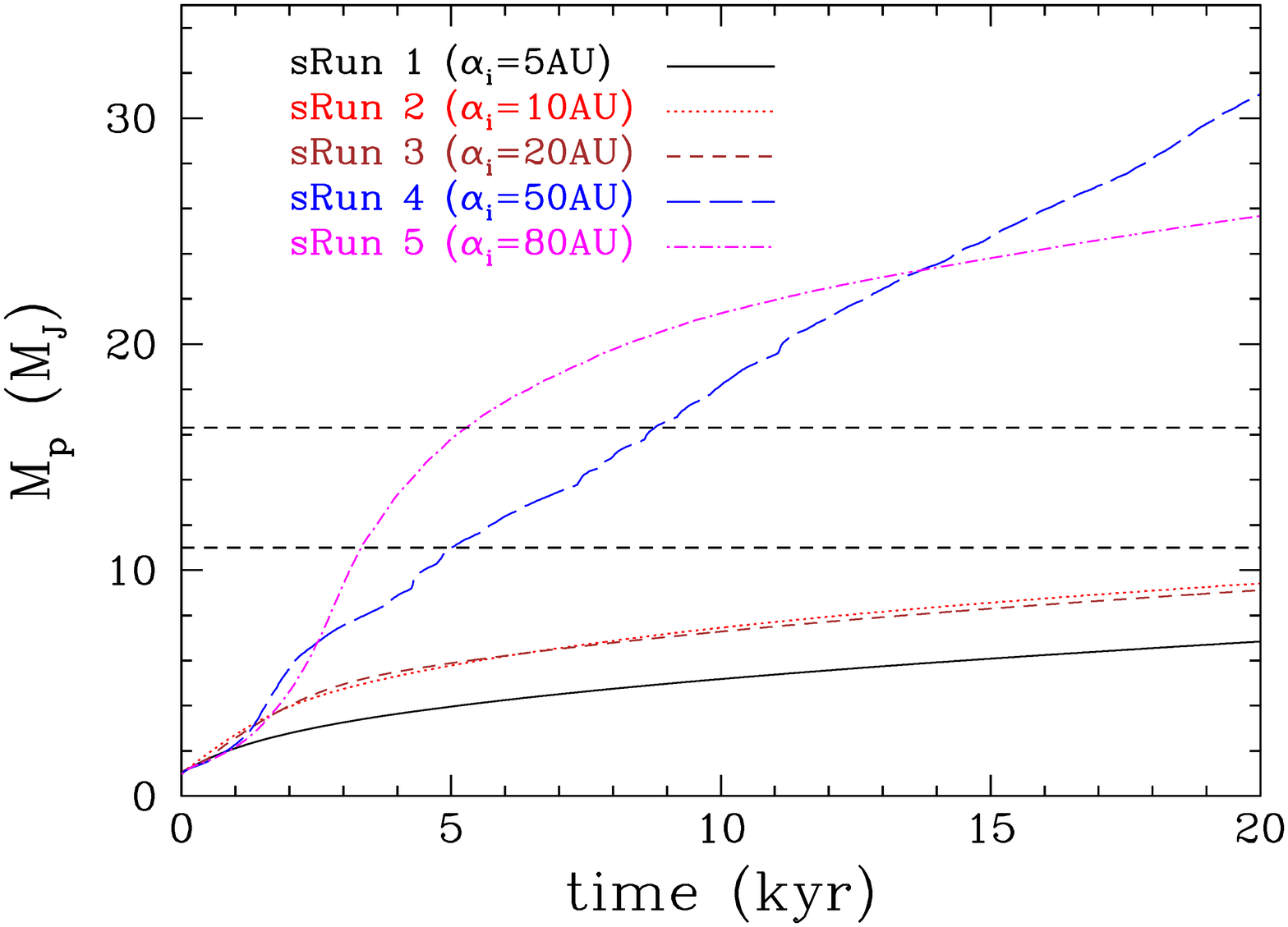}}
\caption{Mass growth of a 1-M$_{\rm J}$ protoplanet placed at different radii within a protostellar disc. Protoplanets that form in the outer disc regions tend to increase in mass considerably, becoming brown dwarfs. Protoplanets in the inner disc region increase in mass but not significantly enough to become brown dwarfs. The horizontal dashed lines correspond to the deuterium-burning mass-limit. }
\label{fig:srun.pmass}
\centerline{\includegraphics[height=0.95\columnwidth,angle=-90]{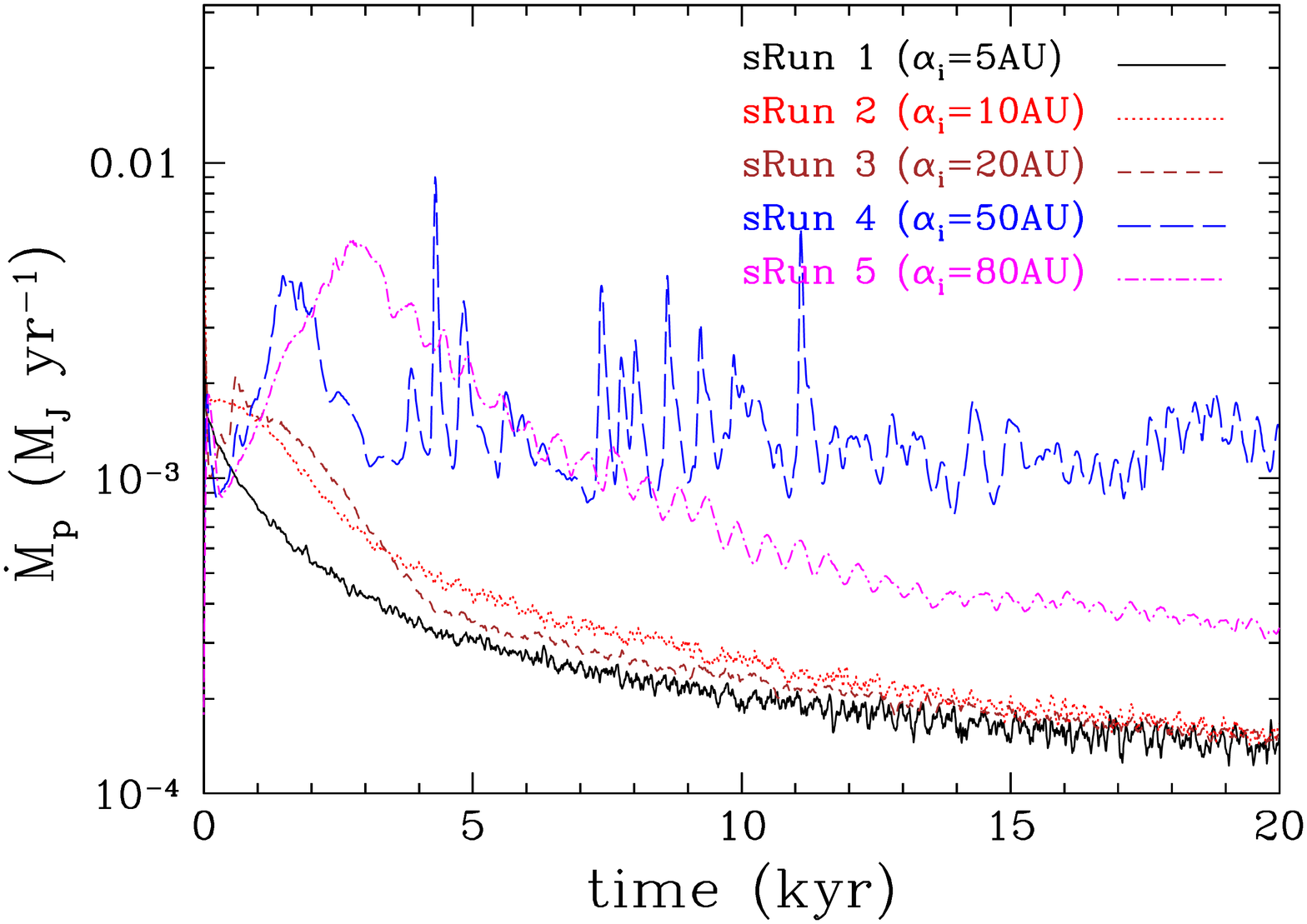}}
\centerline{\includegraphics[height=0.95\columnwidth,angle=-90]{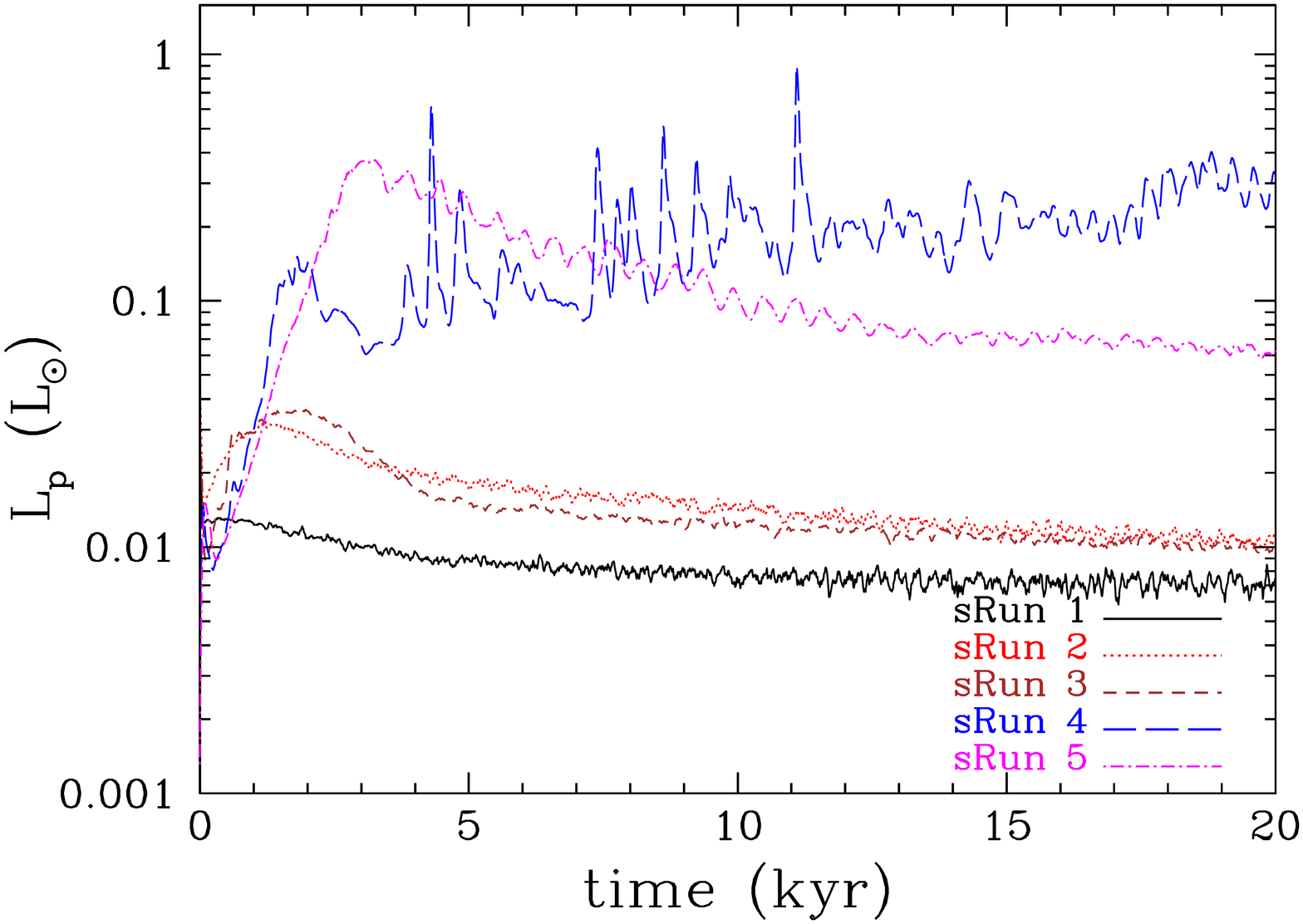}}
\caption{Accretion rate (top) onto 1-M$_{\rm J}$ protoplanet placed at different radii within a protostellar disc, and its corresponding accretion luminosity (bottom).   Protoplanets in the outer disc region continue to accrete gas vigorously  even after they open up a gap in the disc. Protoplanets in the inner disc region accrete significantly less as they reside within a cavity formed around the central star (see Figure~\ref{fig:snapshots2} - top three rows). (Note that in this set of runs the protoplanet's luminosity is not fed back into the disc.)}
\label{fig:srun.pacc}
\label{fig:srun.plum}
\end{figure}
%%%%%%%%%%%%%%%%%%%%%%%%%%%%%%%%%%%%%%%%%%%%%
  
 %%%%%%%%%%%%%%%%%%%%%%%%%%%%%%%%%%%%%%%%%%%%%
\begin{figure}
\centerline{\includegraphics[height=0.95\columnwidth,angle=-90]{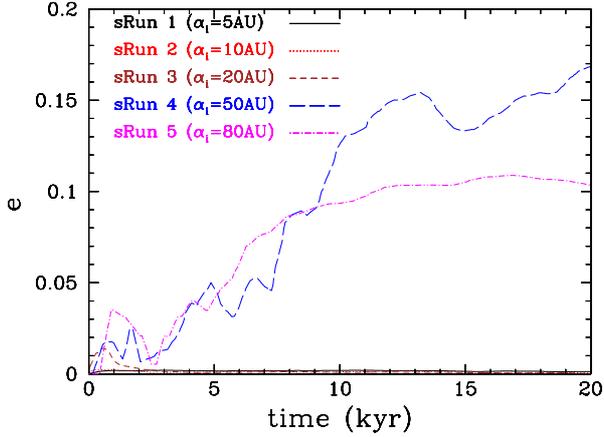}}
\caption{The eccentricity of a 1-M$_{\rm J}$ protoplanet placed at different radii within a protostellar disc.  If the protoplanet is in the outer disc region interactions with the gravitationally unstable gap edges result in eccentricity growth. If the protoplanet is in the inner disc region interactions with a stable disc result in a circular orbit.}
\label{fig:srun.pecc}
\end{figure}
%%%%%%%%%%%%%%%%%%%%%%%%%%%%%%%%%%%%%%%%%%%%%

 %%%%%%%%%%%%%%%%%%%%%%%%%%%%%%%%%%%%%%%%%%%%%
\begin{figure}
\centerline{\includegraphics[height=0.95\columnwidth,angle=-90]{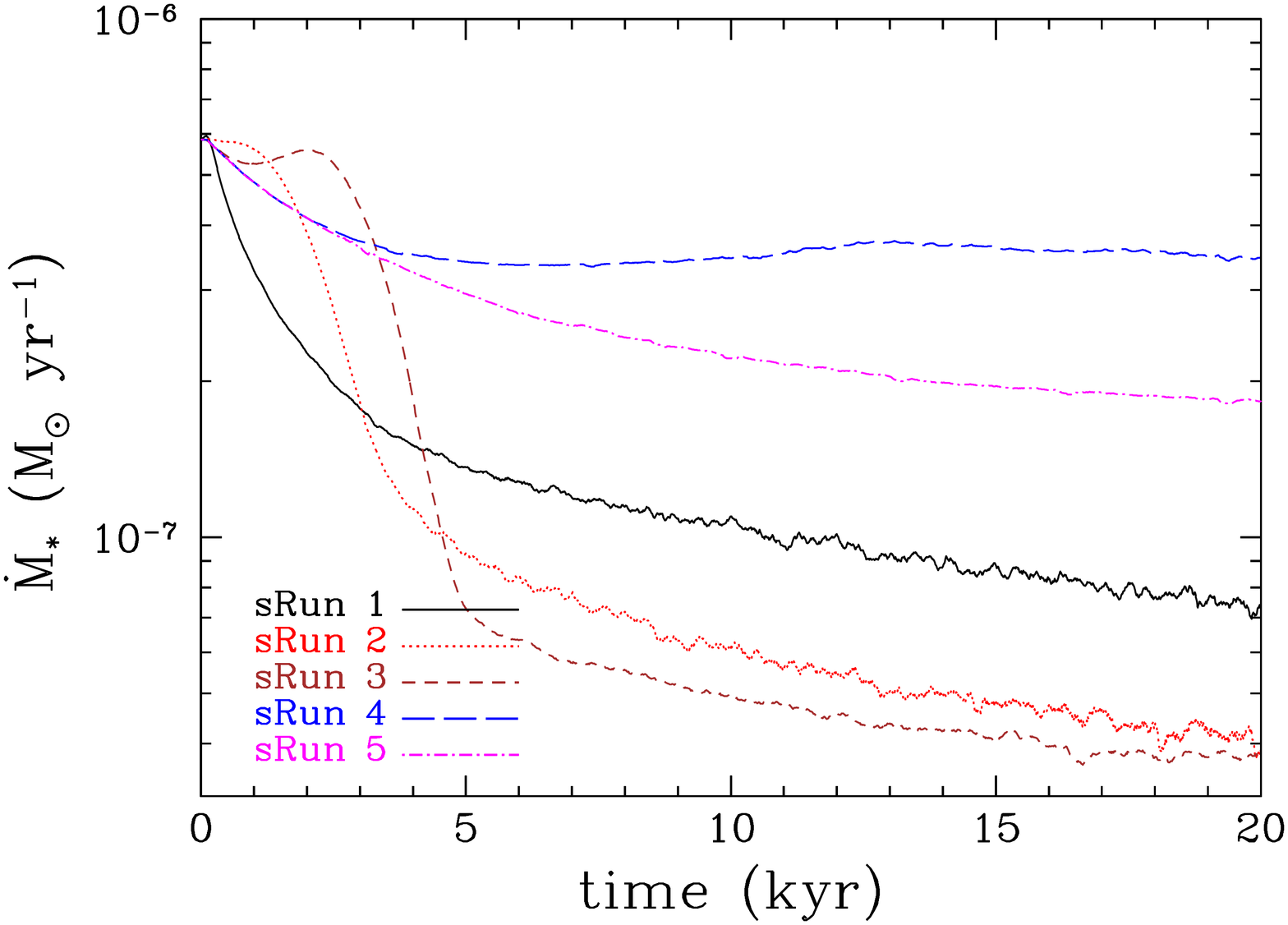}}
\centerline{\includegraphics[height=0.95\columnwidth,angle=-90]{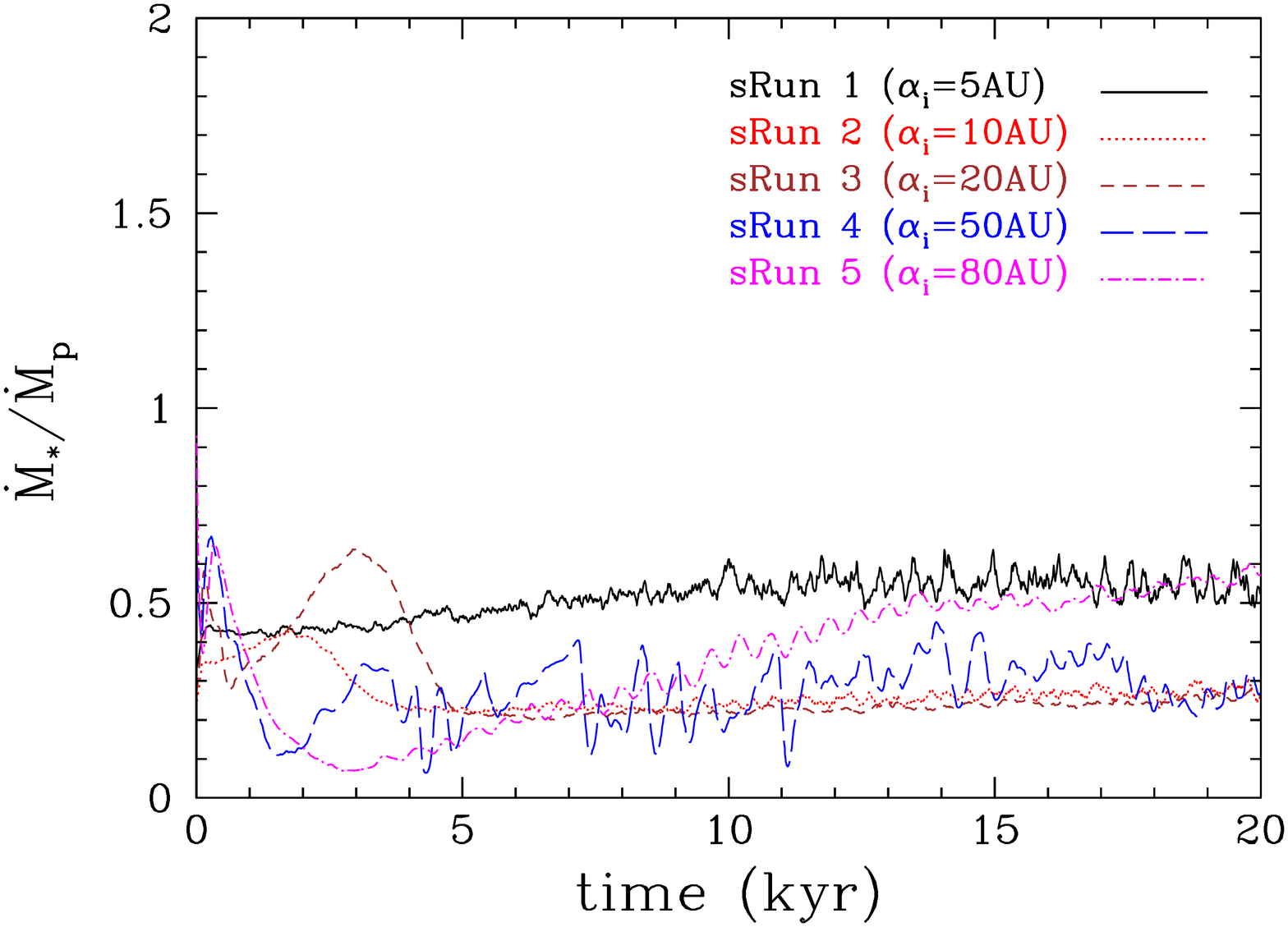}}
\caption{The accretion rate onto the central star (top) and the ratio of the accretion rate onto the star to the accretion rate onto the protoplanet (bottom). The accretion rate onto the star is lower when the protoplanet is in the inner disc region. Generally the star accretion rate is only half (or less than) the accretion rate onto the protoplanet. }
\label{fig:srun.paccstar}
\label{fig:srun.paccratio}
\end{figure}
%%%%%%%%%%%%%%%%%%%%%%%%%%%%%%%%%%%%%%%%%%%%%

\begin{table}
\caption{Evolution of a protoplanet at different orbital radii within a disc  (without radiative feedback): Simulation parameters  (same as in Table~\ref{tab:standardrun}) }
\label{tab:srun}
\centering
\begin{tabular}{@{}cccccc} \hline
\noalign{\smallskip}
Run id  & $M_{p,i}$ 	&  $\alpha_{i,p}$ & $M_{p,f}$ & $\alpha_{f,p}$  &$e_f$\\
	     & (M$_{\rm J}$) &  (AU) & (M$_{\rm J}$)&(AU) &\\
\noalign{\smallskip}
\hline
\noalign{\smallskip}
sRun1 &	   1    &   5 & 6.8 & 6.7 & 0.0013 	 \\
sRun2 &    1    & 10 	 & 9.4	&  8.8 & 0.007\\
sRun3 &    1    & 20 	 & 9.1 	& 9.9 &0.0002 \\
sRun4 &	   1    & 50 & 31	&  53  &0.17\\
sRun5& 	   1    & 80 & 26	&  72 &0.10 \\
 \noalign{\smallskip}
\hline
\end{tabular}
\end{table}

The semi-major axis evolution of the protoplanets is shown in Figure~\ref{fig:srun.psemi}. When the protoplanet is initially placed in the outer disc region (which is characterised by a low Toomre-Q parameter) it migrates inwards on a Type I migration timescale ($\sim10^4$~yr), but once a gap opens up the inward migration stops and it changes into outward migration (see Figure~\ref{fig:tmig_b}). On the other hand, when the protoplanet is placed in the inner disc region ($\stackrel{<}{_\sim} 20$~AU) it migrates inwards initially on a Type I migration timescale ($\sim10^4$~yr) and once the gap is opened up the migration continues to be inwards; however  it slows down and occurs on a timescale typical of Type II migration ($\sim10^5$~yr; see Figure~\ref{fig:tmig_b}). As pointed out in the previous section, a necessary requirement  for outward migration is interactions with a gravitationally unstable outer gap edge. In the inner disc region close to the central star the disc is hotter and therefore stable, so that outward migration does not happen. The only exception is the run in which the protoplanet initially placed at 5 AU; in this run the protoplanet   migrates outwards (but only slightly) throughout the simulated time because it resides near the outer edge of a quickly formed inner cavity around the central star. It accretes less gas than in the other runs; mainly it accretes high angular momentum gas from the outer edge of the cavity, as indicated by the lack of strong inner wake.

We conclude that  the survival of the protoplanet on a relatively wide orbit (assuming that it has formed on this wide orbit) is secured when the disc is massive and cool enough for gravitationally unstable edges to develop. If the protoplanet forms in the outer disc region then its initial inward migration will turn into an outward migration once it opens up a gap in the disc. If the protoplanet forms in the inner disc region its inward migration will slow down considerably (migration timescale of $\stackrel{>}{_\sim} 10^5$~yr; see Table~\ref{tab:srun_migration}) so it can stay on a wide orbit once the disc has dissipated (assuming that the disc dissipates fast enough).

The mass growth of the protoplanet also depends on its location in the disc (see Figure~\ref{fig:srun.pmass}). Protoplanets placed in the outer disc region accrete gas while they open up a gap and they continue to accrete gas at a high rate as they migrate outwards in the disc (Figure~\ref{fig:srun.pacc}, top). They increase in mass considerably becoming brown dwarfs by the end of the simulated 20~kyr, with a mass of $\sim 25-30$~M$_{\rm J}$. Their corresponding accretion luminosities are $\stackrel{>}{_\sim} 0.1$~L$_{\sun}$  \citep[see also][]{Inutsuka:2010a} (Figure~\ref{fig:srun.plum}) and therefore they may be readily observable if observed while they are still young. On the other hand, protoplanets that form in the inner disc region quickly find themselves within a gas-poor cavity formed around the central star and accretion onto them happens at a much lower rate than on protoplanets in the outer disc region  (Figure~\ref{fig:srun.pacc}). Their mass growth is slower and their final mass at the end of the simulation  is below the deuterium burning limit ($8-10$~M$_{\rm J}$). However, their mass continues to increase so they may also end up as brown dwarfs if the disc dissipates slowly. If { the longer term} disc dissipation happens on a shorter timescale than the viscous one  in other ways, e.g. by photoevaporation \citep{Alexander:2006a,Owen:2010a} or by disc winds \citep{Suzuki:2009b,Suzuki:2010a,Suzuki:2014a,Gressel:2015a,Bai:2016a} \citep[see review by][]{Alexander:2014a}, then the mass of the protoplanet will remain below the deuterium burning limit \citep[see discussion by][]{Kratter:2010b}.

The eccentricity of the protoplanet in the different runs is shown in Figure~\ref{fig:srun.pecc}. Protoplanets in the outer disc region interact with the gravitationally unstable gap edges resulting in eccentricity growth ($e\stackrel{>}{_\sim}0.1$). Protoplanets in the inner disc region interact  in a gravitationally stable disc and their orbits remain nearly circular. Therefore, for a protoplanet to to have a significantly eccentric orbit it needs to have formed at a sufficiently large distance from the central star. Protoplanets formed by gravitationally instability naturally form at such large radii when the disc can be both gravitationally unstable and cool fast enough \citep{Stamatellos:2009a, Boley:2009a}. Giant planets are difficult to form by core accretion at such large radii. Therefore, eccentric giant planets on wide orbits in single-planet systems may  have formed due to gravitational instabilities. 

The accretion rate onto the central star also depends on the orbital radius of the protoplanet  (Figure~\ref{fig:srun.paccstar}). A protoplanet in the inner disc region quickly (within a few kyr) forms a cavity  around the central star. During this period  the accretion rate onto the star is high, as the protoplanet drives gas accretion onto the star. After the central cavity opens up the protoplanet starves the star from gas, with the accretion rate dropping considerably. Mass flows through the inner cavity through accretion streams as in a circumbinary disc \citep{Artymowicz:1996a} but accretion happens preferentially onto the protoplanet, the lower mass object of the system as expected: the protoplanet's accretion rate is twice that of the star's (Figure~\ref{fig:srun.paccratio}).

%% CPD DISCS SRUNS %%%%%%%%%%%%%%%%%%%%%%%%%%%%%%%%%%%%%%%
%\begin{figure}
%\enterline{\includegraphics[width=0.8\columnwidth]{gap_b.eps}}
%\label{fig:gap_b}
% \caption{The region around the protoplanet. Same as in Figure~\ref{fig:gap_a} but for the runs listed in  Table~\ref{tab:srun}.}
%\centerline{\includegraphics[width=0.8\columnwidth]{cpd_b.eps}}
%\label{fig:cpd_b}
%\caption{The properties  of the circumplanetary disc in the simulations listed in Table~\ref{tab:srun} at $t=8$~kyr (parameters same as in Figure~\ref{fig:cpd_a}).}
%\centerline{\includegraphics[height=0.8\columnwidth, angle=-90]{srun.pmhill.eps}}
%\label{fig:hill_b}
%\caption{The mass within the protoplanet's Hill radius as a function of time for the simulations listed in Table~\ref{tab:srun}.}
%\end{figure}
%%%%%%%%%%%%%%%%%%%%%%%%%%%%%%%%%%%%%%%%%%%%%

\section{The effect of the protoplanet's orbital radius (with radiative feedback from the protoplanet/star)}

{ We now examine the evolution of a 1-M$_{\rm J}$ protoplanet placed at different orbital radii within a protostellar disc, including the radiative feedback from the protoplanet and from the central star  as described in Section~\ref{sec:methods.rt} (see also Appendix~\ref{sec:star_rt}) (setting $T({\rm 1AU})=250$~K for the pseudo-background temperature due to the central star). These simulations (see Table~\ref{tab:rrun}, Figure~\ref{fig:snapshots3}) probably represent more realistically the evolution of a protoplanet within a young disc. We  use the \cite{Semenov:2003a} opacities for this set of runs.}

%%%%%%%%%%%%%%%%%%%%%%%%%%%%%%%%%%%%%%%%%%%%%
\begin{figure*}
\centerline{
\includegraphics[width=0.9\textwidth]{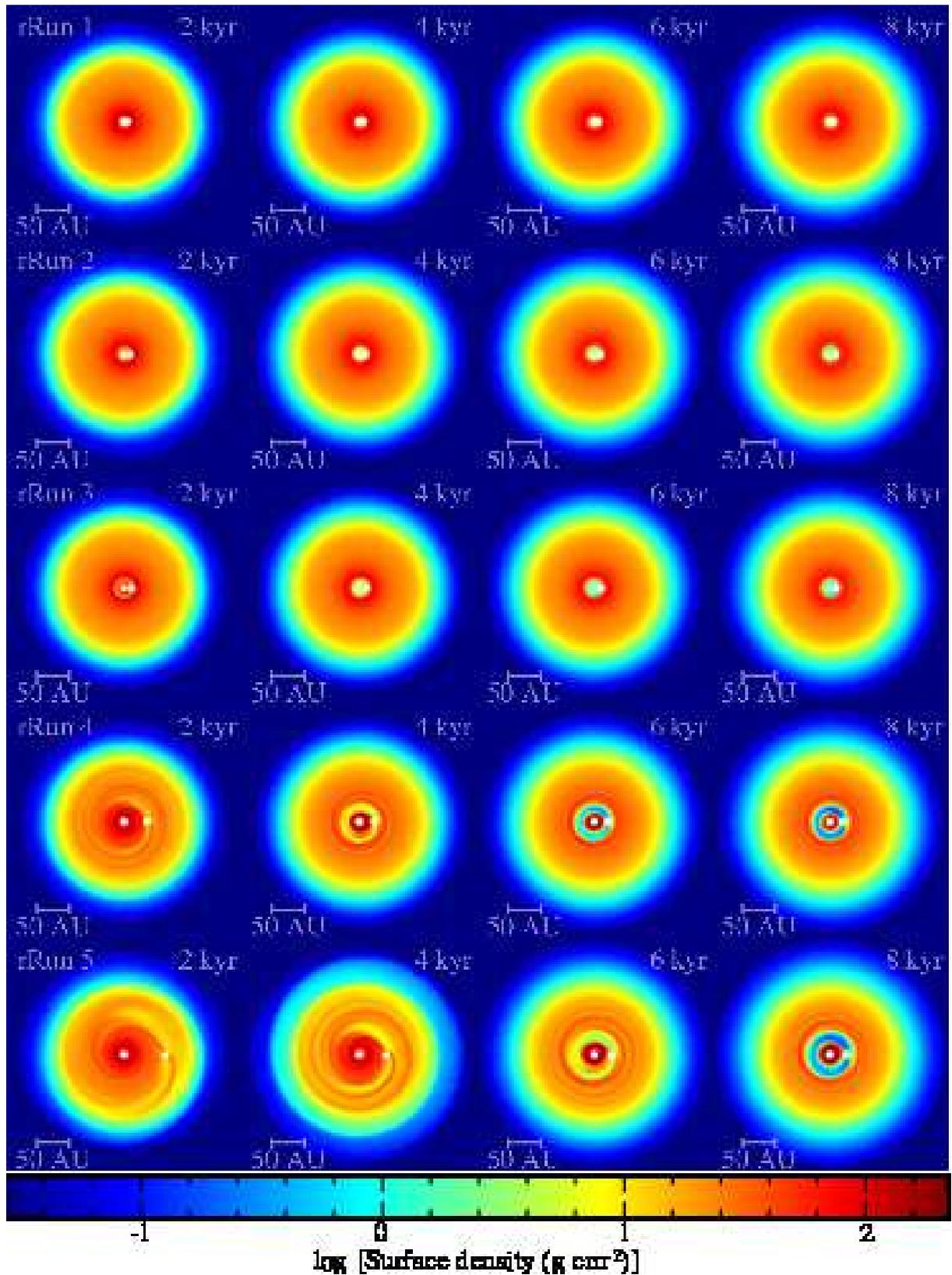}
}
\caption{The evolution of the disc surface density in the 5 simulations in which the radiative feedback from the protoplanet  and the star are included  (see Table~\ref{tab:rrun}). The disc is stable due to the feedback provided by the protoplanet. The protoplanet generally migrates inwards, creating a cavity around the central star (see discussion in text).}
\label{fig:snapshots3}
\end{figure*}
%%%%%%%%%%%%%%%%%%%%%%%%%%%%%%%%%%%%%%%%%%%%%

 \begin{table}
\caption{Evolution of a prototoplanet at different orbital radii within a disc  (with radiative feedback from the star and the protoplanet): Simulation parameters  (same as in Table~\ref{tab:standardrun}) }
\label{tab:rrun}
\centering
\begin{tabular}{@{}cccccc} \hline
\noalign{\smallskip}
Run id  	& $M_{p,i}$ &  $\alpha_{p,i}$ & $M_{p,f}$ & $\alpha_{p,f}$  &$e_f$\\
	    &(M$_{\rm J}$) &  (AU) &  (M$_{\rm J}$)&(AU)& \\
\noalign{\smallskip}
\hline
\noalign{\smallskip}
rRun1  & 1& 5 		&  	8.0	&  	6.6   & 0.0020\\
rRun2  & 1& 10 	& 	11.4	& 	9	& 0.0026\\
rRun3  & 1& 20 	& 	11.1	& 	10	& 0.0008\\
rRun4  & 1& 50	 	& 	15.6	&	18	& 0.0005\\
rRun5  & 1& 80 	& 	18.7	& 	24	& 0.0015\\
\noalign{\smallskip}
\hline
\end{tabular}
\end{table}

The semi-major axis evolution of the protoplanet is shown in Figure~\ref{fig:rrun.psemi}. As in the previous runs the protoplanet migrates inwards while opening up a gap (or a cavity, if its initial semi-major axis is small enough, i.e. $<20$~AU). As in the previous runs without radiative feedback, a protoplanet that orbits close to the central star ($\alpha_{p,i}=5$~AU) migrates slightly outwards. For protoplanets that are initially in the inner disc region, their radiative feedback does not play an important role, and their evolution is quite similar to the case without any feedback, as presented in Section~\ref{sec:sruns}. The edges of the gap opened up by the protoplanet are already hot and gravitationally stable; therefore, the protoplanet continues to migrate inwards but at a slower pace (Figure~\ref{fig:rrun.tmig_c}). On the other hand, radiative feedback from the protoplanet plays an important role for planets that are initially in the outer, relatively cold disc region. This radiative feedback heats the gap edges, stabilising them and ensuring that the protoplanets continue to migrate inwards in contrast to the case without radiative feedback, in which, after the gap opens up, the protoplanet migrate outwards.

 %%%%%%%%%%%%%%%%%%%%%%%%%%%%%%%%%%%%%%%%%%%%%
\begin{figure}
\centerline{\includegraphics[height=0.95\columnwidth,angle=-90]{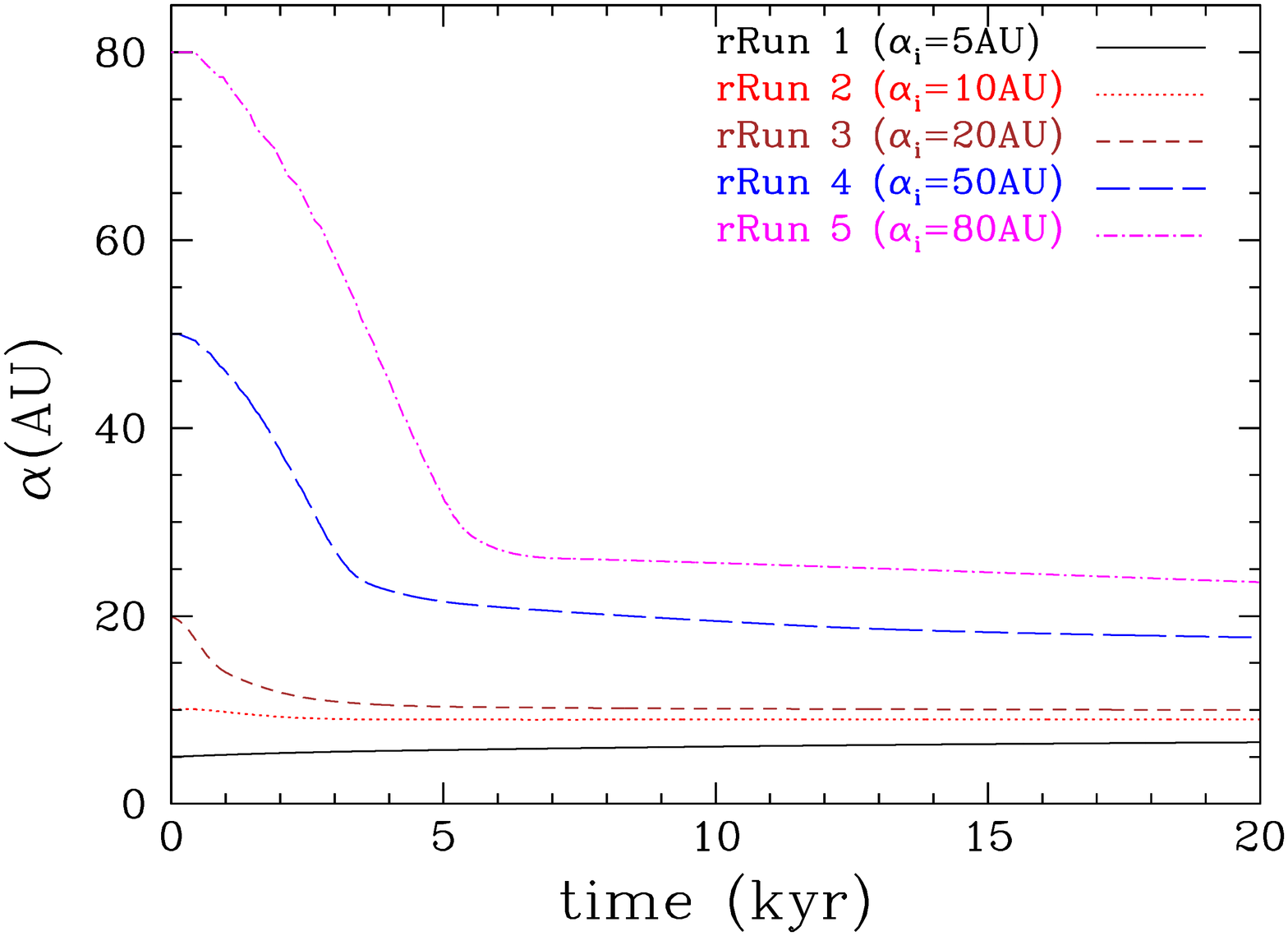}}
\caption{The semi-major axis evolution of an initially 1-M$_{\rm J}$ protoplanet in a 0.1-M$_{\odot}$ disc, including the radiative feedback from the protoplanet.  The protoplanet migrates inwards opening up a gap. Once the gap opens up, the migration slows down.  There is no outward migration irrespective of the orbital radius as the protoplanet's  radiation stabilises the gap edges.}
\label{fig:rrun.psemi}
\centerline{
\includegraphics[height=0.95\columnwidth]{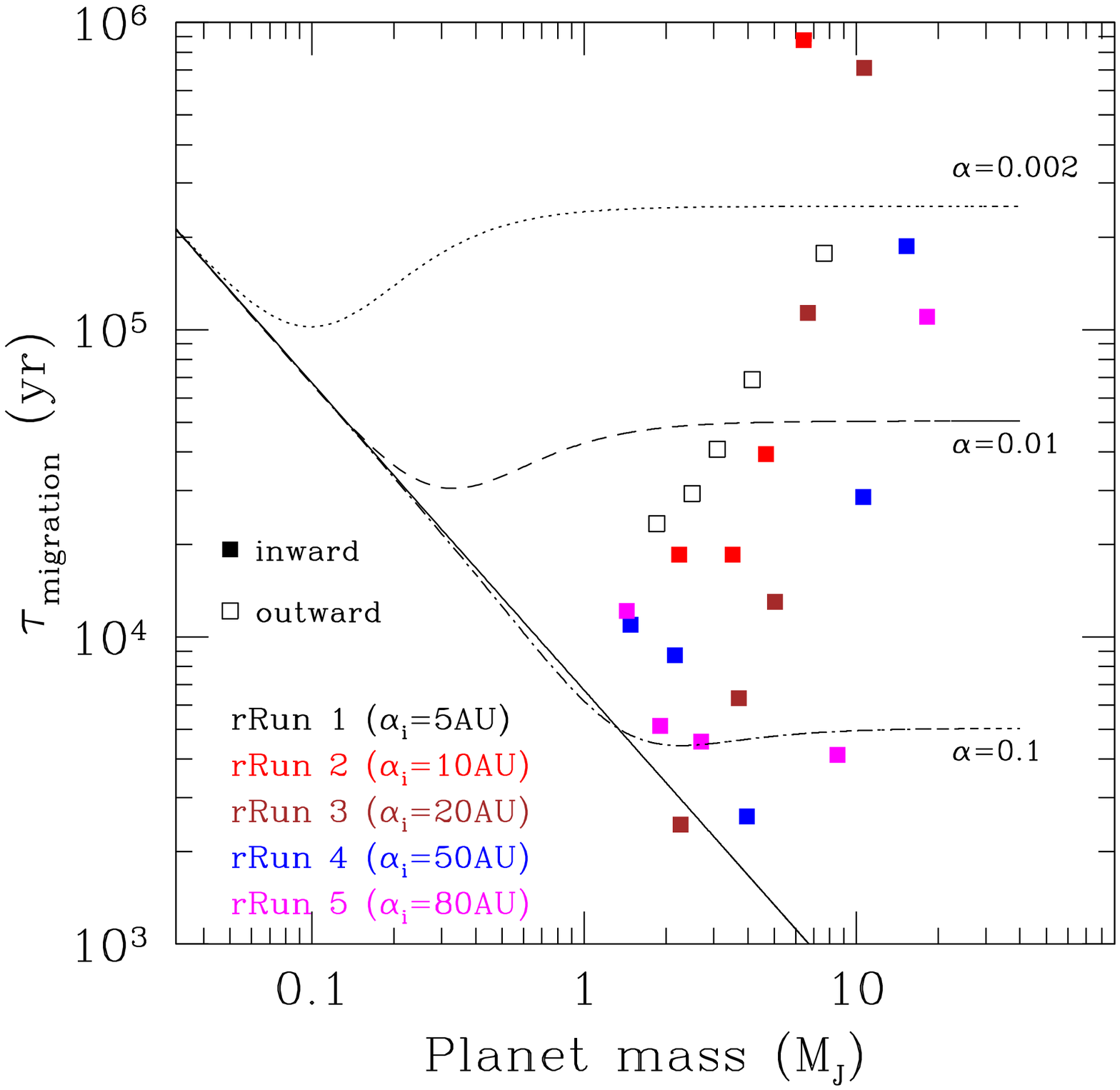}}
\caption{The migration timescales of protoplanets placed at different orbital radii within a protostellar disc, in the runs where both radiative feedback from the protoplanet and the star are included. Protoplanets  migrate inwards, initially  on a Type I migration timescale ($\sim10^4$~yr) and once the gap is opened up on a Type II migration timescale ($\sim10^5$~yr).
 Lines correspond to analytical calculations as in Figure~\ref{fig:tmig_a}. Filled boxes correspond to inward migration and empty boxes to outward migration.}
\label{fig:rrun.tmig_c}
\end{figure}
%%%%%%%%%%%%%%%%%%%%%%%%%%%%%%%%%%%%%%%%%%%%%

 %%%%%%%%%%%%%%%%%%%%%%%%%%%%%%%%%%%%%%%%%%%%%
\begin{figure}
\centerline{\includegraphics[height=0.95\columnwidth,angle=-90]{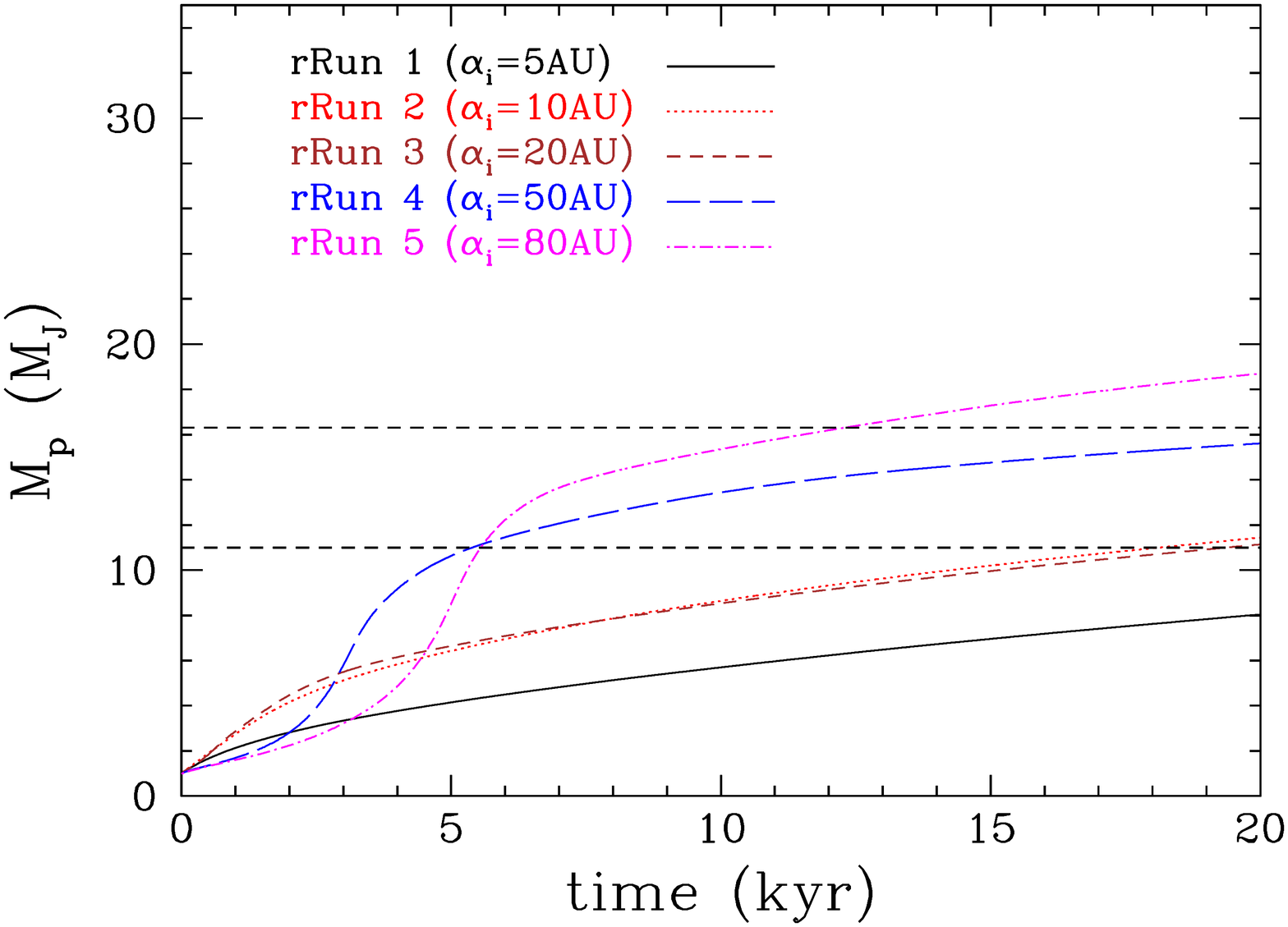}}
\caption{Mass growth of a 1-M$_{\rm J}$ protoplanet placed at different radii within a protostellar disc, when the radiative feedback from the protoplanet and the star are included. The mass growth of the protoplanet is suppressed and the protoplanet remains within the planetary mass regime even if it has formed in the outer disc region. The horizontal dashed lines correspond to the deuterium-burning mass-limit.}
\label{fig:rrun.mass}
\centerline{\includegraphics[height=0.95\columnwidth,angle=-90]{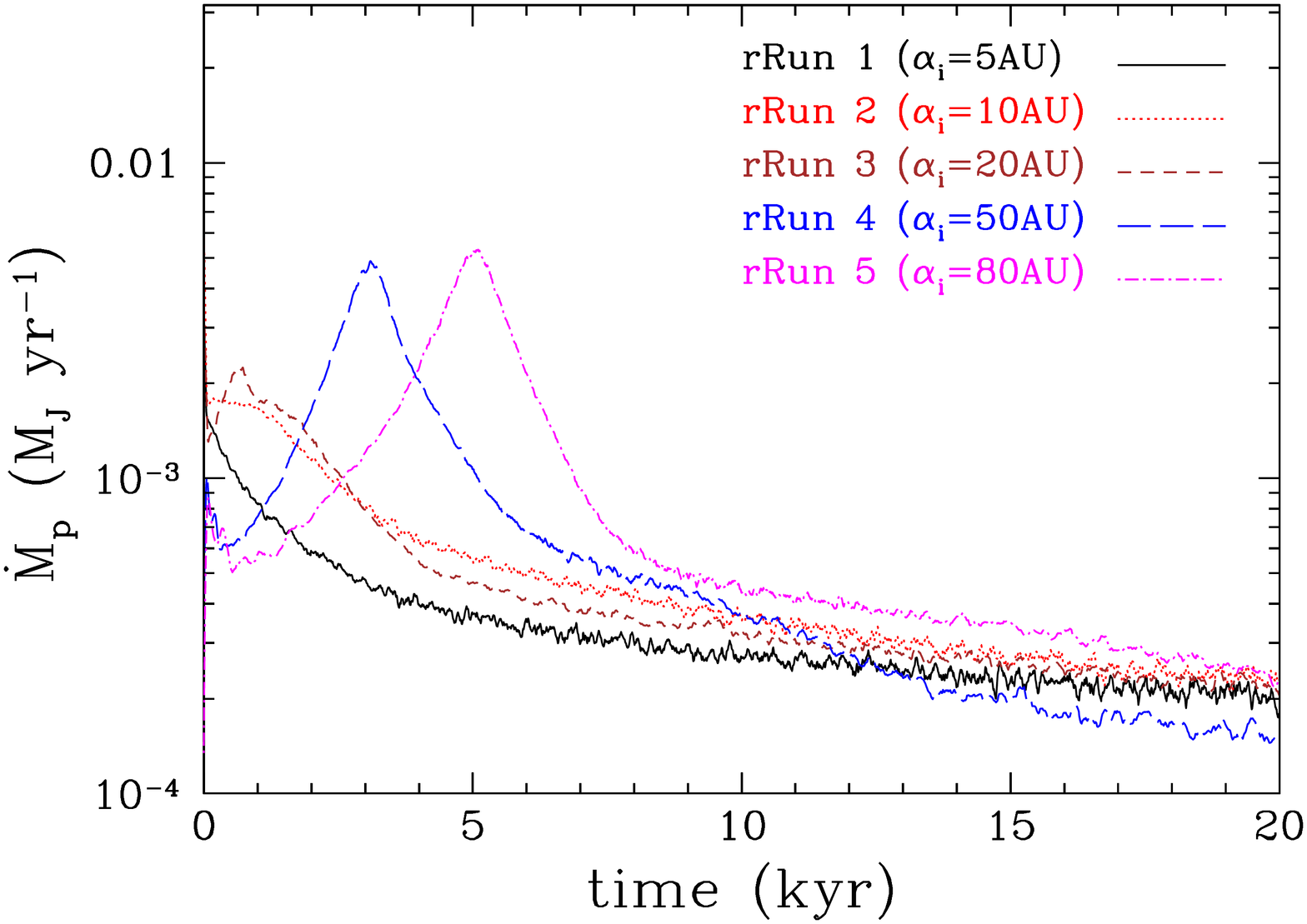}}
\centerline{\includegraphics[height=0.95\columnwidth,angle=-90]{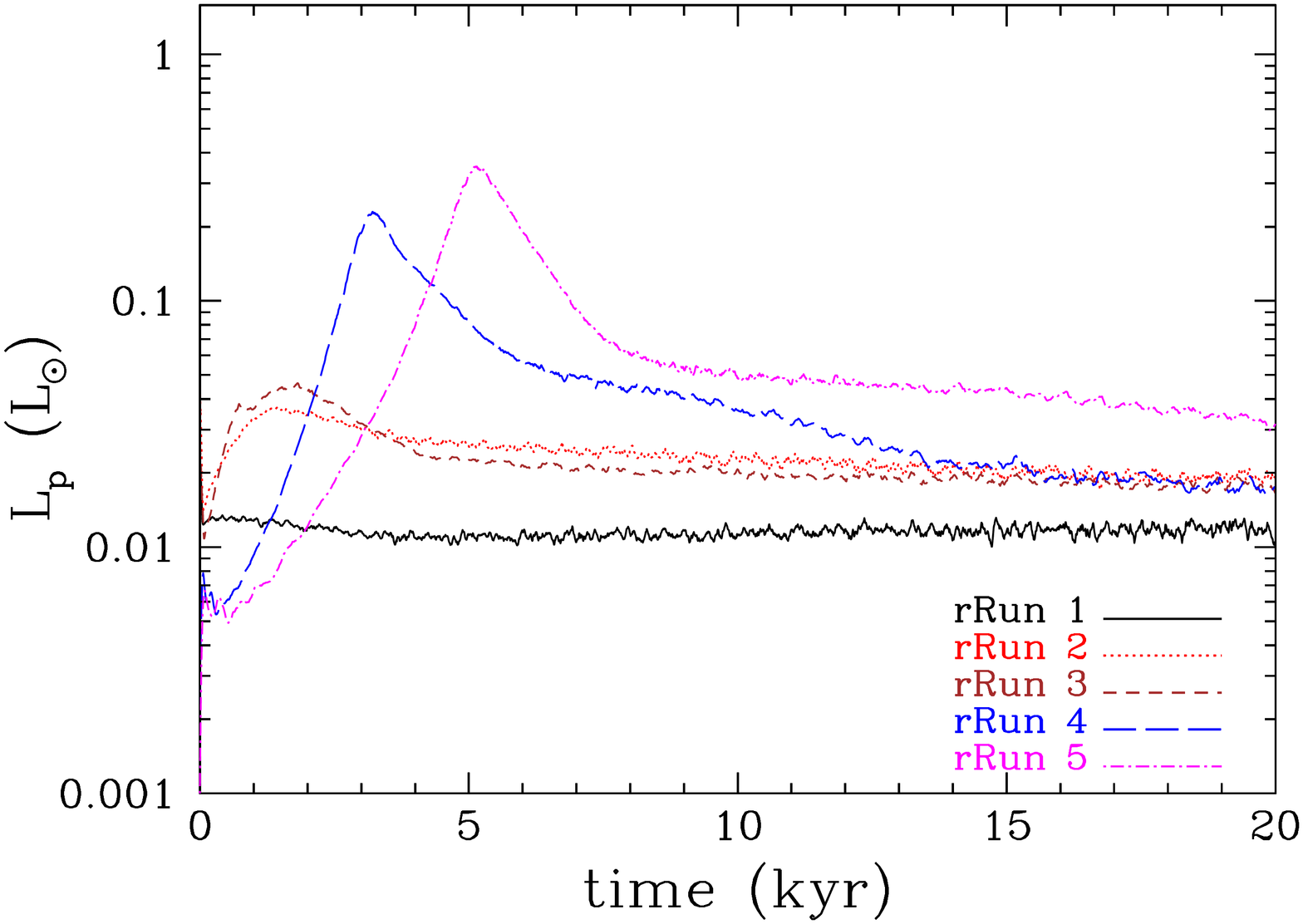}}
\caption{Accretion rate (top) onto a 1-M$_{\rm J}$ protoplanet placed at different radii within a protostellar disc, and its corresponding accretion luminosity (bottom), in the runs where both radiative feedback from the protoplanet and the star are included.   The protoplanets during the gap opening phase exhibit high accretion rates and corresponding accretion luminosities. (Note that in this set of runs the protoplanet's luminosity is  fed back into the disc.)}
\label{fig:rrun.pacclum}
\end{figure}
%%%%%%%%%%%%%%%%%%%%%%%%%%%%%%%%%%%%%%%%%%%%%

%%%%%%%%%%%%%%%%%%%%%%%%%%%%%%%%%%%%%%%%%%%%%
\begin{figure}
\centerline{
\includegraphics[height=0.95\columnwidth,angle=-90]{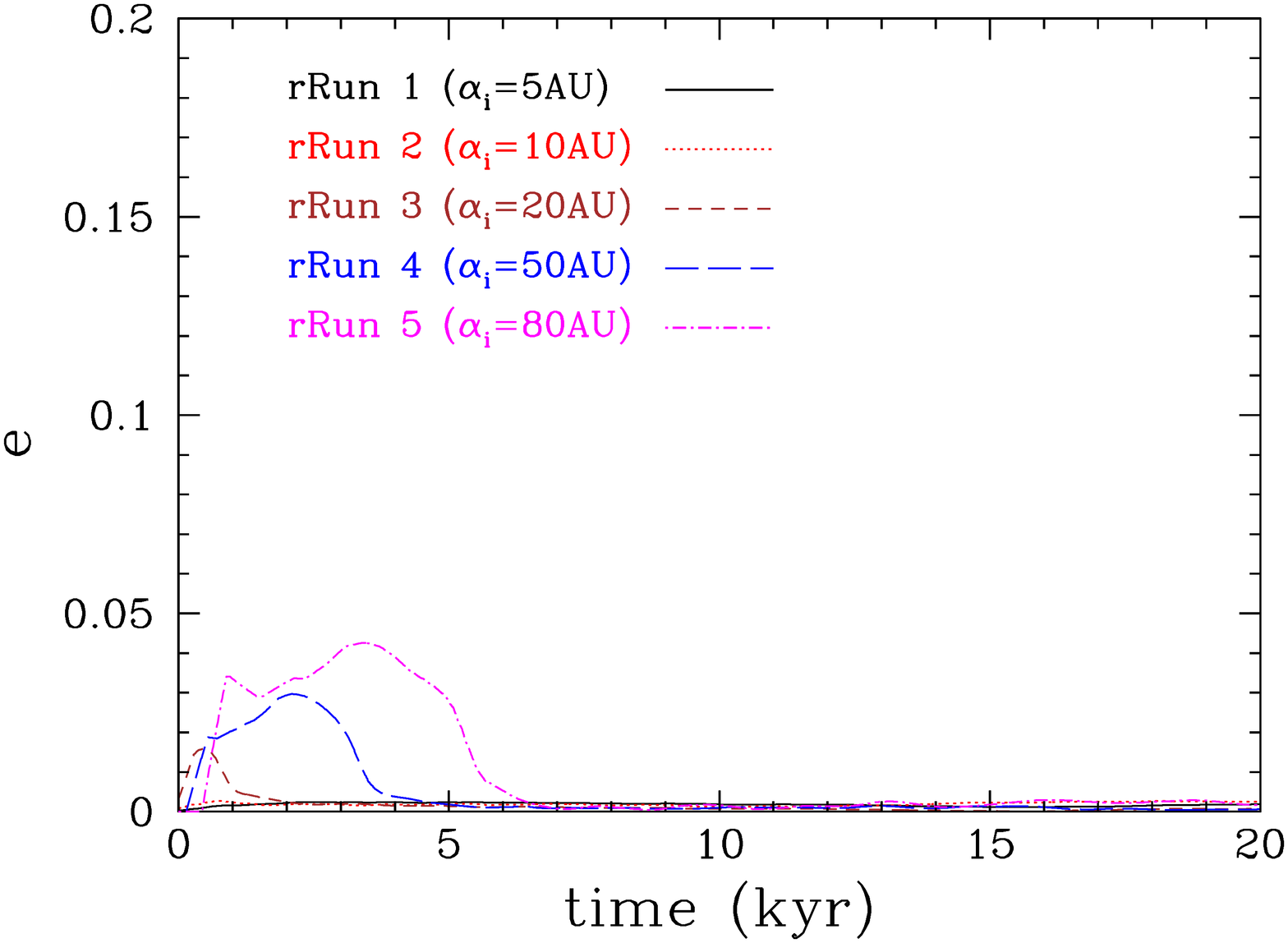}
}
\caption{Eccentricity of a 1-M$_{\rm J}$ protoplanet placed at different radii within a protostellar disc, in the runs where both radiative feedback from the protoplanet and the star are included. The eccentricity increases during the gap opening phase but thereafter quickly dampens and the orbit of the protoplanet becomes circular.}
\label{fig:rrun.pecc}
\centerline{
\includegraphics[height=0.95\columnwidth,angle=-90]{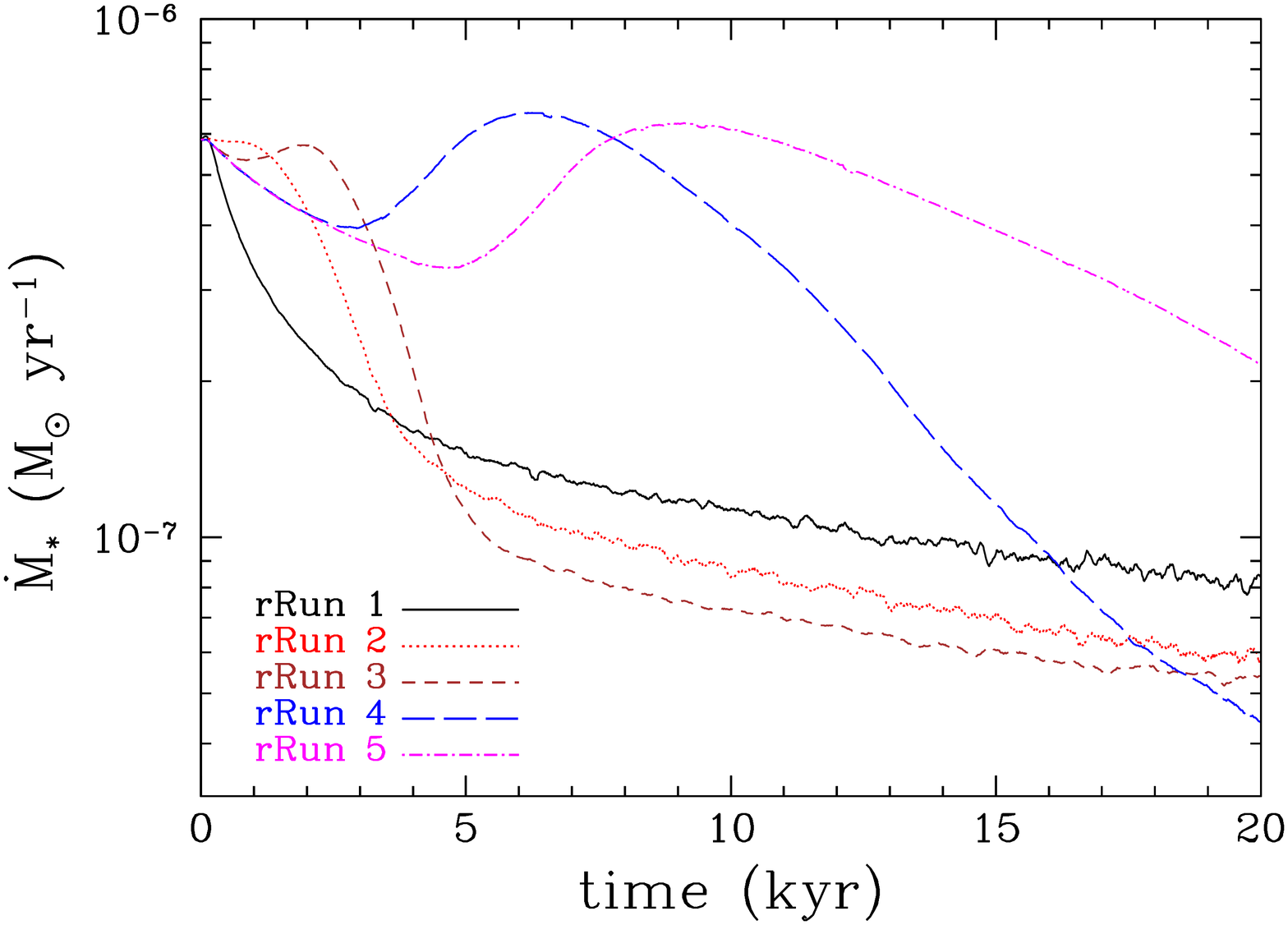}}
\centerline{\includegraphics[height=0.95\columnwidth,angle=-90]{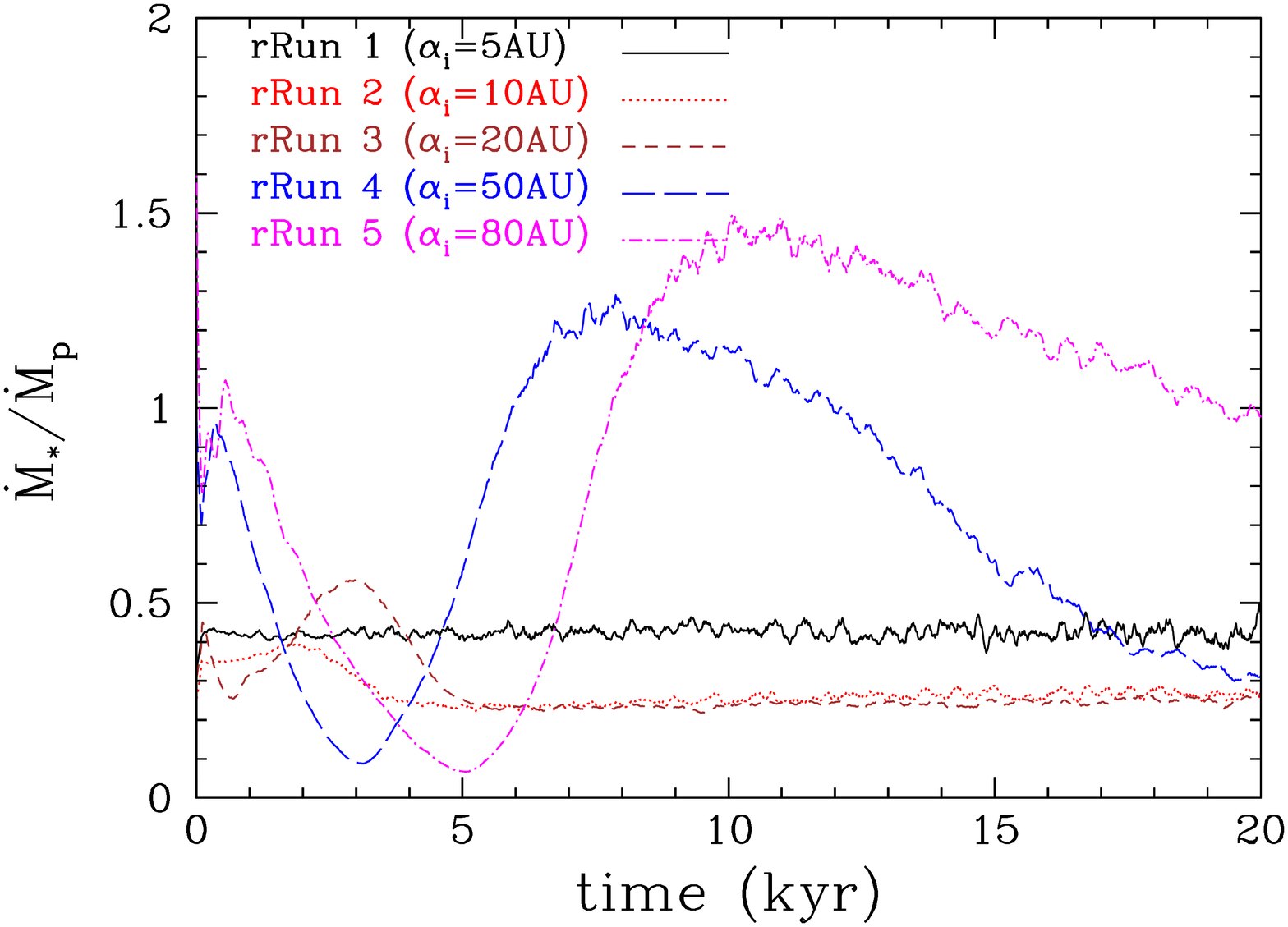}}
\caption{The accretion rate onto the central star (top) and the ratio of the accretion rate onto the star to the accretion rate onto the protoplanet (bottom), in the runs where both radiative feedback from the protoplanet and the star are included. The accretion rate onto the star is lower when the protoplanet is in the inner disc region. }
\label{fig:rrun.paccstar}
\end{figure}
%%%%%%%%%%%%%%%%%%%%%%%%%%%%%%%%%%%%%%%%%%%%%

Similarly to the previous case, we find that interaction with the gravitationally stable gap edges suppresses excessive mass growth so that the protoplanet's mass is near the brown dwarf-planet limit (Figure~\ref{fig:rrun.mass}). Protoplanets that form in the inner disc region find themselves in the inner cavity around the star and their masses are just below the deuterium-burning limit. Protoplanets that form further away from the star go through a longer gap-opening phase as they migrate within the gas-rich disc and they accrete more mass, so that their final mass at the end of the simulation is just above the deuterium burning limit. Due to this longer gap-opening phase, protoplanets in the outer disc region show a strong peak in their accretion rates (Figure~\ref{fig:rrun.pacclum}, top) and high corresponding luminosities, which at the earliest phases of the protoplanets' evolution can reach a few tenths of the solar luminosity (Figure~\ref{fig:rrun.pacclum}, bottom).

The eccentricity of the protoplanet  orbit initially grows during the gap opening phase but thereafter it quickly dampens so that protoplanets at any given radii end up with nearly circular orbits. We find that interactions with a gravitationally stable disc dampen the eccentricity effectively. 

The accretion rate onto the central star is sensitive to the position of the protoplanet in the disc (Figure~\ref{fig:rrun.paccstar}, top). When the protoplanet is in the inner disc region it starves off the young star so that  the  accretion rate onto it is only half the accretion onto the protoplanet (Figure~\ref{fig:rrun.paccstar}, bottom). When the protoplanet is in the outer disc region, the accretion onto the central star is heavily suppressed during the gap opening phase, whereas it is enhanced significantly immediateley afterwards and for a few thousand years (Figure~\ref{fig:rrun.paccstar}, bottom). However,  in all cases the accretion rate onto the protoplanet is higher than the accretion rate onto the star at the end of the simulation.

We conclude that radiative feedback from the protoplanet plays an important role for its orbital evolution and its mass growth only for protoplanets that are formed in outer cold region of a protoplanetary disc. Such protoplanets may have formed by gravitational fragmentation during an early stage of the disc's formation and evolution, while these discs are still relatively massive \cite[e.g.][]{MacFarlane:2017a}.

%% CPD DISCS RRUNS %%%%%%%%%%%%%%%%%%%%%%%%%%%%%%%%%%%%%%%
%\begin{figure}
%\centerline{\includegraphics[width=0.8\columnwidth]{gap_c.eps}}
%\label{fig:gap_c}
%\caption{The region around the protoplanet. Same as in Figure~\ref{fig:gap_a} but for the runs listed in  Table~\ref{tab:rrun}.}
%\centerline{\includegraphics[width=0.8\columnwidth]{cpd_c.eps}}
%\caption{The properties  of the circumplanetary disc in the simulations listed in Table~\ref{tab:rrun} at $t=8$~kyr (parameters same as in Figure~\ref{fig:cpd_a}).}
%\label{fig:cpd_c}
%\centerline{\includegraphics[height=0.8\columnwidth, angle=-90]{rrun.pmhill.eps}}
%\label{fig:hill_c}
%\caption{The mass within the protoplanet's Hill radius as a function of time for the simulations listed in Table~\ref{tab:rrun}.}
%\end{figure}
%%%%%%%%%%%%%%%%%%%%%%%%%%%%%%%%%%%%%%%%%%%%%

 \section{Summary \& Discussion}
 \label{sec:discussion}
 
The final fate of a planet that forms in a protostellar disc is determined by how the two interact with each other. Disc-planet interactions are more critical if the planet forms early-on during its parent disc evolution, while this disc is relatively massive and possibly non-axisymmetric. The effect of the disc is two-fold: (i) it exchanges angular momentum with the planet, to facilitate migration (inward or outward), (ii) it exchanges mass with the planet allowing it to grow \citep[but the opposite may also be possible, see][]{Nayakshin:2017a}.

Here, we presented simulations of the evolution of  a Jupiter-like planet-seed in a relatively-massive disc (0.1~M$_{\sun}$) around a Sun-like star (1~M$_{\sun}$), expanding on the work of \cite{Stamatellos:2015a}. { The disc is massive enough for its gravity to play an important role in the system's evolution, but not massive enough to fragment due to the development of gravitational instabilities ($Q\stackrel{>}{_\sim}1.5$).} We do not restrict the formation mechanism of the planetary seed, although it is rather unlikely that such a protoplanet may have formed by core accretion early-on during the disc's lifetime (i.e. within several $10^4$ yr). On the other hand, such large planet-seeds form naturally fast (on a dynamical timescale) by disc fragmentation. Nevertheless, the results of this paper are independent of the assumed planet formation scenario. 

The evolution of such a planet-seed (protoplanet)  is followed for a relatively short duration corresponding to the disc lifetime (20~kyr; see Appendix~\ref{sec:longrun} for the long term evolution) but even within such a short timescale  disc-planet interactions are important for determining the properties of the protoplanet. It has been suggested that the strong nature of the planet-disc interactions and the inability of the planet to open up a gap in such a massive disc, may lead to rapid inward migration \citep{Michael:2011a, Baruteau:2011a, Malik:2015a} and the demise of the planet as it falls onto its parent star.
The findings of our simulations contradict the results of these  studies as we find that the protoplanet is able to open up a gap in the disc so that its fast inward migration is halted { \citep[as also pointed out in][]{Stamatellos:2015a}}. However, we encounter an alternative problem: the protoplanet grows in mass by accreting gas from the disc,  so that in most cases it becomes a brown dwarf \citep{Stamatellos:2009a,Stamatellos:2011c, Kratter:2010b, Zhu:2012a}.  The final state of the planet depends on where it has formed in the disc, but also on the physics that are at play during its evolution. More specifically, this work demonstrates that radiative heating from the young accreting protoplanet plays an critical role; this radiative feedback is currently not well understood nor constrained \citep{Marleau:2017a,Szulagyi:2017b}. The main results of this work are discussed and placed  in a wider context in the following subsections.

\subsection{Protoplanet mass growth} 
Gas accretion is significant during the initial stages of the protoplanet's evolution, during the gap-opening phase. In most of the simulations presented here, the protoplanet grows to  the planet-brown dwarf mass-limit ($\sim$13~M$_{\rm J}$) within a few thousand years. 
 Once the gap is opened up, accretion occurs mainly from inside the protoplanet's orbit. However, a fraction of it is also accreted from outside the protoplanet's orbit; when the gap edges are gravitationally unstable (low Toomre-Q, i.e $Q\stackrel{<}{_\sim}2$), then an almost equal amount of gas is accreted from outside the protoplanet's orbit. The accretion rate onto the protoplanet is higher  for higher viscosity,  and for more efficient cooling (note that cooling is modulated by the opacity). The role of radiation feedback from the protoplanet is critical as it considerably decreases the accretion rate of gas onto it and therefore its final mass. The effect of radiation feedback from the central star is rather minimal (Appendix~\ref{sec:star_rt}). We also found a dependence of the protoplanet's final mass on its initial position within the disc: a planet that is relatively closer to the central star ($\stackrel{<}{_\sim}20$~AU) quickly opens up a cavity around its parent star and resides near the outer edge of this cavity.  In this case, the star-protoplanet system behaves as a  low mass-ratio binary system attended by a circumbinary disc. In such a system gas flows to the star and the protoplanet only through streams that pass through the Lagrangian points of the system \citep{Artymowicz:1994a,Artymowicz:1996a}, and therefore accretion is slow. These are the only cases (with or without radiative feedback from the protoplanet) that the protoplanet's mass remains within the planetary regime at the end of the simulations.  As the mass loss of the disc due to accretion onto the star and the protoplanet is lower in this case, the disc is expected to live longer, allowing other processes, e.g. photo-evaporation \citep{Alexander:2006a,Owen:2010a} or disc winds \citep{Suzuki:2009b,Suzuki:2010a,Suzuki:2014a,Fromang:2013a,Lesur:2013a,Bai:2013a,Gressel:2015a,Bai:2016a}, to disperse the disc fast enough for  the protoplanet to survive as a planet, not a brown dwarf. Another way to suppress mass growth is  to eject the protoplanet from the disc  (in multiple-planet systems) so that it ends up as a free-floating planet \citep{Li:2015b,Li:2016a,Mercer:2017a}.

\subsection{Protoplanet  Migration}
Disc-planet interactions result in an initial phase of fast inward migration that lasts for a few thousand years, apart from  the runs in which the protoplanet is very close to its parent star ($\stackrel{<}{_\sim}10$~AU); in these runs the protoplanet's orbit changes only slightly as it quickly opens up a cavity around the star. The migration timescale for this initial phase is $\sim 10^4$~yr, i.e. similar to the Type I migration timescale established for planet migration in low-mass discs. This is consistent with previous studies \citep{Michael:2011a, Baruteau:2011a, Malik:2015a}. However, we find  that the protoplanet is always able to open up a gap in the disc, and thereafter the migration pattern diverges from that in previous studies. 

We find that if the gap edges are gravitationally unstable (low Toomre-Q, i.e $Q\stackrel{<}{_\sim}2$), then the protoplanet starts migrating outwards on a timescale $\sim10^5$~yr, as a high fraction of the gas it accretes (almost about 50\%) is  higher angular momentum gas from outside its orbit.  If the gap edges  are gravitationally stable (i.e. in the runs with protoplanet radiative feedback, or when the protoplanet is closer to the star, in a hotter disc region) then the migration continues to be inwards but on a longer, Type II migration timescale, i.e. $\sim 10^5$~yr, with the planet accreting mainly lower-angular momentum gas from inside its orbit. 

Contrary to previous studies we have allowed the protoplanet to grown in mass, and, critically, we have used a more sophisticated treatment of the radiative transfer rather than the commonly used $\beta$-cooling approximation \citep{Baruteau:2011a, Malik:2015a}. The method we use allows the gas to modulate its cooling/heating depending on its properties (density, temperature); this enables gap opening and facilitates enhanced mass growth of the circumplanetary disc and of the planet.

 In the present calculations this ``accretion from outside" that is responsible for initiating outward migration is driven by the gravitational instability in the outer disc. However, it is possible this may be promoted by any mechanism that makes large amplitude fluctuations in the outer disc, such as the magneto-rotational instability \citep[MRI, but see][]{Gressel:2012a} or other instabilities.

\subsection{Protoplanet Eccentricity}
The eccentricity of a protoplanet grows up to 0.05 during the initial gap opening phase. Thereafter, two patterns are seen:  protoplanets that reside within a gap with gravitationally unstable gap edges show  further eccentricity growth up to $0.1-0.2$, whereas the eccentricity of protoplanets within a gap with stable edges dampens, making their orbits nearly circular.  The eccentricity growth is not monotonic but shows a stochastic pattern, characteristic of both periodic and random interactions with condensations present near the gap edges. In the long term runs ($10^5$~yr) we see further eccentricity growth up to $\sim0.3$ (see Appendix~\ref{sec:longrun}). Almost in all cases that  we examined here we see a minimum eccentricity growth of a few 0.01 which may provide the seed for further eccentricity growth \citep{Goldreich:2003a,Duffell:2015a}. Therefore, early protoplanet interactions with a massive, but gravitationally stable disc may  explain in some cases the high eccentricities of giant planets in single-planet systems  \citep{Wright:2011a, Dunhill:2018a}. On the other hand, we see that protoplanets ending up in the inner disc region ($\stackrel{<}{_\sim}20$~AU), tend to have circular orbits.

\subsection{Effect on the parent star}
We find that in most cases the accretion rate onto the protoplanet is higher or at comparable to the accretion rate onto the star. If the protoplanet is close enough to the central star, it creates a cavity, resides within it (at its outer edge), and seemingly starves off the central star from gas, accreting most of the incoming gas. This stage shows similarities with the transition disc phase \citep{Espaillat:2014a}.

\subsection{Observability of protoplanets}

The luminosity of a protoplanet can be quite high during the initial stages of its evolution, as the accretion rate on it is relatively high (typically $\sim10^{-3} {\rm M_J\ yr^{-1}}$, but could be up to 10 times higher during the gap opening phase).  Assuming a relatively large radius for the protoplanet of $R_{\rm acc}=3~{\rm R}_{\sun}$ (which is expected for  a protoplanet that has formed by disc fragmentation), we find a typical accretion luminosity of a few $0.1~L_{\sun}$ \citep[see also][]{Inutsuka:2010a}. Combined with the fact that the protoplanet starves off its central star from gas, it may make its detection easier. 

The high accretion phase lasts only for a few $10^3-10^4$~yr and therefore only  a small fraction of transition discs (assuming that they all are due to planetary companions) are expected to be seen in this high-luminosity phase. If transition discs live for 1~Myr, then, from timescale arguments, we suggest that only a small percentage, $0.01\%$ to $0.1\%$, of transition discs may have observable, high-luminosity protoplanets (or brown dwarfs).

\subsection{Circumplanetary discs}
Protoplanets are attended by circumplanetary discs that are fed with gas from the disc. We find that these discs are hotter than their surroundings and therefore easier to observe.  Their temperatures are a few hundred Kelvin and they are hotter when radiative feedback from the protoplanet is taken into account. If the protoplanet migrates outwards its circumplanetary disc becomes more massive ($\sim1~{\rm M_J}$) and larger $\sim10$~AU.

A protoplanet could be  detected indirectly through its circumplanetary disc  \citep{Szulagyi:2017c}. The circumplanetary disc of a young luminous  protoplanet is hotter than in the case of non-luminous protoplanet (see Figure~\ref{fig:cpd_a}), therefore it would emit significant long-wavelength radiation even if it is not particularly dense. This long-wavelength radiation would escape mostly unattenuated from the system, whereas short-wavelength radiation that is emitted from the protoplanet may be significantly attenuated, especially if the protoplanet orbits  a young embedded protostar (e.g. in a Class 0 object). 

\subsection{Conclusions}

Our study shows that the final fate of a Jupiter-like planet-seed formed early-on during the life of a protostellar disc depends both on the initial properties of the seed (e.g. its initial orbit), and the physical processes that are at play (e.g. the role of the protoplanet's accretion luminosity). The protoplanet may end up as a massive giant planet on a circular orbit close to its parent star or as a low- or high- mass brown dwarf on an eccentric wide orbit. 

The parameter space has not been fully explored. It is expected that the initial disc mass will  play an important role as it will determine to a wide extent whether the disc is close to being gravitationally unstable or not. Additionally, the initial mass of the protoplanet is important; lower-mass planet-seeds (i.e. a few tenths of the mass of Jupiter)  will grow in mass slower and may have a better chance to survive as planets \citep{Nayakshin:2017a,Malik:2015a}. 

We have assumed that a protoplanet has fully formed (i.e. it is a bound object) at the beginning of the simulation. However, if the planet has formed by disc fragmentation then the formation timescale of the clump and how this evolves to a bound object should be taken into account \citep{Boley:2010b,Zhu:2012a}. Previous studies have shown that a large fraction of these pre-protoplanet clumps may be disrupted, with only a few of them surviving \citep{Boley:2010b,Tsukamoto:2015a,Hall:2017a}.

{ The protoplanets is our simulations are represented by sink particles that accrete gas as this approaches within $0.1$~AU from them. The use of sinks in hydrodynamic simulations is needed to avoid small timesteps but it could lead to enhanced gas accretion that may depend on how well the circumplanetary disc is resolved. The study of how the accretion rate onto the protoplanet varies with resolution and with respect to the hydrodynamic method used (e.g. SPH vs grid-based) is important but outside the scope of this paper. However, we note that the accretion rates onto the protoplanets that we obtain in this work are similarly high  (within a factor of a few) as the  accretion rates obtained from previous studies \citep{DAngelo:2008a,Ayliffe:2009a,Zhu:2012a,Gressel:2013b}, which use different numerical methods and disc-planet initial setups.}

The physical processes involved are also important. We have shown that the disc thermodynamics, i.e. how fast the disc heats and cools, affects the accretion rate onto the protoplanet \citep[see also][]{Nayakshin:2017b}, its ability to open up a gap in the disc, and its final mass and orbital radius. The $\beta$-cooling approximation \cite[e.g.][]{Baruteau:2011a}, in which the cooling rate is proportional to the local orbital period, does not properly capture the thermodynamics of the circumstellar and circumplanetary disc,  and may underestimate the mass flow onto the protoplanet and its ability to open up a gap. We also found that the protoplanet accretion luminosity is significant in shaping the properties of the disc in the vicinity of the planet and may stabilize the edges of the gap in which the planet resides, altering its migration pattern. 

Therefore, there are many uncertainties regarding the final fate of a young protoplanet that needs to be taken into account in population synthesis models \citep[e.g.][]{Mordasini:2009a,Forgan:2013a, Nayakshin:2017a} in order for numerical results on planet formation at an early phase during the protostellar disc evolution to be compared with the observed properties of exoplanets. The observational evidence  that only 1-10\% of star host gas giant planets on wide orbits \citep{Brandt:2014a,Galicher:2016a,Vigan:2017a}, corresponds  to the final outcome of the planet formation process which does not necessarily reflect the properties of newly formed planet-seeds in young protostellar discs.
 
\section*{Acknowledgements}

We thank the referee for carefully reading the original manuscript and for providing many suggestions that have improved the paper. DS thanks Sergei Nayakshin, Alex Dunhill and Anthony Mercer for useful discussions. Simulations were performed using the UCLAN HPC facility and the COSMOS Shared Memory system at DAMTP, University of Cambridge operated on behalf of the STFC DiRAC HPC Facility. This equipment is funded by BIS National E-infrastructure capital grant ST/J005673/1 and STFC grants ST/H008586/1, ST/K00333X/1. {\sc Seren} has been developed and maintained by David Hubber, who we thank for his help. Surface density plots were produced using {\sc splash} \citep{Price:2007b}. DS acknowledge support from STFC Grant ST/M000877/1. This work was supported from a Royal Society-Daiwa Foundation International Exchanges award between UCLAN and Nagoya University.

\appendix

\section{Migration timescales and velocities for each set of simulations}

In this Section we present the migration timescales and the corresponding migration velocities for the set of runs in Tables~\ref{tab:standardrun},\ref{tab:srun}, and \ref{tab:rrun}.

\label{sec:tmig.vmig.tables}

\begin{table}
\caption{Migration timescales for the runs in Table~\ref{tab:standardrun} calculated at the times indicated in the table. Negative values correspond to outward migration.}
\label{tab:standard_migration}
\centering
\begin{tabular}{@{}cccccc} \hline
\noalign{\smallskip}
  & \multicolumn{1}{c}{0.7 kyr}& \multicolumn{1}{c}{1.5 kyr} & \multicolumn{1}{c}{2.5 kyr} & \multicolumn{1}{c}{5 kyr } &\multicolumn{1}{c}{18 kyr } \\
\noalign{\smallskip} 
\cline{2-6}
\noalign{\smallskip} 
& \multicolumn{5}{c}{$\tau_{\rm mig}$ (kyr)} \\
\noalign{\smallskip} 
\hline\hline
\noalign{\smallskip}
Run1 & 14 &0.7  & 44 & -140   &-220 \\
Run2 &  11 & 8  &2.4  & 38  & 150 \\
Run3 &  17 &  6.8  & 5.6&-56  &-112  \\
Run4 &  15 &11  & 0.5  &-71& -270 \\
Run5 &    12 & 11  &-31  & -27  & -68 \\
\noalign{\smallskip}
\hline
\end{tabular}
\caption{Migration velocities for the runs in Table~\ref{tab:standardrun}  calculated at the times indicated in the table. Negative values correspond to outward migration.}
\label{tab:standard_vel}
\centering
\begin{tabular}{@{}cccccc} \hline
\noalign{\smallskip}
\multicolumn{1}{c}{} & \multicolumn{1}{c}{0.7 kyr}& \multicolumn{1}{c}{1.5 kyr} & \multicolumn{1}{c}{2.5 kyr} & \multicolumn{1}{c}{5 kyr } &\multicolumn{1}{c}{18 kyr } \\
\noalign{\smallskip} 
\cline{2-6}
\noalign{\smallskip} 
& \multicolumn{5}{c}{ $v_{\rm mig}$ (AU kyr$^{-1}$) } \\
\noalign{\smallskip} 
\hline\hline
\noalign{\smallskip}
Run1 & 3.4  & 6.4  & 0.8 & -0.26   & -0.18\\
Run2 & 4.4 &  5.2  &15 & 0.5  &0.12\\
Run3 &   2.9& 6.7 &6.0  &-0.59  &-0.3 \\
Run4 &   3.2&4.2  &7.7  &-0.5  & -0.13\\
Run5 &   4.1  & 4.2  &-1.4 & -1.6  &-0.8\\
\noalign{\smallskip}
\hline
\end{tabular}
\caption{Migration timescales for the runs in Table~\ref{tab:srun} . These are calculated at the times indicated in the table. Negative values correspond to outward migration.}
\label{tab:srun_migration}
\centering
\begin{tabular}{@{}cccccc} \hline
\noalign{\smallskip}
  & \multicolumn{1}{c}{0.7 kyr}& \multicolumn{1}{c}{1.5 kyr} & \multicolumn{1}{c}{2.5 kyr} & \multicolumn{1}{c}{5 kyr } &\multicolumn{1}{c}{18 kyr } \\
\noalign{\smallskip} 
\cline{2-6}
\noalign{\smallskip} 
& \multicolumn{5}{c}{$\tau_{\rm mig}$ (kyr)} \\
\noalign{\smallskip} 
\hline\hline
\noalign{\smallskip}
sRun1 & -23 			&-30 			& -43 		& -70	 	&-140 	\\
sRun2 &  15 			& 16 			& 43				& $2.0\times 10^3$	& $2.3\times 10^3$	\\
sRun3 &  2.6 			& 6			& 10		&150		&400 	\\
sRun4 &  12 			&11 				& -30		&-27 		& -70 	\\
sRun5 &    12			&5.5 			&34 		& 41 		& -210 	\\
\noalign{\smallskip}
\hline
\end{tabular}
\caption{Migration velocities for the runs in Table~\ref{tab:srun} . These are calculated at the times indicated in the table. Negative values correspond to outward migration.}
\label{tab:srun_vel}
\centering
\begin{tabular}{@{}cccccc} \hline
\noalign{\smallskip}
\multicolumn{1}{c}{} & \multicolumn{1}{c}{0.7 kyr}& \multicolumn{1}{c}{1.5 kyr} & \multicolumn{1}{c}{2.5 kyr} & \multicolumn{1}{c}{5 kyr } &\multicolumn{1}{c}{18 kyr } \\
\noalign{\smallskip} 
\cline{2-6}
\noalign{\smallskip} 
& \multicolumn{5}{c}{ $v_{\rm mig}$ (AU kyr$^{-1}$) } \\
\noalign{\smallskip} 
\hline\hline
\noalign{\smallskip}
sRun1  	& -0.2			& -0.17 			& -0.13		& -0.08  	 	& -0.05\\
sRun2  	& 0.7 		 	& 0.6		&0.21 			& 0.005 		&0.004\\
sRun3 	& 6.7			& 2.2			&1.2 		&0.07	 	&0.025 \\
sRun4  	& 4.1			& 4.2 			&-1.4		&-1.6 		& -0.8\\
sRun5 	& 7.0 			& 14			 	& 2.0		& 1.7	 	&-0.3\\
\noalign{\smallskip}
\hline
\end{tabular}
\end{table}

\begin{table}
\caption{Migration timescales for the runs in Table~\ref{tab:rrun} . These are calculated at the times indicated in the table. Negative values correspond to outward migration.}
\label{tab:rrun_migration}
\centering
\begin{tabular}{@{}cccccc} \hline
\noalign{\smallskip}
  & \multicolumn{1}{c}{0.7 kyr}& \multicolumn{1}{c}{1.5 kyr} & \multicolumn{1}{c}{2.5 kyr} & \multicolumn{1}{c}{5 kyr } &\multicolumn{1}{c}{18 kyr } \\
\noalign{\smallskip} 
\cline{2-6}
\noalign{\smallskip} 
& \multicolumn{5}{c}{$\tau_{\rm mig}$ (kyr)} \\
\noalign{\smallskip} 
\hline\hline
\noalign{\smallskip}
rRun1 & -23  &-29 &-41 & -70   &-180\\
rRun2 &  19  & 19 &40  &  880  & $1.4\times10^4$ \\
rRun3 &  2.4 & 6.3  & 13  &140 &710 \\
rRun4 &  11 &8.7  & 2.6 &29  & 190 \\
rRun5 &    12  & 5.1  & 4.5  & 4.¤ & 110\\
\noalign{\smallskip}
\hline
\end{tabular}
\caption{Migration velocities for the runs in Table~\ref{tab:rrun} . These are calculated at the times indicated in the table. Negative values correspond to outward migration.}
\label{tab:rrun_vel}
\centering
\begin{tabular}{@{}cccccc} \hline
\noalign{\smallskip}
\multicolumn{1}{c}{} & \multicolumn{1}{c}{0.7 kyr}& \multicolumn{1}{c}{1.5 kyr} & \multicolumn{1}{c}{2.5 kyr} & \multicolumn{1}{c}{5 kyr } &\multicolumn{1}{c}{18 kyr } \\
\noalign{\smallskip} 
\cline{2-6}
\noalign{\smallskip} 
& \multicolumn{5}{c}{ $v_{\rm mig}$ (AU kyr$^{-1}$) } \\
\noalign{\smallskip} 
\hline\hline
\noalign{\smallskip}
rRun1  & -0.21  & -0.18  & -0.13 & -0.008   & -0.04\\
rRun2  & 0.5  & 0.5  &0.2 & 0.01 &$6\times10^4$\\
rRun3  & 6.9 & 2.1 &0.9  &0.09  &0.014 \\
rRun4  & 4.5  &5.0 &13  &0.8  & 0.1\\
rRun5 & 6.6  & 15.0  & 15.0  & 8.4  &0.2\\
\noalign{\smallskip}
\hline
\end{tabular}
\end{table}

\section{The effect of radiation from the central star on the direction of migration}
\label{sec:star_rt}

The effect of radiative feedback from the protoplanet is important only when the energy that is fed back to the disc alters its dynamical state. In this section we examine how important radiative feedback from the protoplanet is, when the central star is also heating the disc.

%%%%%%%%%%%%%%%%%%%%%%%%%%%%%%%%%%%%%%%%%%%%%
\begin{figure}
\centerline{\includegraphics[height=0.97\columnwidth,angle=-90]{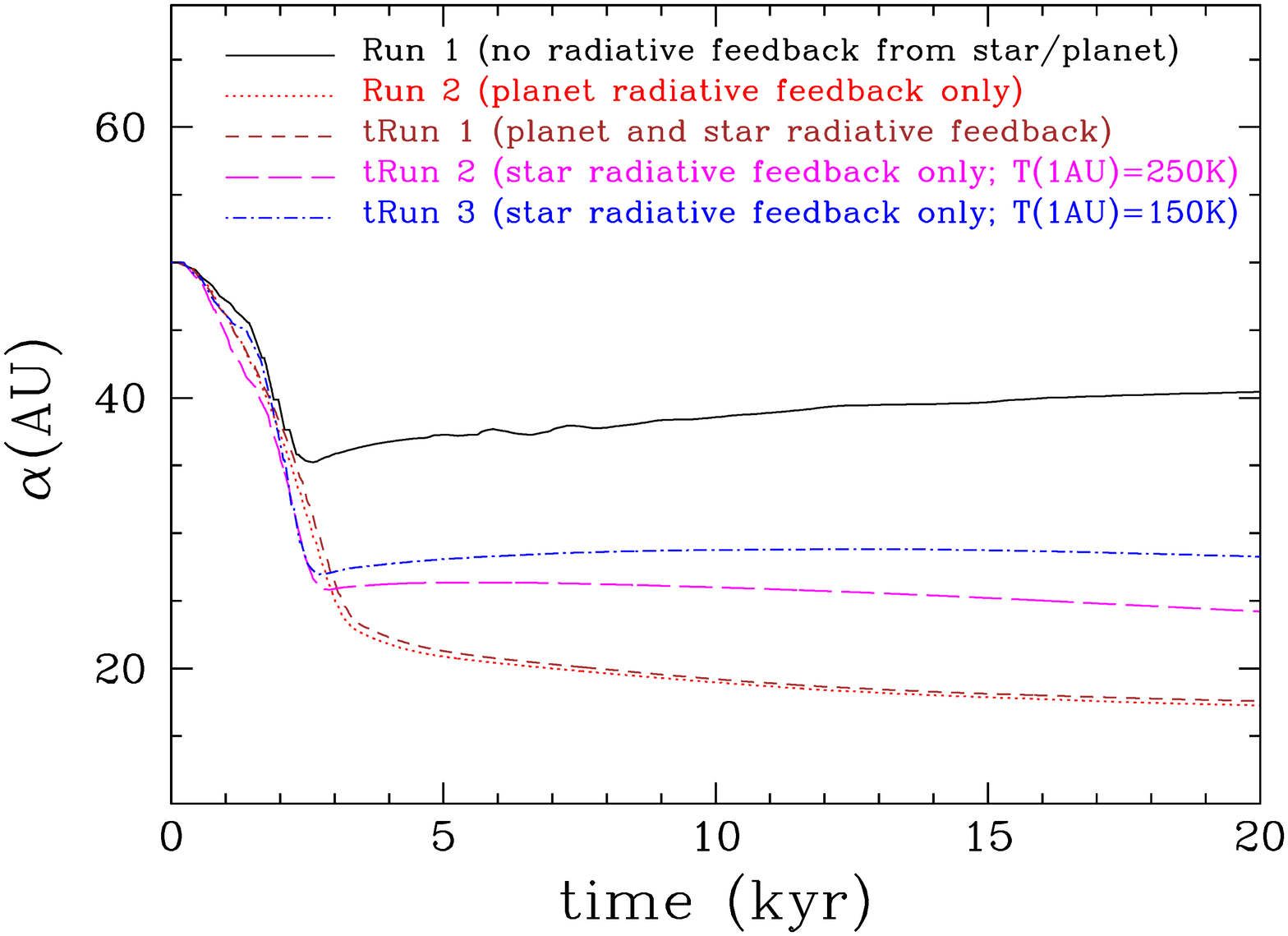}}
\centerline{\includegraphics[height=0.97\columnwidth,angle=-90]{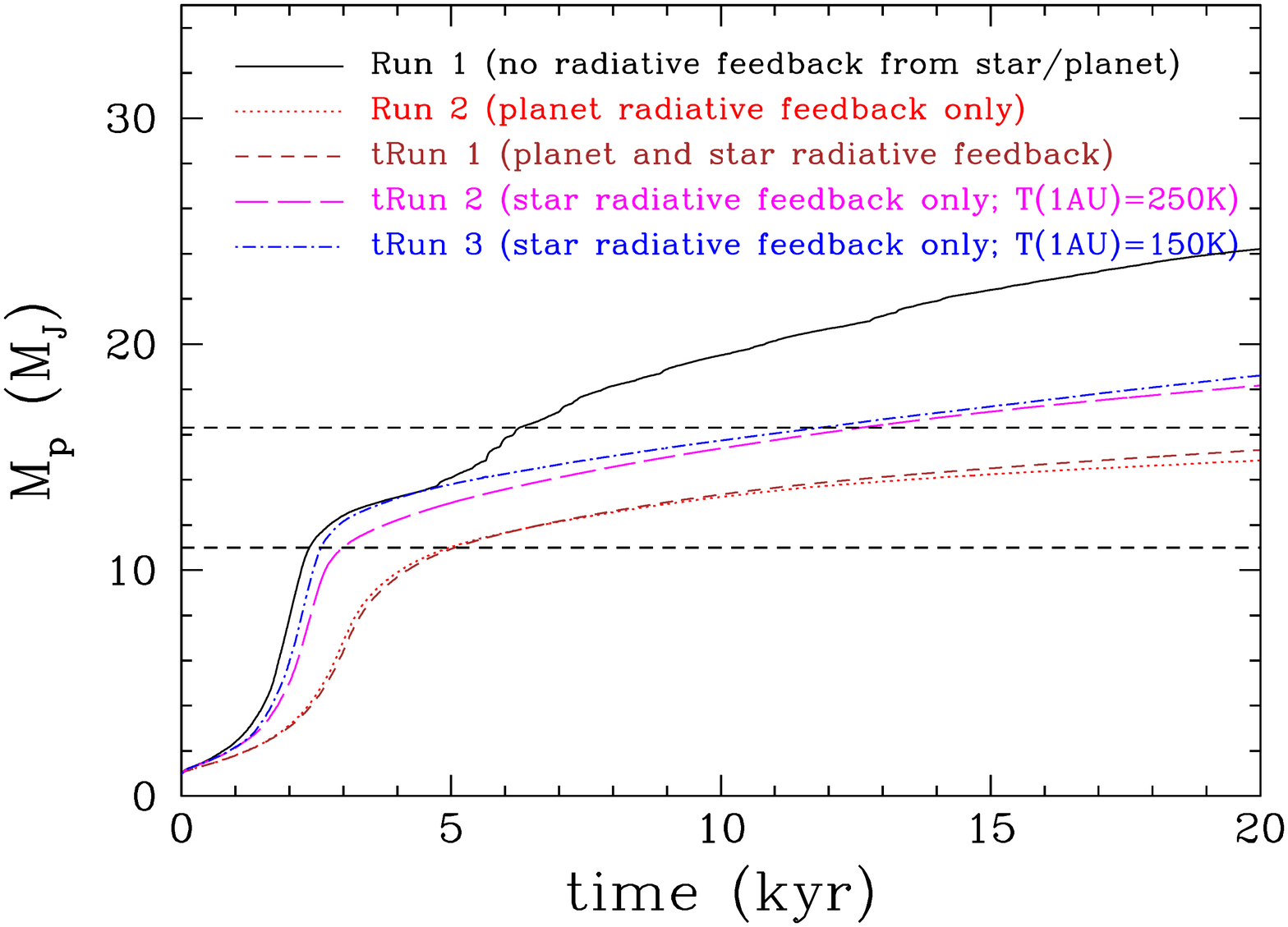}}
\centerline{\includegraphics[height=0.97\columnwidth,angle=-90]{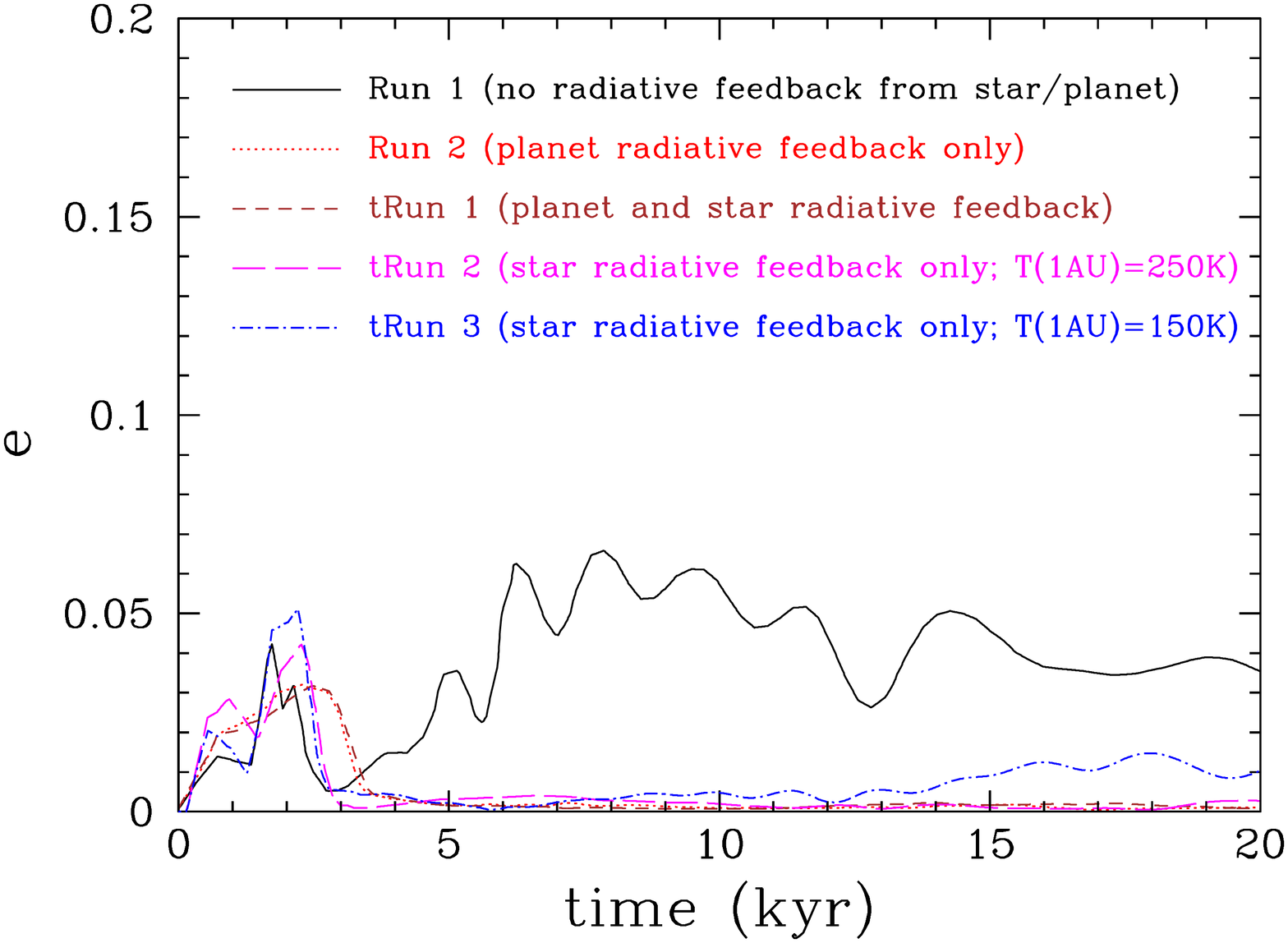}}
\caption{The mass (top), the semi-major axis (middle), and the eccentricity (bottom) of a protoplanet embedded in a disc in 5 simulations with different types of radiative feedback from the star and the protoplanet, as marked on the graph (see text for details).  Feedback from the protoplanet dominates over the stellar feedback.}
\label{fig:truns}
\end{figure}
%%%%%%%%%%%%%%%%%%%%%%%%%%%%%%%%%%%%%%%%%%%%%

We compare 5 simulations simulations of an initially 1-M$_{\rm J}$ protoplanet placed at 50~AU from the central star in  a 0.1-M$_{\odot}$ disc: (i) without any radiative feedback (Run1; same as in Section~\ref{sec:runs}), (ii) with protoplanet radiative feedback only (Run2; same as in Section~\ref{sec:runs}), (iiii) with protoplanet and star radiative feedback (tRun1; T(1AU)=250~K), and  (iv) only with stellar radiative feedback (tRun2 with T(1AU)=250~K; tRun3 with  T(1AU)=150~K).  The protoplanet's orbital radius, mass, eccentricity are shown in Figure~\ref{fig:truns}. 

We see by comparing  the runs with stellar feedback (tRun 2, tRun3) with the run without any feedback  (Run1) that the  protoplanet needs more time to open up a gap because the disc is hotter. Therefore the protoplanet migrates further inward in these runs. When the stellar feedback is not high enough (tRun 3) the protoplanet, after opening up the gap, migrates outwards, whereas when the feedback is stronger (tRun 2) the migration continues inwards but at a much longer timescale. In any case the mass growth of the protoplanet is slower when the stellar feedback is included (Figure~\ref{fig:truns}, middle). The eccentricity of the protoplanet is lower (Figure~\ref{fig:truns}, middle), but if the feedback is not strong enough there is a small eccentricity growth (tRun3).

We also see that there is almost no difference in the evolution of the protoplanet when stellar radiative heating is added on top of the protoplanet radiative heating (compare Run2 and tRun1 in Figure~\ref{fig:truns}.) The effect of the protoplanet feedback is to heat and therefore stabilize the gap edges, so that  the feedback from the central star has no additional effect. We conclude that radiative heating from the protoplanet plays a more critical role in the protoplanet's evolution  than heating from the central star

\section{Long-term protoplanet evolution}
\label{sec:longrun}

The simulations presented so far have followed the evolution of the protoplanet in the disc only for 20~kyr, i.e. for a relatively short duration compared to the estimated disc lifetimes (a few Myr). In this section, we present simulations in which we follow 3 representative simulations for much longer, i.e. 100~kyr (see Figure~\ref{fig:longrun}). The details of the simulations are shown in the graphs. These simulations are computationally time-consuming and cannot be performed for all the runs presented in this paper.

In the run with protoplanet radiative feedback (Figure~\ref{fig:longrun}; blue lines) the inward migration continues at a very slow rate and the protoplanet eventually ends up at an orbit with a semi-major axis of about 10~AU from the central star. Its mass growth also continues and at the end of the simulation it has become a low-mass brown dwarf (mass $\sim25~{\rm M_J}$).  There is no eccentricity growth in this run and the protoplanets orbit remain circular, as radiative feedback from the protoplanet stabilises the gap edges. The outcome of this simulation is a low-mass brown dwarf at a close-orbit near its star.

In the two runs without radiative feedback from the protoplanet (Figure~\ref{fig:longrun}; black and red lines), the protoplanet continues to migrate outwards at a slow pace for the run with the \cite{Bell:1994a} opacities (black line) and at a faster pace  for the run with the \cite{Semenov:2003a} opacities (which are generally larger). In the latter case, the protoplanet has reached a semi-major axis of $\sim90$~AU at the end of the simulation. However, in both runs the protoplanet has grown considerably in mass to become a mid to high-mass brown dwarf (mass $\sim40-55~{\rm M_J}$). The objects in the two runs sustain  significant eccentricity growth reaching $e\sim 0.3$ at the end of the simulation. The outcomes of these two simulations are high-mass brown dwarfs on wide, eccentric orbits around their stars. Therefore, for a protoplanet to end up as a planet it either needs to accrete at a much lower rate or the disc needs to dissipate within a relatively short timescale.

%%%%%%%%%%%%%%%%%%%%%%%%%%%%%%%%%%%%%%%%%%%%%
\begin{figure}
\centerline{\includegraphics[height=0.95\columnwidth,angle=-90]{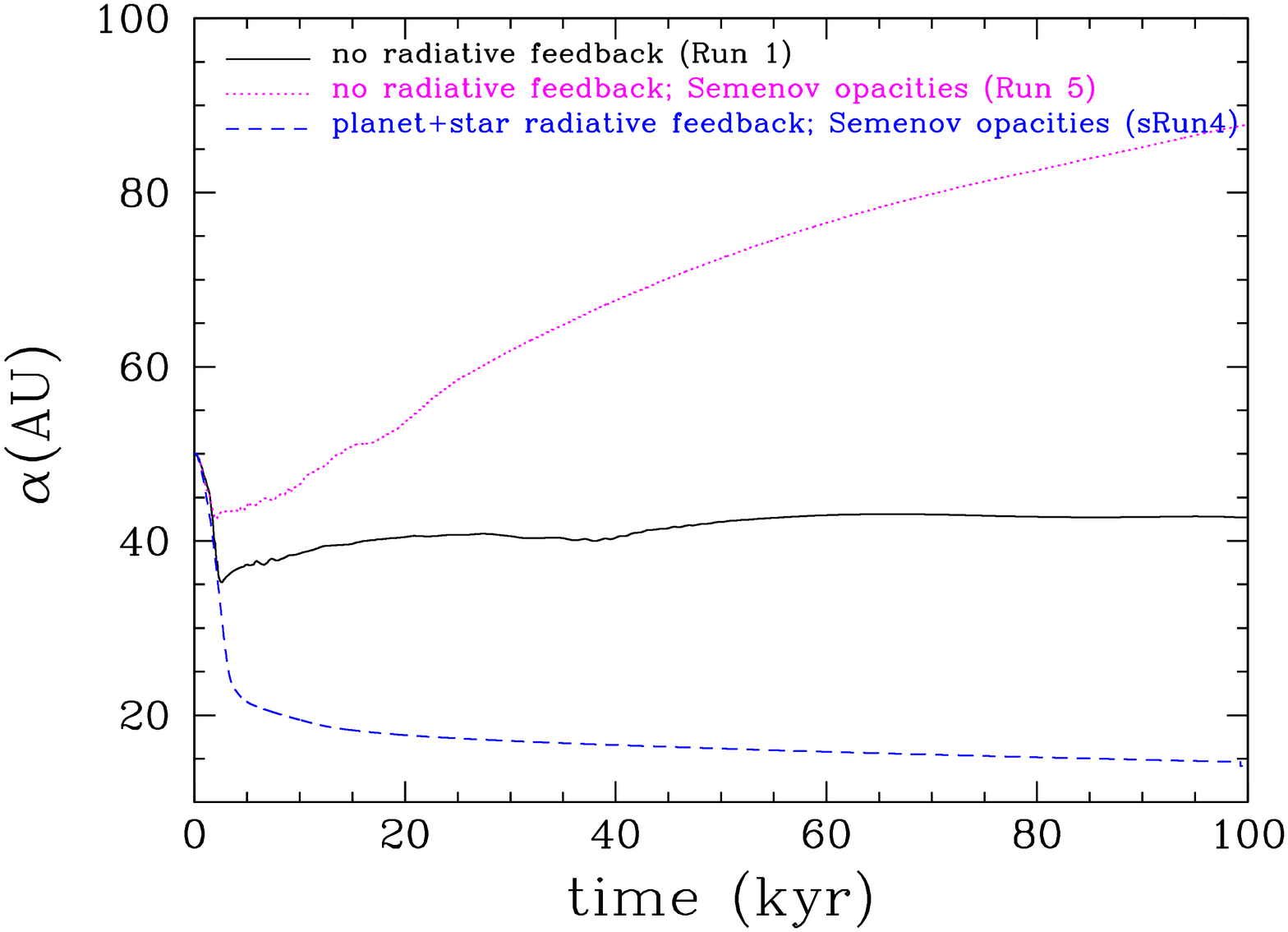}}
\centerline{\includegraphics[height=0.95\columnwidth,angle=-90]{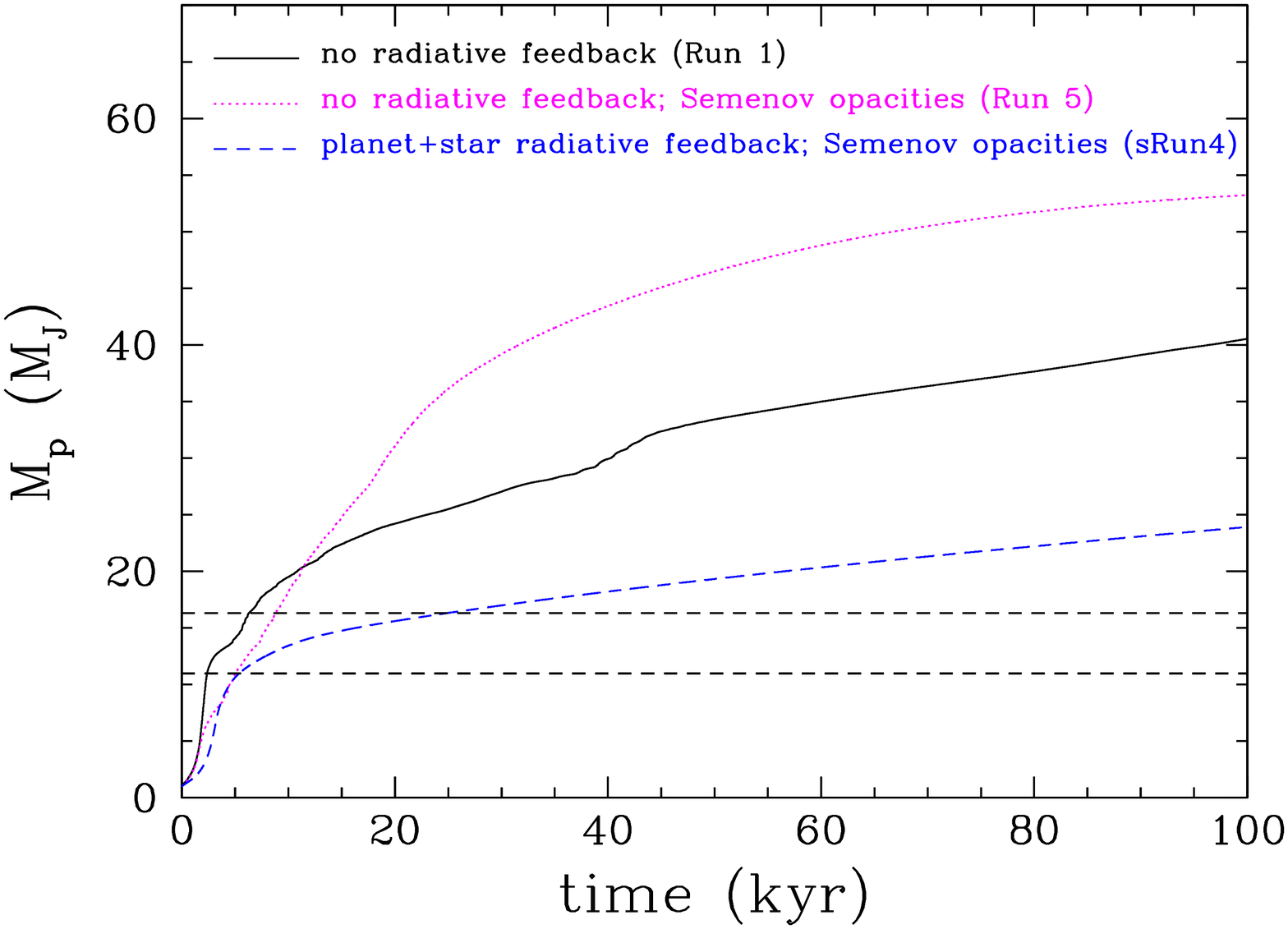}}
\centerline{\includegraphics[height=0.95\columnwidth,angle=-90]{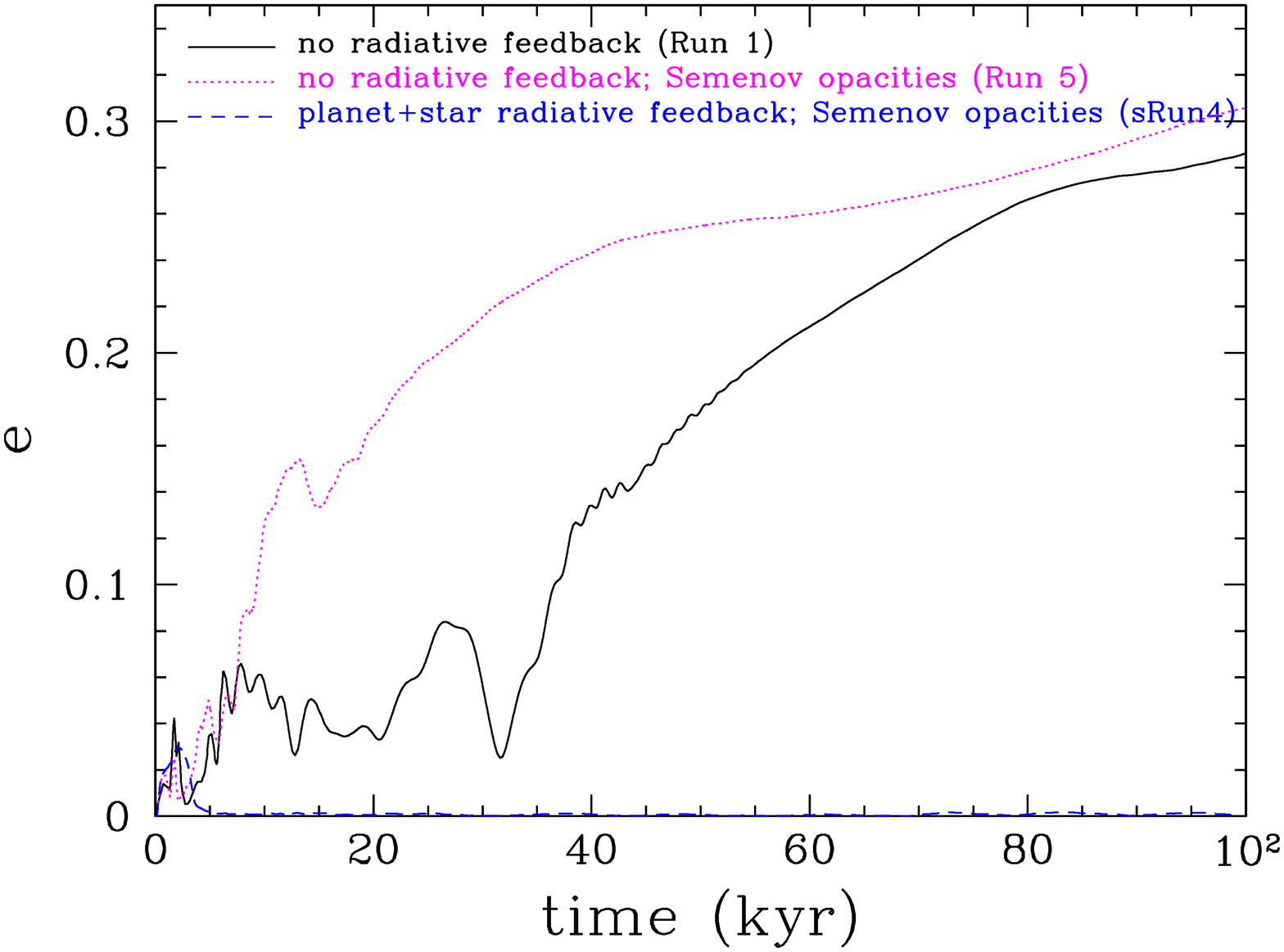}}
\caption{The long term evolution of a protoplanet in a disc for 4 representative simulations. Semi-major axis (top), protoplanet mass (middle),  and eccentricity (bottom) are plotted against time. }
\label{fig:longrun}
\end{figure}
%%%%%%%%%%%%%%%%%%%%%%%%%%%%%%%%%%%%%%%%%%%%%

\bibliography{/Users/dimitris/bibliography}{}

% Don't change these lines
\bsp	% typesetting comment
\label{lastpage}
\end{document}